\documentclass[11pt]{article}
\usepackage{epsfig}
\usepackage{amssymb}
\usepackage{amsbsy}
\newcommand{\sect}[1]{ \section{#1} \setcounter{equation}{0} }

\newcommand{\pslash}{p \! \! \! /} 
\newcommand{\qslash}{q \! \! \! /}

\newcommand{\half}{\mbox{\small{$\frac{1}{2}$}}}

\newcommand{\MSbar}{\overline{\mbox{MS}}} 
\newcommand{\MSbars}{\overline{\mbox{\footnotesize{MS}}}}

\newcommand{\MOMg}{\mbox{MOMg}}
\newcommand{\MOMgs}{\mbox{\footnotesize{MOMg}}}
\newcommand{\MOMh}{\mbox{MOMh}}
\newcommand{\MOMhs}{\mbox{\footnotesize{MOMh}}}
\newcommand{\MOMq}{\mbox{MOMq}}
\newcommand{\MOMqs}{\mbox{\footnotesize{MOMq}}}

\newcommand{\MOMmgs}{\mbox{\footnotesize{MOMmg}}}

\newcommand{\MOMmhs}{\mbox{\footnotesize{MOMmh}}}

\newcommand{\MOMmqs}{\mbox{\footnotesize{MOMmq}}}
\newcommand{\iMOMx}{\mbox{iMOMi}}
\newcommand{\iMOMxs}{\mbox{\footnotesize{iMOMi}}}
\newcommand{\iMOMg}{\mbox{iMOMg}}
\newcommand{\iMOMgs}{\mbox{\footnotesize{iMOMg}}}
\newcommand{\iMOMh}{\mbox{iMOMh}}
\newcommand{\iMOMhs}{\mbox{\footnotesize{iMOMh}}}
\newcommand{\iMOMq}{\mbox{iMOMq}}
\newcommand{\iMOMqs}{\mbox{\footnotesize{iMOMq}}}
\newcommand{\iMOMmx}{\mbox{iMOMmi}}
\newcommand{\iMOMmxs}{\mbox{\footnotesize{iMOMmi}}}
\newcommand{\iMOMmg}{\mbox{iMOMmg}}
\newcommand{\iMOMmgs}{\mbox{\footnotesize{iMOMmg}}}
\newcommand{\iMOMmh}{\mbox{iMOMmh}}
\newcommand{\iMOMmhs}{\mbox{\footnotesize{iMOMmh}}}
\newcommand{\iMOMmq}{\mbox{iMOMmq}}
\newcommand{\iMOMmqs}{\mbox{\footnotesize{iMOMmq}}}

\newcommand{\Nc}{N_{\!c}}
\newcommand{\Nf}{N_{\!f}}
\newcommand{\NF}{N_{\!F}}
\newcommand{\NA}{N_{\!A}}
\newcommand{\Nda}{N^d_{\!A}}
\newcommand{\Noda}{N^o_{\!A}}

\begin{document}
\title{Renormalization of QCD in the interpolating momentum subtraction scheme
at three loops}
\author{J.A. Gracey \& R.M. Simms, \\ Theoretical Physics Division, \\ 
Department of Mathematical Sciences, \\ University of Liverpool, \\ P.O. Box 
147, \\ Liverpool, \\ L69 3BX, \\ United Kingdom.} 
\date{} 
\maketitle 

\vspace{5cm} 
\noindent 
{\bf Abstract.} We introduce a more general set of kinematic renormalization
schemes than the original momentum (MOM) subtraction schemes of Celmaster and
Gonsalves. These new schemes will depend on a parameter $\omega$ which tags the
external momentum of one of the legs of the $3$-point vertex functions in 
Quantum Chromodynamics (QCD). In each of the three new schemes we renormalize
QCD in the Landau and maximal abelian gauges and establish the three loop 
renormalization group functions in each gauge. As an application we evaluate 
two critical exponents at the Banks-Zaks fixed point and demonstrate that their
values appear to be numerically scheme independent in a subrange of the 
conformal window.

\vspace{-17.5cm}
\hspace{13.5cm}
{\bf LTH 1152}

\newpage

\sect{Introduction.}

The renormalization of a renormalizable quantum field theory is a technical
exercise which first requires the evaluation of the Feynman diagrams of the 
relevant divergent $n$-point Green's functions of the theory to a specific 
order in perturbation theory. The machinery to subsequently render the theory 
finite is well-established and is completed by encoding the computed 
renormalization constants in the fundamental renormalization group functions.
These functions, such as the $\beta$-function which relates to the 
renormalization of the coupling constant, allows one to determine the behaviour
and value of all the Green's functions with the renormalization scale and 
energy. Of course this has to be tempered by noting that any values can only be
reliable within the confines of the perturbative approximation made or 
equivalently the range of validity of the loop expansion. However, in principle
with sufficient orders in perturbation theory any estimates should be 
reasonably reliable. Indeed the last known term of the perturbative series can
sometimes be used as a way of estimating errors. In outlining the general 
process of renormalization several more technical issues lurk within the 
procedure. One of these major areas is that of how the renormalization 
constants are determined. There are two main aspects to achieving this. First, 
one has to specify the point where the renormalization constants are to be 
defined. By this we mean the momentum configuration of the external legs of the
divergent $n$-point functions. In other words the values of the square of each 
external momentum have to be specified. Clearly there are infinitely many 
possibilities for such momenta values but there are a subset which have to be 
avoided. These are where the sum of a subset of the external momenta is zero. 
Termed an exceptional momentum configuration such momenta values can lead to 
infrared problems in the evaluation or running of the final value of the 
Green's function. The second general feature of renormalization is that once 
the renormalization point is specified one has to specify the prescription to 
defining the renormalization constant associated with each Green's function. 
This is known as the renormalization scheme. Again there are infinitely many of
ways of achieving this. The most commonly used scheme is the modified minimal 
subtraction ($\MSbar$) scheme, \cite{1}. It is a variation on the original 
minimal subtraction scheme denoted by MS, \cite{2,3}. In the MS scheme the 
renormalization constants are determined by removing only the divergences with 
respect to the regulator. The $\MSbar$ scheme is a variant on this where not 
only are the poles removed but also a specific finite part which is 
$\ln(4\pi e^{-\gamma})$ where $\gamma$ is the Euler-Mascheroni constant, 
\cite{1}. The removal of this extra piece appears to improve the convergence of
the series for the Green's function, \cite{1}. 

One major benefit of the $\MSbar$ scheme is that the evaluation of (massless)
Feynman diagrams can be completed to very high loop order. Particularly
impressive has been the progress in determining the $\beta$-function and 
other renormalization group functions of Quantum Chromodynamics (QCD) to
{\em five} loop order, \cite{4,5,6,7}, in the $\MSbar$ scheme. This together 
with other five loop results, \cite{8,9,10,11}, built on the earlier one to 
four loop $\MSbar$ $\beta$-function results of \cite{12,13,14,15,16,17,18} over
a period of around forty years. At this stage we make a side remark in relation
to the general renormalization process. We noted that the renormalization 
scheme involves the divergent part of a Green's function which is quantified by
a regulator. The specific regularization was not stated at that point as its 
nature is irrelevant to the scheme definition. For instance, there are several 
main regularizations such as Pauli-Villars cutoff, spacetime lattice 
regularization and dimensional regularization. Each has particular attributes 
best suited to the problem of interest. For instance, spacetime lattice 
regularization is particularly useful for studying infrared physics numerically
where perturbation theory is not applicable. Although that regularization 
breaks Lorentz invariance this can be addressed in order to obtain reliable
physics results. Also it is not an easy regularization to use for analytic 
perturbative results and only a few loop orders have ever been determined 
analytically. Equally a standard cutoff approach is only useful for a few loop 
orders and has the drawback of breaking gauge invariance. To circumvent these 
technical constraints high order loop computations are efficiently carried out 
using dimensional regularization where the critical (integer) dimension of the 
quantum field theory is replaced by a continuum spacetime dimension $d$. This 
is an analytic continuation with the regulator introduced as a small 
perturbation from the critical dimension. For QCD we then have 
$d$~$=$~$4$~$-$~$2\epsilon$ where $\epsilon$ is a complex variable whose
magnitude is very much less than unity and is the regulator. Unlike the other 
regularizations gauge and Lorentz symmetry are not broken.

One feature of the $\MSbar$ scheme which is maybe not immediately obvious but
which is exploited in the higher loop computations is that the correct $\MSbar$
renormalization of a Green's function emerges even at an exceptional
momentum configuration. For instance, the determination of the coupling
constant renormalization requires the evaluation of one of the three $3$-point
functions in QCD. The $\MSbar$ divergences can be extracted by nullifying one
of the external momenta. This relegates the evaluation of $3$-point functions
to the level of a $2$-point function which is significantly easier to 
determine. This observation has been beneficial to finding the QCD 
$\beta$-function at various loop orders. In this sense $\MSbar$ is regarded as 
a non-kinematical renormalization scheme. It carries no information within the 
renormalization constants with respect to the location of the subtraction 
point. By contrast the renormalization constants of a kinematic scheme contain 
data corresponding to that point. Several kinematic renormalization schemes 
have been used to study QCD. For example, there is an on-shell scheme of 
\cite{19} and the momentum subtraction (MOM) schemes of \cite{20,21}. In the 
latter scheme a non-exceptional momentum configuration is chosen to evaluate 
the $3$-point vertex functions of QCD. The second aspect of the MOM scheme 
definition is that at that subtraction point the divergences as well as the 
finite parts of the $2$- and $3$-point functions are absorbed into the 
renormalization constants. Specifically the subtraction point for the MOM 
scheme is defined as the point where the squares of the external momenta are 
all equal, \cite{20,21}. This is known as the symmetric (subtraction) point. 
For QCD it leads to three separate MOM schemes known as $\MOMg$, $\MOMh$ and 
$\MOMq$ corresponding to schemes based on the triple gluon, ghost-gluon and 
quark-gluon vertices respectively. In \cite{20,21} the two loop renormalization
group functions were determined to two loops. More recently this was extended 
to three loops in \cite{22} due to the advance in the determination of the two 
loop $3$-point integrals for non-exceptional momenta configurations, 
\cite{23,24,25,26}.

At this point one natural question arises which is to do with convergence. It
could be the case that the value of a Green's function, for instance, appears 
to converge quicker in one scheme than another at the same loop order. If one 
knew the full series then there would be no difference in the value of the 
Green's function at the same evaluation point. However, with a truncated series
the numerical values of the coefficients of the coupling constant differ in 
different schemes. Also, a Green's function itself is not a physical quantity 
and a more proficient way of seeing the scheme dependence is by computing a 
renormalization group invariant. One simple and accessible example of such a 
quantity is the set of critical exponents at a phase transition. Critical 
exponents are related to the underlying renormalization group functions 
themselves since they are the values of the latter at the fixed point or zero
of the $\beta$-function. In general the main fixed point is the Wilson-Fisher 
fixed point in the $d$-dimensional version of the quantum field theory, 
\cite{27,28,29,30}. For kinematic schemes these have been studied in QCD in
detail at three loops in \cite{31}, for example. A second fixed point, however,
is available in QCD which is the Banks-Zaks fixed point, \cite{32}. It which 
has been of intense interest recently,
\cite{33,34,35,36,37,38,39,40,41,42,43,44,45,46,47,48,49,50}, from the point of
view of studying various non-kinematical schemes and responding to the new
higher loop QCD data. The Banks-Zaks fixed point is a strictly four dimensional
phenomenon and exists in the conformal window defined by the two loop
$\beta$-function. For the $SU(3)$ colour group the window is
$9$~$\leq$~$\Nf$~$\leq$~$16$, \cite{32}, where $\Nf$ is the number of massless 
quarks. At the Banks-Zaks fixed point one can determine critical exponents and 
there has been theoretical analysis into extracting accurate estimates which 
have been shown to be scheme independent,
\cite{33,34,35,36,37,38,39,40,41,42,43,44,45,46,47,48,49,50}. Detailed analyses
in the main used the $\MSbar$ data of \cite{4,5,6,7,8,9,11}. However, in 
\cite{31} the critical exponents for the three MOM schemes of QCD were examined
with the aim of seeing to what extent the renormalization group invariance held
as a function of $\Nf$ in the conformal window. This is not a trivial exercise
because of the different structure of the renormalization group functions. For 
instance, the classes of numbers appearing in the $\MSbar$ scheme 
renormalization group functions are the rationals and the Riemann zeta function
evaluated for integers $n$~$\geq$~$3$. By contrast the MOM scheme functions in 
addition to rationals and Riemann zeta's involve polylogarithms reflecting the 
kinematical information of the subtraction point. Therefore, analytically it is 
difficult to ascertain the true scheme independence of the results for the 
Banks-Zaks critical exponents. While this was resolved numerically in \cite{31}
for the region $12$~$<$~$\Nf$~$\leq$~$16$ it was not clear if the consistency 
for the MOM schemes was a coincidence or not. For instance, there may have been
something implicitly related to the $\MSbar$ scheme in the choice of 
renormalizing at the fully symmetric point for the vertex functions.

Therefore, we have chosen to re-examine the problem of scheme independence of
the Banks-Zaks critical exponents in a {\em new} set of MOM related schemes
which we will term the interpolating MOM (iMOM) schemes. There will be an iMOM
scheme for each of the three $3$-point vertices of QCD which will be termed
$\iMOMg$, $\iMOMh$ and $\iMOMq$ in direct parallel to the earlier MOM ones with
the set denoted by $\iMOMx$. The renormalization group functions in the 
iMOM schemes will depend on a parameter $\omega$ which is restricted to 
$0$~$<$~$\omega$~$<$~$4$. It tags one of the external momenta of the $3$-point 
vertices and the concept was introduced in \cite{51} for the specific case of 
the quark mass operator renormalization only. The QCD Lagrangian itself was not 
treated in our iMOM scheme defined here since the application in \cite{51} was 
to assist with matching to a lattice gauge theory computation where the 
coupling constant was renormalized in the $\MSbar$ scheme. By contrast here we 
will actually renormalize the QCD Lagrangian itself by defining the scheme 
originally and determining all the renormalization group functions. The earlier
MOM construction of \cite{20,21} will correspond to $\omega$~$=$~$1$. By 
allowing for a parameter we will be able to quantify where and when the 
divergence from renormalization group invariance of the Banks-Zaks critical 
exponents begins in the conformal window. In practical terms our approach for 
the exponents will still be numerical and focus on the specific values of 
$\omega$~$=$~$\half$ and $2$, although the full analytic renormalization will 
be for arbitrary $\omega$. These two values will be sufficient to band the MOM 
$\omega$~$=$~$1$ value and gauge the tolerance on the exponents. Alternatively 
one could regard these two values as a method of error estimation for truncated
renormalization group invariant or physical quantities. One popular way of 
assessing the effect of higher loop behaviour in physical quantities is to 
evaluate the quantity as a function of the scale $\mu$. Then the values at 
$\half \mu$ and $2 \mu$ are regarded as the error bounds. One disadvantage of 
this is that the lower value may be beyond the region of perturbative validity 
and hence unreliable for perturbative error estimates. We will introduce a new 
approach here which will have a scheme motivation. With the parameter $\omega$ 
acting as a variation on the subtraction point its variation between $\half$ 
and $2$ would be a better measure of the errors. In other words it tracks the 
effect of the vertex subtraction {\em within} the graphs constituting the 
truncated series of the quantity of interest. The exponents we will compute 
will demonstrate the effectiveness of this different interpolating approach and
we suggest it would be useful to investigate other QCD quantities with it.

One property of the $\beta$-function in kinematic schemes is that it is gauge
dependent, \cite{20,21}. However, in general the gauge parameter of a linear
covariant gauge can be regarded as a second coupling constant. So at 
criticality the renormalization group function of the gauge parameter has to
be zero which corresponds to the Landau gauge. Therefore all the MOM and iMOM
scheme data will be in that gauge. However, there is a second covariant gauge 
which is of interest called the maximal Abelian gauge (MAG). It is based on 
gauge fixing the gluon in the abelian subgroup of the colour group differently 
from the other gluons. The MAG was introduced in \cite{52,53,54} to study 
abelian monopole condensation since this was believed to be a potential 
mechanism for colour confinement, \cite{55,56,57}. Subsequently it has been 
studied in that context, \cite{55,56,57}, and shown to be renormalizable, 
\cite{58,59,60,61,62,63}. The renormalization group functions are available at 
three loops in $\MSbar$, \cite{64}, and the MOM schemes, \cite{65,66}. 
Therefore, it seems natural to examine the Banks-Zaks fixed point in the iMOM 
schemes in the MAG context as well and we will carry out that analysis in 
parallel with the Landau gauge. The aim is to quantify how far the gauge 
independence of the Banks-Zaks exponents extends into the conformal window. 
This will also be to the same three loop order as the linear covariant gauge in
order to have as comprehensive overview of the scheme and gauge dependence of 
the strictly four dimensional exponents.

The article is organized as follows. We review the formalism required to
evaluate and renormalize $3$-point functions at the non-exceptional momentum
configuration in section $2$ for the two gauges we will consider. The 
definition of the iMOM scheme is given there as well before we record all the
renormalization group functions in section $3$. Subsequently section $4$ is 
devoted to the determination of the two critical exponents of interest at the
Banks-Zaks fixed point. We present our conclusions in section $5$. An appendix
records the tensor bases and projection matrices for each of the three vertices
at the interpolating substraction point.

\sect{Formalism.}

In order to renormalize QCD in the interpolating MOM schemes in both gauges we 
have to introduce a wide body of formalism, such as notation and conventions, 
as well as the computational tools required for the whole process. There are
common aspects of the renormalization for both the Landau and MAG gauges which 
can be outlined together. These will centre on the $2$- and $3$-point functions
or self-energy and vertex Green's functions respectively. The $2$-point 
functions are relatively straightforward to treat in the sense that with the 
massless fields we use here there is only one scale which is the external 
momentum. Therefore, in each of the $\iMOMx$ schemes we choose to define the 
wave function renormalization constants at a point $p^2$~$=$~$-$~$\mu^2$ where 
$\mu$ is the mass scale introduced when we dimensionally regularize in 
$d$~$=$~$4$~$-$~$2\epsilon$ dimensions. This scale is necessary to ensure the 
coupling constant is dimensionless in $d$-dimensions.  As the evaluation of the
$2$-point functions is a straightforward exercise for which we use the 
{\sc Mincer} algorithm, \cite{67,68}, we devote the remainder of the present 
discussion to the vertex function computation. For these the situation is more 
involved in that there are two independent external momenta and so one has to 
be careful in specifying the point where the three Green's functions are 
renormalized. For the present discussion we will focus on the canonical linear 
covariant gauge and then indicate the modification to the formalism to 
accommodate the MAG. First, to be more concrete the three vertex functions we 
consider are  
\begin{eqnarray}
\left. \langle A^a_{\mu}(p)A^b_{\nu}(q)A^c_{\sigma}(r)\rangle \right|_\omega
&=& \left. f^{abc} ~\Sigma^{\mbox{\small{ggg}}}_{\mu \nu \sigma}(p,q) 
\right|_\omega \nonumber \\
\left. \langle \psi^i_I (p) \bar{\psi}^j_J (q)A^c_{\sigma}(r)\rangle 
\right|_\omega
&=& \left. T^c_{IJ} \delta^{ij} 
~\Sigma^{\mbox{\small{qqg}}}_{ \sigma}(p,q) \right|_\omega \nonumber \\
\left. \langle c^a (p) \bar{c}^b (q)A^c_{\sigma}(r)\rangle \right|_\omega 
&=& \left. f^{abc} ~\Sigma^{\mbox{\small{ccg}}}_{ \sigma}(p,q) \right|_\omega 
\label{vertfun}
\end{eqnarray}
where $A^a_\mu$, $c^a$ and $\psi^i$ are the gluon, Faddeev-Popov ghost and 
quark fields respectively. The indices have the ranges 
$1$~$\leq$~$a$~$\leq$~$\NA$ and $1$~$\leq$~$I$~$\leq$~$\NF$ for the linear 
covariant gauge with $\NF$ and $\NA$ corresponding to the dimension of the 
fundamental and adjoint representations respectively. Throughout we will use 
similar notation to \cite{22,66}. In (\ref{vertfun}) we have indicated the 
momenta of the external legs which are $p$, $q$ and $r$ but we take the first 
two as the independent ones and set 
\begin{equation}
r ~=~ -~ p ~-~ q  ~.
\end{equation}
For simplicity we have factored off the colour group structure in each case in
(\ref{vertfun}) since to two loops there are no other colour tensors. The 
definition of the $\iMOMx$ schemes do not depend on these tensors. The 
amplitude which remains is a Lorentz tensor which although it depends in 
general on the two independent external momenta are restricted as indicated to 
the particular external momentum configuration of the $\iMOMx$ subtraction 
point. In particular the squared external momenta are constrained to satisfy 
\begin{equation}
p^2 ~=~ q^2 ~=~ -~ \mu^2 ~~,~~ r^2 ~=~ -~ \omega \mu^2 
\label{imomdef1}
\end{equation}
where $\omega$ is our interpolating parameter. These lead to 
\begin{equation}
pq ~=~ \left[ 1 ~-~ \frac{\omega}{2} \right] \mu^2 ~~,~~
pr ~=~ qr ~=~ -~ \omega \mu^2 ~.
\label{imomdef2}
\end{equation}
The latter relations introduce restrictions on the range of validity of
$\omega$ such that $0$~$<$~$\omega$~$<$~$4$. The lower bound would correspond
to an infrared divergence and the supremum leads to collinear singularities.
The original MOM configuration of Celmaster and Gonsalves, \cite{20,21},
corresponds to $\omega$~$=$~$1$ which will be used as an internal check 
throughout. In order to carry out the renormalization of a Green's function two
aspects have to be considered. The first is the specification of the values of 
the external momenta. For the $\iMOMx$ schemes we have already indicated this 
with (\ref{imomdef1}) and (\ref{imomdef2}). The second is the prescription to 
define the renormalization constants. We have not phrased this as the way to
remove the divergences with respect to the regularization as any scheme has to
do this at the very least. The crucial part is the treatment of the 
non-divergent pieces. As the subtraction point is specified the complicated
function of the external momenta which ordinarily is present within a Green's 
function is reduced to a particular value of this function. While in our case 
there are no internal masses upon which this function could also depend, there
will be dependence on the parameter $\omega$. Therefore the finite part of any 
renormalization constant would correspond to a particular number or for iMOM
involve a parameter which sweeps over a range of numbers. For the $\iMOMx$
schemes the subtraction prescription is that the renormalization constants for
the $2$- and $3$-point functions are chosen so that at the subtraction point 
there are no $O(a)$ corrections where $a$~$=$~$g^2/(16\pi^2)$. This is within 
the spirit of the original Celmaster and Gonsalves scheme for the symmetric 
point case where $\omega$~$=$~$1$, \cite{20,21}. Therefore as noted there are 
three different $\iMOMx$ schemes for each of the linear covariant gauge and MAG
which derive from each $3$-point vertex. 

The next stage of the vertex function evaluation is the evaluation of the
Lorentz tensor amplitudes. For an $\iMOMx$ renormalization this means that the 
amplitudes have to computed to the finite parts. The first stage is to 
decompose these into a set of scalar amplitudes for each vertex by the 
projection method discussed in \cite{22,66}. For each vertex we define these 
scalar functions by  
\begin{eqnarray}
\left. \Sigma^{\mbox{\small{ggg}}}_{\mu \nu \sigma}(p,q) \right|_\omega &=&
\sum_{k=1}^{14} \mathcal{P}^{\mbox{\small{ggg}}}_{(k) \mu \nu \sigma} (p,q)
\Sigma^{\mbox{\small{ggg}}}_{(k)}(p,q) \nonumber \\
\left. \Sigma^{\mbox{\small{qqg}}}_{\sigma}(p,q) \right|_\omega &=&
\sum_{k=1}^{6} \mathcal{P}^{\mbox{\small{qqg}}}_{(k) \sigma} (p,q)
\Sigma^{\mbox{\small{qqg}}}_{(k)}(p,q) \nonumber \\
\left. \Sigma^{\mbox{\small{ccg}}}_{\sigma}(p,q) \right|_\omega &=&
\sum_{k=1}^{2} \mathcal{P}^{\mbox{\small{ccg}}}_{(k) \sigma} (p,q)
\Sigma^{\mbox{\small{ccg}}}_{(k)}(p,q) 
\label{ampdecomp}
\end{eqnarray}
where $\mathcal{P}^{\mbox{\small{ggg}}}_{(k) \mu \nu \sigma} (p,q)$,
$\mathcal{P}^{\mbox{\small{qqg}}}_{(k) \sigma} (p,q)$ and
$\mathcal{P}^{\mbox{\small{ccg}}}_{(k) \sigma} (p,q)$ are the independent
Lorentz tensors which can be built out of the independent external momenta and
tensors such as $\eta_{\mu\nu}$ for each vertex. The evaluation at 
(\ref{imomdef1}) and (\ref{imomdef2}) is understood on the right hand side of
(\ref{ampdecomp}). In addition for the quark-gluon vertex the spinor indices of
the quark fields have to be attached to $\gamma$-matrices. This means that we 
include combinations of $\gamma$-matrices in the set of objects from which the 
basis tensors of the vertex decomposition are selected. As we dimensionally 
regularize in order to evaluate the Feynman integrals we use generalized 
$\gamma$-matrices, \cite{69,70,71,72,73}, which are denoted by 
$\Gamma_{(n) \, \mu_1 \ldots \mu_n}$ and defined by
\begin{equation}
\Gamma_{(n)}^{\mu_1 \ldots \mu_n} ~=~ \gamma^{[\mu_1} \ldots \gamma^{\mu_n]} ~.
\end{equation}
An overall factor of $1/n!$ is understood in the definition of this totally
antisymmetric object. The benefits of using this is that these matrices span
the spinor space of the dimensionally regularized theory as well as giving a
partition since 
\begin{equation}
\mbox{tr} \left( \Gamma_{(m)}^{\mu_1 \ldots \mu_m}
\Gamma_{(n)}^{\nu_1 \ldots \nu_n} \right) ~ \propto ~ \delta_{mn}
I^{\mu_1 \ldots \mu_m \nu_1 \ldots \nu_n}
\label{gentr}
\end{equation}
where $I^{\mu_1 \ldots \mu_m \nu_1 \ldots \nu_n}$ is the generalized identity
tensor. For each vertex function the explicit Lorentz tensors are defined in 
the Appendix. The important step is the isolation of the scalar amplitudes. 

We illustrate this for the triple gluon vertex where for each amplitude we have 
\begin{eqnarray}
\left. \frac{}{} f^{abc} ~\Sigma^{\mbox{\small{ggg}}}_{(k)}(p,q) \right|_\omega
&=& \left. \mathcal{M}_{kl}^{\mbox{\small{ggg}}} \left(
\mathcal{P}^{\mbox{\small{ggg}} ~\mu \nu \sigma}_{(l)} (p,q)
\langle A^a_{\mu}(p)A^b_{\nu}(q)A^c_{\sigma}(-p-q)\rangle \right) 
\right|_\omega ~.
\end{eqnarray}
Here $\mathcal{M}_{kl}^{\mbox{\small{ggg}}}$ is the projection matrix which
depends on $\omega$ and $d$. It is derived from the related matrix
$\mathcal{N}_{kl}^{\mbox{\small{ggg}}}$ which is defined by 
\begin{equation}
{\cal N}^{\mbox{\small{ggg}}}_{kl} ~=~ 
\left. \left[ {\cal P}^{\mbox{\small{ggg}}}_{(k) \, \mu \nu \sigma}(p,q)
{\cal P}^{{\mbox{\small{ggg}}} ~\, \mu \nu \sigma}_{(l)}(p,q) \right]
\right|_\omega
\end{equation}
and is a symmetric matrix with $d$ and $\omega$ dependent entries. Finally, the
symmetric matrix $\mathcal{M}_{kl}^{\mbox{\small{ggg}}}$ is the inverse of  
$\mathcal{N}_{kl}^{\mbox{\small{ggg}}}$, \cite{22,66}. The process for the 
remaining vertices is similar and we have provided the projection matrices in 
the Appendix. The only caveat is that for the construction of the quark-gluon 
projection matrix a trace over the spinor indices is also taken. As the 
decomposition (\ref{ampdecomp}) is based on the Lorentz sector we will use the 
same projection matrices for the MAG.

The final Green's function we will consider is that from which we can extract 
the renormalization of the quark mass operator $\bar{\psi} \psi$. In particular
we will evaluate $\left. \left\langle \psi(p) [\bar{\psi} \psi](r) 
\bar{\psi}(q) \right\rangle \right|_\omega$ also to the finite part. The 
procedure to achieve this is completely parallel to that outlined for the 
vertex functions and in particular the quark-gluon vertex. The reason for the 
connection with that specific vertex is the open spinor indices. For the
projection there are two independent tensors in the basis which actually
partitions into two separate sectors due to (\ref{gentr}). Aside from the unit 
matrix $\Gamma_{(0)}$ the second basis element involves $\Gamma_{(2)\,\mu\nu}$ 
and is given in the Appendix together with the $\omega$ dependent projection 
matrix. 

The outcome of the projection process is to relegate the Green's functions to
a sum over Lorentz scalar amplitudes for each gauge. So far this process has 
been general and not appealed to the specific Feynman diagrams which comprise 
the vertex functions. In practice we generate all the graphs by the 
{\sc Fortran} based {\sc Qgraf} package, \cite{74}. The electronic 
representations of the graphs are then individually passed through the 
projection algorithm once the colour, spinor, flavour and Lorentz indices have 
been appended. The consequence is that the amplitude for each Feynman graph is 
a sum of Feynman integrals which has scalar products of the external and 
internal momenta. To two loops the majority of these scalar products can be 
rewritten in terms of the propagators of a topology which allows one to naively
reduce the powers of various propagators in an integral. However, at two loops
for a $3$-point function there is no more than one irreducible scalar product. 
While this would ordinarily be a difficulty in carrying out the integration 
there is an elegant algorithm designed to accommodate this problem which was 
developed by Laporta, \cite{75}. In essence the set of propagators of a 
topology is completed by including an additional propagator in the two loop 
$3$-point case but more can be added for other Green's functions and loops 
which is a process termed completion. Although this extension puts the 
irreducible numerator propagator on the same footing as the other propagators 
of the original integral the new integral cannot correspond to a bona fide 
Feynman graph. This is not a problem as the Laporta algorithm, \cite{75}, at 
large involves integrating by parts completed Feynman integrals. Then the huge 
set of resultant linear equations, which as it turns out are redundant, are
solved. The end product is that all the integrals contributing to a Feynman 
graph of the original Green's function can be written as a sum over a 
relatively small set of what is termed master integrals. Their $\epsilon$ 
expansion has to be determined by explicit evaluation. For our two loop 
$3$-point function the master integrals have been known for some time, 
\cite{23,24,25,26}. In terms of practically implementing the procedure we have 
outlined we have used both versions of the {\sc Reduze} package, \cite{76,77},
which encodes the Laporta algorithm. One advantage is that the ouput of the 
reduction to masters can be converted into the symbolic manipulation language 
{\sc Form}, \cite{78,79}. This was used intensively with the whole process of 
evaluating each of our Green's functions to two loops prior to the 
renormalization in a fully automatic way. The latter is achieved by the method 
of \cite{80} by computing all the Green's functions in terms of the bare 
coupling constant and gauge parameter. Their renormalized counterparts are
introduced by the canonical rescaling. The renormalization constants are fixed 
in a particular scheme by the procedure given earlier where all the one loop 
Green's functions are converted to renormalized variables first before 
repeating the exercise at two loops. This is carried out first for all the wave
function, coupling constant and gauge parameter renormalizations. Then to 
extract the renormalization constant for the quark mass operator
$\bar{\psi} \psi$ the Green's function 
$\left. \left\langle \psi(p) [\bar{\psi} \psi](r) \bar{\psi}(q) \right\rangle 
\right|_\omega$ is also evaluated in terms of the bare coupling and gauge 
parameter. These then are rescaled in a particular $\iMOMx$ scheme and the 
operator renormalization constant for that scheme is determined by ensuring 
that there are no $O(a)$ corrections at the subtraction point. This process is 
repeated separately for the other two schemes. 

To two loops the iMOM master integrals have been deduced from 
\cite{23,24,25,26} and discussed in \cite{66} for the renormalization of the 
quark mass operator as a function of $\omega$ used for lattice matching. 
Therefore, we summarize key aspects required for the present work. It 
transpires that several complicated functions arise in the finite parts of the 
one and two loop master integrals. These are, \cite{24,25}, 
\begin{eqnarray}
\Phi_1(x,y) &=& \frac{1}{\lambda} \left[ 2 \mbox{Li}_2(-\rho x)
+ 2 \mbox{Li}_2(-\rho y)
+ \ln \left( \frac{y}{x} \right)
\ln \left( \frac{(1+\rho y)}{(1+\rho x)} \right)
+ \ln(\rho x) \ln(\rho y) + \frac{\pi^2}{3} \right] \nonumber \\
\Phi_2(x,y) &=& \frac{1}{\lambda} \left[ 6 \mbox{Li}_4(-\rho x)
+ 6 \mbox{Li}_4(-\rho y) + 3 \ln \! \left( \frac{y}{x} \right)
\left[ \mbox{Li}_3(-\rho x) - \mbox{Li}_3(-\rho y) \right] 
+ \frac{\pi^2}{12} \ln^2 \! \left( \frac{y}{x} \right) + \frac{7\pi^4}{60} 
\right. \nonumber \\
&& \left. ~~~
+ \frac{1}{2} \ln^2 \left( \frac{y}{x} \right)
\left[ \mbox{Li}_2(-\rho x) + \mbox{Li}_2(-\rho y) \right]
+ \frac{1}{4} \ln^2(\rho x) \ln^2(\rho y)
+ \frac{\pi^2}{2} \ln(\rho x) \ln(\rho y)
\right]
\nonumber \\
\Omega_2(x,y) &=& 3 \ln \left( \frac{y}{x} \right)
\left[ \mbox{Li}_2(-\rho x) - \mbox{Li}_2(-\rho y) \right] 
- \frac{1}{2} \ln^2 \left( \frac{y}{x} \right)
\left[ \ln(1+\rho x) + \ln(1+\rho y) \right]
\nonumber \\
&& +~ 6 \mbox{Li}_3(-\rho x) + 6 \mbox{Li}_3(-\rho y)
+ \frac{1}{2} \left[ \pi^2 + \ln(\rho x) \ln(\rho y) \right]
\left[ \ln(\rho x) + \ln(\rho y) \right] 
\end{eqnarray}
which are symmetric in $x$ and $y$. Here $\mbox{Li}_n(z)$ is the polylogarithm 
function and the other functions are defined by 
\begin{equation}
\lambda(x,y) ~=~ 
\sqrt{\left[ x^2 - 2 x y + y^2 - 2 x - 2 y + 1 \right]} ~~~,~~~ 
\rho(x,y) ~=~ \frac{2}{[1-x-y+\lambda(x,y)]} ~.
\end{equation}
The variables $x$ and $y$ are specific to the $3$-point function with non-zero
external $p$, $q$ and $r$ since, \cite{51},
\begin{equation}
x ~=~ \frac{p^2}{r^2} ~~~,~~~
y ~=~ \frac{q^2}{r^2} ~~~,~~~
r^2 ~=~ -~ \omega \mu^2 ~.
\end{equation}
Within the interpolating setup two specific argument combinations emerge for
$\Phi_i(x,y)$ and $\Omega_2(x,y)$ with $(x,y)$ being either $(1,\omega)$ or 
$(\omega,\omega)$. 

As our focus will be on the renormalization of QCD in various schemes we need
to recall several aspects of the formalism. If the subscript 
${}_{\mbox{\footnotesize{o}}}$ denotes a bare quantity then the relation 
between such an object and its renormalized counterpart for a linear covariant
gauge is given by 
\begin{equation}
A^{a \, \mu}_{\mbox{\footnotesize{o}}} ~=~ \sqrt{Z_A} \, A^{a \, \mu} ~,~
c^a_{\mbox{\footnotesize{o}}} ~=~ \sqrt{Z_c} \, c^a ~,~
\psi_{\mbox{\footnotesize{o}}} ~=~ \sqrt{Z_\psi} \psi ~,~
g_{\mbox{\footnotesize{o}}} ~=~ \mu^\epsilon Z_g \, g ~,~
\alpha_{\mbox{\footnotesize{o}}} ~=~ Z^{-1}_\alpha Z_A \, \alpha 
\end{equation}
where the constant of proportionality is the renormalization constant, $\alpha$
is the gauge parameter of the linear covariant gauge fixing and the mass scale
$\mu$ ensures the dimensionlessness of the coupling constant. The definition of
the renormalization constants for the MAG is somewhat different and can be 
found in \cite{63}. If we introduce the shorthand notation
\begin{equation}
{\cal O} ~=~ \bar{\psi} \psi
\end{equation}
for the quark mass operator then we have
\begin{equation}
{\cal O}_{\mbox{\footnotesize{o}}} ~=~ Z_{\cal O} {\cal O}
\end{equation}
for both gauges. However, in defining these relations it is worth stressing 
that there are an infinite number of different renormalized fields and 
variables. This is because the renormalized objects are in a particular scheme.
With the interpolating momentum subtraction scheme the parameter $\omega$ will 
play the role of potentially running over a range of different possible 
schemes. Although we will focus primarily on two values. The key point is that
any set of renormalization constants will depend on variables such as the 
coupling constant defined with respect to a scheme. Another set will depend on 
the variables in a different scheme but the two sets can be related through
properties of the renormalization group. We can illustrate this formally for
the linear covariant gauges in the $\iMOMx$ schemes. From the defining 
relations of renormalized variables we can deduce  
\begin{equation}
g_{\iMOMxs}(\mu) ~=~ \frac{Z_g}{Z_g^{\iMOMxs}} \, g(\mu) ~~~,~~~
\alpha_{\iMOMxs}(\mu) ~=~ \frac{Z_A Z_\alpha^{\iMOMxs}}
{Z_A^{\iMOMxs} Z_\alpha} \, \alpha(\mu)
\label{paramcon}
\end{equation}
for the coupling constant and gauge parameter. We have been careful to label
the quantities in the $\iMOMx$ schemes and this label means that the variables
in the particular object are in that scheme. We take the convention that 
unlabelled quantities are in the $\MSbar$ scheme which is regarded as the 
reference scheme. There is no specific reason why we take $\MSbar$ as the
reference aside from the fact that it is the most widely used as well as being
the scheme for which has the renormalization group functions are known to the
highest loop order. For completeness we note that the anomalous dimensions of
the fields and gauge parameter in any scheme are related to their
renormalization constants by 
\begin{eqnarray}
\gamma_A(a,\alpha) &=& \beta(a,\alpha) \frac{\partial}{\partial a} \ln Z_A ~+~
\alpha \gamma_\alpha(a,\alpha) \frac{\partial}{\partial \alpha} \ln Z_A
\nonumber \\
\gamma_\alpha(a,\alpha) &=& \left[ \beta(a,\alpha) \frac{\partial}{\partial a}
\ln Z_\alpha ~-~ \gamma_A(a,\alpha) \right] \left[ 1 ~-~ \alpha
\frac{\partial}{\partial \alpha} \ln Z_\alpha \right]^{-1} \nonumber \\
\gamma_c(a,\alpha) &=& \beta(a,\alpha) \frac{\partial}{\partial a} \ln Z_c ~+~
\alpha \gamma_\alpha(a,\alpha) \frac{\partial}{\partial \alpha} \ln Z_c
\nonumber \\
\gamma_\psi(a,\alpha) &=& \beta(a,\alpha) \frac{\partial}{\partial a}
\ln Z_\psi ~+~ \alpha \gamma_\alpha(a,\alpha) \frac{\partial}{\partial \alpha}
\ln Z_\psi \nonumber \\
\gamma_{\cal O}(a,\alpha) &=& -~ \beta(a,\alpha) \frac{\partial}{\partial a}
\ln Z_{\cal O} ~-~ \alpha \gamma_\alpha(a,\alpha) 
\frac{\partial}{\partial \alpha} \ln Z_{\cal O} 
\label{rgedef} 
\end{eqnarray}
where the same conventions as \cite{22,65,66} are used. We have included an 
$\alpha$ dependence in the $\beta$-function as it is gauge dependent in 
general. In the $\MSbar$ scheme it can be shown that the gauge dependence is 
absent, \cite{2,3}. Our apparently non-standard relation for 
$\gamma_\alpha(a,\alpha)$ is because we have not made any assumption on the 
form of $Z_\alpha$. In a linear covariant gauge in our convention
$Z_\alpha$~$=$~$1$ but this is not true in general. In particular in the MAG
the corresponding parameter of the off-diagonal gauge fixing is not unity, 
\cite{58,59,60,61,62,63}.

Once the renormalization constants have been determined in a scheme, for
example, then the renormalization group functions can be determined in that
scheme from (\ref{rgedef}). However, there is a second way which expressions 
can be deduced without evaluating (\ref{rgedef}). This requires the 
renormalization group functions in another or what we will term a base scheme. 
Here this will be the $\MSbar$ scheme and the $\iMOMx$ schemes are the ones for
which we wish to deduce the renormalization group functions. To achieve this
we define the respective conversion functions by
\begin{equation}
C^{\iMOMxs}_\phi(a,\alpha) ~=~ \frac{Z_\phi^{\iMOMxs}}{Z_\phi} ~~~,~~~
C^{\iMOMxs}_\alpha(a,\alpha) ~=~ \frac{Z_\alpha^{\iMOMxs} Z_A}
{Z_\alpha Z_A^{\iMOMxs}}
\label{condef} 
\end{equation}
where $\phi$ will be used to represent one of the three fields or quark mass 
operator. The arguments of the conversion functions are in the base or $\MSbar$
scheme. This is an important observation since the $\iMOMx$ renormalization 
constants on the right hand side have their variables in that scheme. Naively 
evaluating these functions by taking the explicit renormalization constants 
will lead to a divergent function with respect to $\epsilon$. To obtain finite 
expressions the variables of the $\iMOMx$ scheme have to be converted to their 
$\MSbar$ counterparts. To find the relations between the coupling constant and 
gauge parameter in the different schemes the relations (\ref{paramcon}) are 
solved recursively order by order in perturbation theory. Once the maps are 
known to the necessary loop order then one can explicitly find the conversion 
functions to the same order. Equipped with these the renormalization group 
functions in the $\iMOMx$ scheme for the linear covariant gauge are given by 
\begin{eqnarray}
\beta^{\iMOMxs} ( a_{\iMOMxs}, \alpha_{\iMOMxs} ) &=&
\left[ \beta(a)
\frac{\partial a_{\iMOMxs}}{\partial a} ~+~
\alpha \gamma_\alpha (a, \alpha)
\frac{\partial a_{\iMOMxs}}{\partial \alpha}
\right]_{ \MSbars \rightarrow \iMOMxs } \nonumber \\
\gamma_\phi^{\iMOMxs} ( a_{\iMOMxs}, \alpha_{\iMOMxs} )
&=& \left[ \gamma_\phi (a)
+ \beta (a) \frac{\partial ~}{\partial a} \ln C_\phi^{\iMOMxs} (a,\alpha) 
\right. \nonumber \\
&& \left. ~+~ \alpha \gamma_\alpha (a,\alpha) 
\frac{\partial ~}{\partial \alpha}
\ln C_\phi^{\iMOMxs} (a,\alpha) \right]_{ \MSbars \rightarrow \iMOMxs } 
\label{rgemap}
\end{eqnarray}
from the renormalization group equation. The restriction on each equation 
indicates that once the right hand side has been computed, which necessarily
will be in the base scheme variables, they have to be mapped to the $\iMOMx$
variables. This mapping is the (perturbative) inverse of the one used in 
deriving the conversion functions themselves. While this process is equivalent
to the direct evaluation of the renormalization group functions using 
(\ref{rgedef}) the advantage of the conversion function approach is that if the
conversion functions are known at $L$ loops and the renormalization group
functions of the base theory are available at $(L+1)$ loops then the latter
formalism produces the renormalization group functions of the new scheme at
$(L+1)$ loops without having to perform an explicit $(L+1)$ loop 
renormalization in the new scheme. We will benefit from this observation by 
using it to construct the {\em three} loop $\iMOMx$ renormalization group 
functions.  

We close this section by briefly discussing the definition of the MAG and the
relevant aspects of the renormalization of QCD in this gauge with respect to
the iMOM schemes. More background details to our conventions can be found in 
\cite{64,65,66}. In essence the gauge is defined by treating the diagonal 
colour fields separately from the off-diagonal ones, \cite{58,59,60,61,62,63}. 
First we write the group valued gluon field ${\cal A}$ as 
\begin{equation}
{\cal A}_\mu ~=~ A^{\bar{a}}_\mu T^{\bar{a}} ~+~ A^{\bar {i}}_\mu T^{\bar {i}}
\end{equation}
where the group generators $T^a$ are treated differently. Our notation varies
slightly from that of \cite{64} given that we are treating QCD in two different
gauges. The diagonal fields are labelled by indices $\bar{i}$, $\bar{j}$ and 
$\bar{k}$ where by diagonal we mean those gluons whose associated group 
generators commute with each other. For $SU(3)$ there are two such gluons. The 
remaining gluons are termed off-diagonal and denoted by barred lowercase Roman 
letters. For a general Lie group the indices have the ranges 
$1$~$\leq$~$a$~$\leq$~$\NA$, $1$~$\leq$~$\bar{a}$~$\leq$~$\Noda$ and 
$1$~$\leq$~$\bar{i}$~$\leq$~$\Nda$ where $\Nda$ is the dimension of the 
diagonal sector and $\Noda$ is the dimension of the off-diagonal sector. The 
sum of $\Noda$ and $\Nda$ is $\NA$. The gauge fixing for each sector is 
different with the diagonal gluons fixed in the Landau gauge, 
\cite{58,59,60,61,62,63}. By contrast the off-diagonal gluons have a gauge 
fixing similar to the linear covariant gauge. However with the different 
treatment of the gluons the resultant MAG gauge fixed Lagrangian involves a 
significantly larger number of interactions which includes gluon-ghost and 
purely quartic ghost interactions. In addition to the two sets of gluons being 
gauge fixed differently there are two sets of Faddeev-Popov ghosts, 
$c^{\bar{a}}$ and $c^{\bar{i}}$, which are responsible for the additional type 
of interactions. Consequently one has a much larger number of Feynman graphs to
evaluate when examining the $3$-point vertices. One of the complications in 
dealing with the enlarged basis of terms in the Lagrangian is that the group 
theory used in evaluating the Feynman graphs becomes more involved. This has 
been detailed in \cite{64} and we refer the interested reader to those papers. 
We have used the same group theory routines in carrying out our MAG 
computations here as were used in \cite{64,65,66}.

While the MAG has additional interactions and more structure to handle for a
computation its renormalization has several interesting features deriving from
the Slavnov-Taylor identities constructed in \cite{63}. The first is that the
diagonal gluons effectively act as a background field similar to the background
field gauge \cite{81,82,83,84,85}. For instance, the renormalization of the 
diagonal gluon is in one-to-one correspondence with the coupling constant 
renormalization which is parallel to the situation for the background gluon of 
the background field gauge. However for the MAG in the context of the iMOM 
renormalization we derive each coupling constant renormalization from the 
$3$-point functions directly and extend the MOM approach of \cite{65,66}. This 
therefore retains the spirit of the MOM approach in that the iMOM vertex 
renormalization incorporates kinematical information which will involve 
$\omega$ and hence allows us to explore the dependence on $\omega$. Another 
consequence of the Slavnov-Taylor identities is that we need only consider the 
three vertices defined by the purely off-diagonal gluons as the basis for our 
MAG iMOM schemes as in the MOM study of \cite{65,66}. Vertex functions with a 
diagonal gluon will not lead to any iMOM schemes. Therefore our focus in the 
MAG will be on the parallel Green's functions to (\ref{vertfun}) except that 
the adjoint indices of the linear covariant gauge are replaced by their barred 
counterparts. The colour and Lorentz decomposition are completely similar in 
the MAG. Therefore we do not need to add anything to this part of the 
discussion except to note that we have applied the same projection algorithms 
to the respective MAG vertices. Therefore after this point we will not need to 
distinguish between barred and unbarred adjoint colour indices in the 
discussion. Once all the renormalization constants have been extracted for the 
MAG we use the parallel definitions to (\ref{rgedef}) for the off-diagonal 
gluon and ghosts as well as the other anomalous dimensions. One complication is
that like the MOM results of \cite{65,66} the explicit expressions will depend
on $\Noda$ and $\Nda$ for instance. Equally there are parallel definitions for 
the corresponding conversion functions (\ref{condef}) and (\ref{rgemap}). The 
only caveat is that we need to distinguish our linear covariant gauge formalism
from the corresponding MAG expressions. We do this by denoting MAG iMOM schemes
in general by $\iMOMmx$ where the three MAG schemes will be denoted by 
$\iMOMmg$, $\iMOMmh$ and $\iMOMmq$ for the triple off-diagonal gluon, 
off-diagonal gluon-ghost and off-diagonal gluon-quark schemes respectively. 

\sect{Renormalization group functions.}

We now discuss the results of our renormalization in both gauges. In this 
context it is worthwhile displaying one of the renormalization group functions 
in analytic form as it illustrates several important features. For instance the 
$\beta$-function for the $\iMOMh$ scheme for the $SU(3)$ group in the Landau
gauge is 
\begin{eqnarray}
\left. \beta^{\iMOMhs}(a,0) \right|^{SU(3)} &=&
[2 \Nf - 33] \frac{a^2}{3} ~+~ \frac{2}{3} [19 \Nf - 153] a^3 \nonumber \\
&& +~ \left[ 
24192 \ln^2(\omega) \omega^4 \Nf^2
- 1728 \ln^2(\omega) \omega^5 \Nf^2 
- 110592 \ln^2(\omega) \omega^3 \Nf^2
\right. \nonumber \\
&& \left. ~~~~
+ 165888 \ln^2(\omega) \omega^2 \Nf^2
- 71928 \ln^2(\omega) \omega^5 \Nf
+ 270864 \ln^2(\omega) \omega^4 \Nf
\right. \nonumber \\
&& \left. ~~~~
+ 787968 \ln^2(\omega) \omega^3 \Nf
- 2882304 \ln^2(\omega) \omega^2 \Nf
+ 1657260 \ln^2(\omega) \omega^5
\right. \nonumber \\
&& \left. ~~~~
- 11055528 \ln^2(\omega) \omega^4
+ 17107200 \ln^2(\omega) \omega^3
+ 2395008 \ln^2(\omega) \omega^2
\right. \nonumber \\
&& \left. ~~~~
+ 2304 \ln(\omega) \Phi_{(1)\omega,\omega} \omega^5 \Nf^2
- 31104 \ln(\omega) \Phi_{(1)\omega,\omega} \omega^4 \Nf^2
\right. \nonumber \\
&& \left. ~~~~
+ 148608 \ln(\omega) \Phi_{(1)\omega,\omega} \omega^3 \Nf^2
- 285696 \ln(\omega) \Phi_{(1)\omega,\omega} \omega^2 \Nf^2
\right. \nonumber \\
&& \left. ~~~~
+ 165888 \ln(\omega) \Phi_{(1)\omega,\omega} \omega \Nf^2
- 3132 \ln(\omega) \Phi_{(1)\omega,\omega} \omega^5 \Nf
\right. \nonumber \\
&& \left. ~~~~
+ 309096 \ln(\omega) \Phi_{(1)\omega,\omega} \omega^4 \Nf
- 2256336 \ln(\omega) \Phi_{(1)\omega,\omega} \omega^3 \Nf
\right. \nonumber \\
&& \left. ~~~~
+ 5011200 \ln(\omega) \Phi_{(1)\omega,\omega} \omega^2 \Nf
- 2923776 \ln(\omega) \Phi_{(1)\omega,\omega} \omega \Nf
\right. \nonumber \\
&& \left. ~~~~
- 575586 \ln(\omega) \Phi_{(1)\omega,\omega} \omega^5
+ 3367980 \ln(\omega) \Phi_{(1)\omega,\omega} \omega^4
\right. \nonumber \\
&& \left. ~~~~
- 3228984 \ln(\omega) \Phi_{(1)\omega,\omega} \omega^3
- 4904064 \ln(\omega) \Phi_{(1)\omega,\omega} \omega^2
\right. \nonumber \\
&& \left. ~~~~
+ 3079296 \ln(\omega) \Phi_{(1)\omega,\omega} \omega
- 3456 \ln(\omega) \omega^5 \Nf^2
+ 24192 \ln(\omega) \omega^4 \Nf^2
\right. \nonumber \\
&& \left. ~~~~
- 27648 \ln(\omega) \omega^3 \Nf^2
- 55296 \ln(\omega) \omega^2 \Nf^2
- 141696 \ln(\omega) \omega^5 \Nf
\right. \nonumber \\
&& \left. ~~~~
+ 1461024 \ln(\omega) \omega^4 \Nf
- 4886784 \ln(\omega) \omega^3 \Nf
+ 5239296 \ln(\omega) \omega^2 \Nf
\right. \nonumber \\
&& \left. ~~~~
+ 505440 \ln(\omega) \omega^5
- 6286896 \ln(\omega) \omega^4
+ 26034048 \ln(\omega) \omega^3
\right. \nonumber \\
&& \left. ~~~~
- 35894016 \ln(\omega) \omega^2
+ 30456 \Omega_{(2)\omega,\omega} \omega^5 \Nf
- 339552 \Omega_{(2)\omega,\omega} \omega^4 \Nf
\right. \nonumber \\
&& \left. ~~~~
+ 1254528 \Omega_{(2)\omega,\omega} \omega^3 \Nf
- 1534464 \Omega_{(2)\omega,\omega} \omega^2 \Nf
- 502524 \Omega_{(2)\omega,\omega} \omega^5
\right. \nonumber \\
&& \left. ~~~~
+ 5602608 \Omega_{(2)\omega,\omega} \omega^4
- 20699712 \Omega_{(2)\omega,\omega} \omega^3
+ 25318656 \Omega_{(2)\omega,\omega} \omega^2
\right. \nonumber \\
&& \left. ~~~~
- 4608 \Omega_{(2)1,\omega} \omega^5 \Nf^2
+ 55296 \Omega_{(2)1,\omega} \omega^4 \Nf^2
- 221184 \Omega_{(2)1,\omega} \omega^3 \Nf^2
\right. \nonumber \\
&& \left. ~~~~
+ 294912 \Omega_{(2)1,\omega} \omega^2 \Nf^2
+ 102168 \Omega_{(2)1,\omega} \omega^5 \Nf
- 1302480 \Omega_{(2)1,\omega} \omega^4 \Nf
\right. \nonumber \\
&& \left. ~~~~
+ 5523552 \Omega_{(2)1,\omega} \omega^3 \Nf
- 7824384 \Omega_{(2)1,\omega} \omega^2 \Nf
+ 124416 \Omega_{(2)1,\omega} \omega \Nf
\right. \nonumber \\
&& \left. ~~~~
- 431244 \Omega_{(2)1,\omega} \omega^5
+ 6436584 \Omega_{(2)1,\omega} \omega^4
- 30921264 \Omega_{(2)1,\omega} \omega^3
\right. \nonumber \\
&& \left. ~~~~
+ 48812544 \Omega_{(2)1,\omega} \omega^2
- 2052864 \Omega_{(2)1,\omega} \omega
- 4374 \Phi^2_{(1)\omega,\omega} \omega^5 \Nf
\right. \nonumber \\
&& \left. ~~~~
+ 34992 \Phi^2_{(1)\omega,\omega} \omega^4 \Nf
- 110808 \Phi^2_{(1)\omega,\omega} \omega^3 \Nf
+ 209952 \Phi^2_{(1)\omega,\omega} \omega^2 \Nf
\right. \nonumber \\
&& \left. ~~~~
- 217728 \Phi^2_{(1)\omega,\omega} \omega \Nf
+ 124416 \Phi^2_{(1)\omega,\omega} \Nf
+ 72171 \Phi^2_{(1)\omega,\omega} \omega^5
\right. \nonumber \\
&& \left. ~~~~
- 577368 \Phi^2_{(1)\omega,\omega} \omega^4
+ 1828332 \Phi^2_{(1)\omega,\omega} \omega^3
- 3464208 \Phi^2_{(1)\omega,\omega} \omega^2
\right. \nonumber \\
&& \left. ~~~~
+ 3592512 \Phi^2_{(1)\omega,\omega} \omega
- 2052864 \Phi^2_{(1)\omega,\omega}
+ 7488 \Phi_{(1)\omega,\omega} \omega^5 \Nf^2
\right. \nonumber \\
&& \left. ~~~~
- 89856 \Phi_{(1)\omega,\omega} \omega^4 \Nf^2
+ 376704 \Phi_{(1)\omega,\omega} \omega^3 \Nf^2
- 617472 \Phi_{(1)\omega,\omega} \omega^2 \Nf^2
\right. \nonumber \\
&& \left. ~~~~
+ 276480 \Phi_{(1)\omega,\omega} \omega \Nf^2
- 46224 \Phi_{(1)\omega,\omega} \omega^5 \Nf
+ 618192 \Phi_{(1)\omega,\omega} \omega^4 \Nf
\right. \nonumber \\
&& \left. ~~~~
- 2741472 \Phi_{(1)\omega,\omega} \omega^3 \Nf
+ 4091904 \Phi_{(1)\omega,\omega} \omega^2 \Nf
\right. \nonumber \\
&& \left. ~~~~
- 235008 \Phi_{(1)\omega,\omega} \omega \Nf
- 443880 \Phi_{(1)\omega,\omega} \omega^5
+ 5942808 \Phi_{(1)\omega,\omega} \omega^4
\right. \nonumber \\
&& \left. ~~~~
- 28479600 \Phi_{(1)\omega,\omega} \omega^3
+ 56215296 \Phi_{(1)\omega,\omega} \omega^2
- 35894016 \Phi_{(1)\omega,\omega} \omega
\right. \nonumber \\
&& \left. ~~~~
+ 11664 \Phi_{(2)\omega,\omega} \omega^5 \Nf
- 159408 \Phi_{(2)\omega,\omega} \omega^4 \Nf
+ 832032 \Phi_{(2)\omega,\omega} \omega^3 \Nf
\right. \nonumber \\
&& \left. ~~~~
- 1982880 \Phi_{(2)\omega,\omega} \omega^2 \Nf
+ 1804032 \Phi_{(2)\omega,\omega} \omega \Nf
+ 124416 \Phi_{(2)\omega,\omega} \Nf
\right. \nonumber \\
&& \left. ~~~~
- 192456 \Phi_{(2)\omega,\omega} \omega^5
+ 2630232 \Phi_{(2)\omega,\omega} \omega^4
- 13728528 \Phi_{(2)\omega,\omega} \omega^3
\right. \nonumber \\
&& \left. ~~~~
+ 32717520 \Phi_{(2)\omega,\omega} \omega^2
- 29766528 \Phi_{(2)\omega,\omega} \omega
- 2052864 \Phi_{(2)\omega,\omega}
\right. \nonumber \\
&& \left. ~~~~
- 1296 \Phi_{(2)1,\omega} \omega^6 \Nf
+ 106272 \Phi_{(2)1,\omega} \omega^5 \Nf
- 808704 \Phi_{(2)1,\omega} \omega^4 \Nf
\right. \nonumber \\
&& \left. ~~~~
+ 1700352 \Phi_{(2)1,\omega} \omega^3 \Nf
- 331776 \Phi_{(2)1,\omega} \omega^2 \Nf
+ 21384 \Phi_{(2)1,\omega} \omega^6
\right. \nonumber \\
&& \left. ~~~~
- 1753488 \Phi_{(2)1,\omega} \omega^5
+ 13343616 \Phi_{(2)1,\omega} \omega^4
- 28055808 \Phi_{(2)1,\omega} \omega^3
\right. \nonumber \\
&& \left. ~~~~
+ 5474304 \Phi_{(2)1,\omega} \omega^2
- 6144 \zeta_3 \omega^5 \Nf^2
- 123136 \omega^5 \Nf^2
\right. \nonumber \\
&& \left. ~~~~
+ 73728 \zeta_3 \omega^4 \Nf^2
+ 1477632 \omega^4 \Nf^2
- 294912 \zeta_3 \omega^3 \Nf^2
\right. \nonumber \\
&& \left. ~~~~
- 5910528 \omega^3 \Nf^2
+ 393216 \zeta_3 \omega^2 \Nf^2
+ 7880704 \omega^2 \Nf^2
\right. \nonumber \\
&& \left. ~~~~
- 147456 \zeta_3 \omega^5 \Nf
+ 4157856 \omega^5 \Nf
+ 1715040 \zeta_3 \omega^4 \Nf
\right. \nonumber \\
&& \left. ~~~~
- 49894272 \omega^4 \Nf
- 6689088 \zeta_3 \omega^3 \Nf
+ 199577088 \omega^3 \Nf
\right. \nonumber \\
&& \left. ~~~~
+ 8939520 \zeta_3 \omega^2 \Nf
- 266102784 \omega^2 \Nf
- 746496 \zeta_3 \omega \Nf
\right. \nonumber \\
&& \left. ~~~~
+ 4105728 \zeta_3 \omega^5 
- 23466672 \omega^5
- 48370608 \zeta_3 \omega^4 
\right. \nonumber \\
&& \left. ~~~~
+ 281600064 \omega^4
+ 190659744 \zeta_3 \omega^3 
- 1126400256 \omega^3
\right. \nonumber \\
&& \left. ~~~~
- 254555136 \zeta_3 \omega^2 
+ 1501867008 \omega^2
\right. \nonumber \\
&& \left. ~~~~
+ 12317184 \zeta_3 \omega 
\right] 
\frac{a^4}{6912 \omega^2 [\omega-4]^2} ~+~ O(a^5)
\end{eqnarray}
where we have introduced the shorthand notation 
\begin{eqnarray}
\Phi_{(n)\,1,\omega} &=& \Phi_{(n)} \left(1,\omega\right) ~~~,~~~
\Phi_{(n)\,\omega,\omega} ~=~
\Phi_{(n)} \left( \frac{1}{\omega},\frac{1}{\omega} \right) \nonumber \\
\Omega_{(n)\,1,\omega} &=& \Omega_{(n)} \left(1,\omega\right) ~~~,~~~
\Omega_{(n)\,\omega,\omega} ~=~
\Omega_{(n)} \left( \frac{1}{\omega},\frac{1}{\omega} \right) ~.
\end{eqnarray}
Analytic expressions for this $\beta$-function for an arbitrary colour group
together with all the other renormalization group functions and conversion 
functions in both gauges are included in an attached data file which also
contains the six $\MSbar$ vertex functions. In addition the decomposition of 
the vertex functions into the tensor basis are also provided in the data file. 
While we have presented the $SU(3)$ expression the arbitrary colour result in 
the Landau gauge shares the same property that the one and two loop terms are 
in agreement with the scheme independent parts which were first computed in 
\cite{12,13,14,15}. For a non-zero $\alpha$ the two loop term is in fact 
$\alpha$ dependent as one would expect. This is because $\alpha$ can be 
regarded as a second coupling constant and the theorem which shows that the 
$\beta$-function is scheme independent at two loops only applies to a theory 
with a single coupling constant. The main difference of the $\beta$-function, 
however, is in the scheme dependent part which is clearly $\omega$ dependent. 
Moreover, it contains poles at $\omega$~$=$~$0$ and $4$ as expected from the 
kinematics of the external momenta of the $3$-point functions. The same feature
occurs in all the other renormalization group functions in this and the other 
schemes in the scheme dependent terms for an arbitrary colour group. Therefore,
in terms of validity our results are restricted to the range 
$0$~$<$~$\omega$~$<$~$4$. 

The remainder of our results will be given in numerical form for both gauges as
this is the most straightforward way to compare the same expressions in 
different schemes as well as to see the effect of varying $\omega$ within a 
renormalization group function. As noted earlier we will focus specifically on 
the cases of $\omega$~$=$~$\half$ and $2$ so that one can gauge how scheme
independent the renormalization group invariant critical exponents are. While 
the difference in these values of $\omega$ from the MOM case of unity are not 
the same we note that the difference in $|\ln \omega|$ from zero for these
$\omega$ values is equivalent. In order to obtain numerical information we 
evaluate each of the special functions which appear at both values of $\omega$ 
and these are 
\begin{eqnarray}
\Phi_{(1)} (2,2) &=& 1.45542479 ~~~,~~~ 
\Phi_{(2)} (2,2) ~=~ 3.32388397 \nonumber \\ 
\Phi_{(2)} (1,\half) &=& 8.59797371 ~~~,~~~ 
\Omega_{(2)} (2,2) ~=~ 8.69451627 \nonumber \\ 
\Omega_{(2)} (1,\half) &=& 2.86508328 
\end{eqnarray}
for the case when $\omega$~$=$~$\half$ and
\begin{eqnarray}
\Phi_{(1)} (\half,\half) &=& 3.66386238 ~~~,~~~ 
\Phi_{(2)} (\half,\half) ~=~ 11.86733462 \nonumber \\ 
\Phi_{(2)} (1,2) &=& 4.54422692 ~~~,~~~ 
\Omega_{(2)} (\half,\half) ~=~ -~ 1.35231402 \nonumber \\            
\Omega_{(2)} (1,2) &=& 7.88849843
\end{eqnarray}
for $\omega$~$=$~$2$. The first stage in the construction of the three loop
renormalization group functions is the determination of the various conversion
functions in different schemes. 

As our main focus with be on two specific values of $\omega$ we concentrate on 
these cases. Moreover, all the numerical results presented in the remainder of 
this section are for the $SU(3)$ colour group. The respective coupling constant
mapping functions for arbitrary $\alpha$ are 
\begin{eqnarray}
\left. C_g^{\iMOMgs}(a,\alpha) \right|_{\omega=\half} &=&
1 + \left[ -~ 0.088593 \alpha^3 
+ 0.145415 \alpha^2 
+ 0.973734 \alpha
+ 1.545158 \Nf 
\right. \nonumber \\
&& \left. ~~~~~~
- 12.931045 \right] a 
\nonumber \\
&&
+~ \left[ 0.007849 \alpha^6 
+ 0.073901 \alpha^5 
- 0.813066 \alpha^4 
- 0.462374 \Nf \alpha^3 
\right. \nonumber \\
&& \left. ~~~~
+ 0.838289 \alpha^3 
+ 0.005079 \Nf \alpha^2 
- 6.028249 \alpha^2 
+ 2.747737 \Nf \alpha 
\right. \nonumber \\
&& \left. ~~~~
-~ 30.527809 \alpha 
+ 0.275213 \Nf^2 
+ 34.734827 \Nf 
- 228.297272 \right] a^2 \nonumber \\
&& +~ O(a^3) \nonumber \\
\left. C_g^{\iMOMhs}(a,\alpha) \right|_{\omega=\half} &=&
1 + \left[ - 0.295179 \alpha^2 
- 1.225290 \alpha 
+ 0.555556 \Nf 
- 9.326914 \right] a 
\nonumber \\
&&
+~ \left[ 0.127510 \alpha^4 
- 0.479424 \alpha^3 
- 0.455080 \Nf \alpha^2 
- 1.607853 \alpha^2 
\right. \nonumber \\
&& \left. ~~~~
-~ 0.363822 \Nf \alpha 
- 18.6313969 \alpha 
- 0.1543210 \Nf^2 
+ 27.1598858 \Nf 
\right. \nonumber \\
&& \left. ~~~~
-~ 195.1062372 \right] a^2 ~+~ O(a^3) \nonumber \\
\left. C_g^{\iMOMqs}(a,\alpha) \right|_{\omega=\half} &=&
1 + \left[ -~ 0.103925 \alpha^2 
+ 0.191374 \alpha 
+ 0.555556 \Nf 
- 7.559667 \right] a 
\nonumber \\
&&
+~ \left[ -~ 0.020540 \alpha^4 
- 0.760731 \alpha^3 
- 0.047946 \Nf \alpha^2 
- 0.025079 \alpha^2 
\right. \nonumber \\
&& \left. ~~~~
+~ 1.106752 \Nf \alpha 
- 8.869723 \alpha 
- 0.154321 \Nf^2 
+ 26.411323 \Nf 
\right. \nonumber \\
&& \left. ~~~~
-~ 130.701756 \right] a^2
~+~ O(a^3) \nonumber \\
\left. C_g^{\iMOMmgs}(a,\alpha) \right|_{\omega=\half} &=&
1 + \left[ 0.048472 \alpha^2 + 0.930452 \alpha + 1.545158 \Nf - 9.614714 
\right] a \nonumber \\
&& + \left[ -~ 0.093437 \alpha^4 - 0.778673 \alpha^3 + 0.001693 \alpha^2 \Nf
- 2.387913 \alpha^2  
\right. \nonumber \\ 
&& \left. ~~~~
+ 2.447811 \alpha \Nf - 10.719814 \alpha + 0.275213 \Nf^2 
+ 36.437144 \Nf \right. \nonumber \\
&& \left. ~~~~ - 231.335376 \right] a^2 ~+~ O(a^3) \nonumber \\
\left. C_g^{\iMOMmhs}(a,\alpha) \right|_{\omega=\half} &=&
1 + \left[ - 0.125000 \alpha^2 - 1.084141 \alpha + 0.555556 \Nf 
- 9.922960 \right] a \nonumber \\
&& + \left[ 0.023437 \alpha^4 - 0.536449 \alpha^3 - 0.208333 \alpha^2 \Nf 
- 3.642248 \alpha^2 \right. \nonumber \\
&& \left. ~~~~ - 0.179774 \alpha \Nf - 30.505406 \alpha - 0.154321 \Nf^2
+ 29.832673 \Nf \right. \nonumber \\
&& \left. ~~~~ - 226.633850 \right] a^2 ~+~ O(a^3) \nonumber \\
\left. C_g^{\iMOMmqs}(a,\alpha) \right|_{\omega=\half} &=&
1 + \left[ - 0.034642 \alpha^2 + 0.168280 \alpha + 0.555556 \Nf 
- 6.663671 \right] a \nonumber \\ 
&& + \left[ - 0.002282 \alpha^4 - 0.203870 \alpha^3 - 0.015982 \alpha^2 \Nf  
- 0.5936004 \alpha^2 \right. \nonumber \\ 
&& \left. ~~~~ + 0.760445 \alpha \Nf - 3.103542 \alpha - 0.154321 \Nf^2
+ 25.879816 \Nf \right. \nonumber \\ 
&& \left. ~~~~ - 128.426739 \right] a^2 ~+~ O(a^3)
\end{eqnarray}
for $\omega$~$=$~$\half$ and
\begin{eqnarray}
\left. C_g^{\iMOMgs}(a,\alpha) \right|_{\omega=2} &=&
1 + \left[ -~ 0.167112 \alpha^3 
+ 0.148595 \alpha^2 
+ 2.246555 \alpha 
+ 1.925389 \Nf 
\right. \nonumber \\
&& \left. ~~~~~~
-~ 13.345319 \right] a 
\nonumber \\
&&
+~ \left[ 0.027927 \alpha^6 
+~ 0.138338 \alpha^5 
- 1.346201 \alpha^4 
- 0.844818 \Nf \alpha^3 
\right. \nonumber \\
&& \left. ~~~~
+~ 1.116340 \alpha^3 
- 0.119722 \Nf \alpha^2 
- 8.125459 \alpha^2 
+ 5.258434 \Nf \alpha 
\right. \nonumber \\
&& \left. ~~~~
-~ 47.432029 \alpha 
+ 0.961103 \Nf^2 
+ 29.304598 \Nf 
- 192.265972 \right] a^2 \nonumber \\
&& +~ O(a^3) \nonumber \\
\left. C_g^{\iMOMhs}(a,\alpha) \right|_{\omega=2} &=&
1 + \left[ -~ 0.115070 \alpha^2 
- 0.876052 \alpha 
+ 0.555556 \Nf 
- 9.135824 \right] a 
\nonumber \\
&&
+~ \left[ -~ 0.013920 \alpha^4 
- 1.243069 \alpha^3 
- 0.311896 \Nf \alpha^2 
- 2.121830 \alpha^2 
\right. \nonumber \\
&& \left. ~~~~
+~ 0.122619 \Nf \alpha 
- 19.093491 \alpha 
- 0.154321 \Nf^2 
+ 27.357866 \Nf 
\right. \nonumber \\
&& \left. ~~~~
-~ 186.353580 \right] a^2 ~+~ O(a^3) \nonumber \\
\left. C_g^{\iMOMqs}(a,\alpha) \right|_{\omega=2} &=&
1 + \left[ 0.375000 \alpha^2 
+ 2.587278 \alpha 
+ 0.555556 \Nf 
- 9.565419 \right] a 
\nonumber \\
&&
+~ \left[ -~ 0.070312 \alpha^4 
+ 0.254398 \alpha^3 
+ 0.278426 \Nf \alpha^2 
+ 2.718626 \alpha^2 
\right. \nonumber \\
&& \left. ~~~~
+~ 1.999510 \Nf \alpha 
- 9.199750 \alpha 
- 0.154321 \Nf^2 
+ 29.206664 \Nf 
\right. \nonumber \\
&& \left. ~~~~
-~ 131.140706 \right] a^2 ~+~ O(a^3) \nonumber \\
\left. C_g^{\iMOMmgs}(a,\alpha) \right|_{\omega=2} &=&
1 + \left[ 0.049532 \alpha^2 + 2.032448 \alpha + 1.925389 \Nf
- 8.842253 \right] a \nonumber \\
&& + \left[ - 0.158617 \alpha^4 - 1.135650 \alpha^3 - 0.039907 \alpha^ 2 \Nf
- 2.056887 \alpha^2 \right. \nonumber \\
&& \left. ~~~~ + 4.543830 \alpha \Nf - 1.192456 \alpha + 0.961103 \Nf^2
+ 33.612910 \Nf \right. \nonumber \\
&& \left. ~~~~ - 189.116753 \right] a^2 ~+~ O(a^3) \nonumber \\
\left. C_g^{\iMOMmhs}(a,\alpha) \right|_{\omega=2} &=&
1 + \left[ - 0.125000 \alpha^2 - 0.888591 \alpha + 0.555556 \Nf
- 11.044349 \right] a \nonumber \\
&& + \left[ 0.023437 \alpha^4 - 0.840201 \alpha^3 - 0.208333 \alpha^2 \Nf
- 2.145247 \alpha^2 \right. \nonumber \\
&& \left. ~~~~ + 0.318758 \alpha \Nf - 27.105424 \alpha - 0.154321 \Nf^2
+ 31.291364 \Nf \right. \nonumber \\
&& \left. ~~~~ - 204.314912 \right] a^2 ~+~ O(a^3) \nonumber \\
\left. C_g^{\iMOMmqs}(a,\alpha) \right|_{\omega=2} &=&
1 + \left[ 0.125000 \alpha^2 + 1.817289 \alpha + 0.555556 \Nf
- 8.476166 \right] a \nonumber \\
&& + \left[ - 0.007812 \alpha^4 + 0.073625 \alpha^3 + 0.092809 \alpha^2 \Nf
+ 3.101975 \alpha^2 \right. \nonumber \\
&& \left. ~~~~ + 1.384360 \alpha \Nf + 9.787599 \alpha - 0.154321 \Nf^2
+ 28.679908 \Nf \right. \nonumber \\
&& \left. ~~~~ - 137.007566 \right] a^2 ~+~ O(a^3)
\end{eqnarray}
for $\omega$~$=$~$2$. We recall that in all the conversion functions the 
coupling constant and gauge parameter variables are the $\MSbar$ ones. The 
corresponding expressions for the quark mass operator are more compact in that  
\begin{eqnarray}
\left. C_{\bar{\psi}\psi}^{\iMOMxs}(a,\alpha) \right|_{\omega=\half} &=&
1 + \left[ -~ 0.363050 \alpha - 2.422484 \right] a \nonumber \\
&& +~ \left[ -~ 0.685057 \alpha^2 + 0.554007 \alpha + 5.987945 \Nf 
- 64.755701 \right] a^2 \nonumber \\
&& +~ O(a^3) \nonumber \\ 
\left. C_{\bar{\psi}\psi}^{\iMOMxs}(a,\alpha) \right|_{\omega=2} &=&
1 + \left[ 1.109242 \alpha + 1.994391 \right] a \nonumber \\
&& +~ \left[ 3.726210 \alpha^2 + 13.658375 \alpha + 1.080306 \Nf 
+ 38.567136 \right] a^2 \nonumber \\
&& +~ O(a^3) \nonumber \\ 
\left. C_{\bar{\psi}\psi}^{\iMOMmxs}(a,\alpha) \right|_{\omega=\half} &=&
1 + \left[ -~ 0.272288 \alpha - 2.422484 \right] a \nonumber \\ 
&& + \left[ -~ 0.198147 \alpha^2 - 1.898939 \alpha + 5.987945 \Nf 
- 65.436420 \right] a^2 \nonumber \\
&& +~ O(a^3) \nonumber \\
\left. C_{\bar{\psi}\psi}^{\iMOMmxs}(a,\alpha) \right|_{\omega=2} &=&
1 + \left[ 0.831931 \alpha + 1.994391 \right] a \nonumber \\ 
&& + \left[ 1.524041 \alpha^2 + 17.315197 \alpha + 1.080306 \Nf 
+ 40.646964 \right] a^2 \nonumber \\
&& +~ O(a^3) ~.
\end{eqnarray}
The same feature emerges here as in the $\omega$~$=$~$1$ situation in that in 
each scheme the conversion function is the same for a particular value of 
$\omega$. This is also the case for the various wave function conversion
functions which are  
\begin{eqnarray}
C_A^{\iMOMxs}(a,\alpha) &=&
1 + \left[ 9 \alpha^2 C_A + 18 \alpha C_A + 97 C_A - 80 \Nf T_F \right] 
\frac{a}{36} \nonumber \\
&& +~ \left[ 810 \alpha^3 C_A^2 + 2430 \alpha^2 C_A^2 
+ 5184 \zeta_3 \alpha C_A^2 + 2817 \alpha C_A^2 - 2880 \alpha C_A \Nf T_F
\right. \nonumber \\
&& \left. ~~~~
-~ 7776 \zeta_3 C_A^2 + 83105 C_A^2 - 20736 \zeta_3 C_A \Nf T_F
- 69272 C_A \Nf T_F 
\right. \nonumber \\
&& \left. ~~~~
+~ 41472 \zeta_3 C_F \Nf T_F - 47520 C_F \Nf T_F
+ 12800 \Nf^2 T_F^2 \right] \frac{a^2}{2592} ~+~ O(a^3) \nonumber \\
C_c^{\iMOMxs}(a,\alpha) &=&
1 + C_A a \nonumber \\
&& +~ \left[ -~ 36 \zeta_3 \alpha^2 C_A + 72 \alpha^2 C_A 
+ 72 \zeta_3 \alpha C_A - 21 \alpha C_A - 180 \zeta_3 C_A + 1943 C_A 
\right. \nonumber \\
&& \left. ~~~~
-~ 760 \Nf T_F \right]
\frac{C_A a^2}{192} ~+~ O(a^3) \nonumber \\
C_\psi^{\iMOMxs}(a,\alpha) &=&
1 - \alpha C_F a \nonumber \\
&& +~ \left[ -~ 9 \alpha^2 C_A + 8 \alpha^2 C_F + 24 \zeta_3 \alpha C_A
- 52 \alpha C_A + 24 \zeta_3 C_A - 82 C_A + 5 C_F 
\right. \nonumber \\
&& \left. ~~~~
+~ 28 \Nf T_F \right]
\frac{C_F a^2}{8} ~+~ O(a^3) ~.
\end{eqnarray}
They are totally equivalent to the MOM conversion functions of \cite{20,21,22}.
This is not unexpected as the renormalization of the $2$-point functions here 
was carried out using the MOM renormalization criterion. As the corresponding 
conversion functions in the MAG are more involved but are also equivalent to 
those of the MOM scheme, \cite{65,66}, we record their explicit $\iMOMmx$ 
numerical values which are,
\begin{eqnarray}
C_A^{\iMOMmxs}(a,\alpha) &=&
1 + \left[ 0.250000 \alpha^2 + \alpha - 1.111111 \Nf + 5.083333 \right] a
\nonumber \\
&& + \left[ 0.809362 \alpha^3 + 6.131144 \alpha^2 - 1.111111 \alpha \Nf
+ 11.092087 \alpha
\right. \nonumber \\
&& \left. ~~~
+ 1.234568 \Nf^2 - 45.843483 \Nf + 164.706865 \right] a^2 ~+~ O(a^3)
\nonumber \\
C_c^{\iMOMmxs}(a,\alpha) &=&
1 + 5.000000 a \nonumber \\
&& + \left[ 0.937500 \alpha^2 + 21.048378 \alpha - 9.895833 \Nf + 150.659355
\right] a^2 ~+~ O(a^3) \nonumber \\
C_\psi^{\iMOMmxs}(a,\alpha) &=&
1 - \alpha a \nonumber \\
&& + \left[ - 0.625000 \alpha^2 - 18.825317 \alpha + 2.333333 \Nf - 31.214206
\right] a^2 ~+~ O(a^3) ~.
\end{eqnarray}
Equipped with the coupling constant and mass operator conversion functions we
can now determine the respective {\em three loop} $\iMOMx$ and $\iMOMmx$
renormalization group functions. First, in order to compare we recall both the 
$\MSbar$ and MOM scheme $\beta$-functions in numerical form which are,
\cite{12,13,14,15,16,20,21,22,65,66},
\begin{eqnarray}
\beta^{\MSbars}(a) &=& 
\left[ 0.666667 \Nf - 11.000000 \right] a^2 
+ \left[ 12.666667 \Nf - 102.000000 \right] a^3
\nonumber \\
&& + \left[ - 6.018518 \Nf^2 + 279.611111 \Nf - 1428.500000 \right] a^4
\nonumber \\
&& +~ O(a^5) \nonumber \\
\beta^{\MOMgs}(a,0) &=& 
\left[ 0.666667 \Nf - 11.000000 \right] a^2 
+ \left[ 12.666667 \Nf - 102.000000 \right] a^3 
\nonumber \\
&& + \left[ - 2.658115 \Nf^3 + 67.089536 \Nf^2 - 0.565929 \Nf - 1570.984380 
\right] a^4 \nonumber \\
&& +~ O(a^5) \nonumber \\ 
\beta^{\MOMhs}(a,0) &=& 
\left[ 0.666667 \Nf - 11.000000 \right] a^2 
+ \left[ 12.666667 \Nf - 102.000000 \right] a^3 
\nonumber \\
&& + \left[ - 21.502818 \Nf^2 + 617.647154 \Nf - 2813.492948 
\right] a^4 ~+~ O(a^5) \nonumber \\ 
\beta^{\MOMqs}(a,0) &=& 
\left[ 0.666667 \Nf - 11.000000 \right] a^2 
\nonumber \\
&& + \left[ 12.666667 \Nf - 102.000000 \right] a^3 
\nonumber \\
&& + \left[ - 22.587812 \Nf^2 + 588.654845 \Nf - 1843.65273 
\right] a^4 ~+~ O(a^5) \nonumber \\ 
\beta^{\MOMmgs}(a,0) &=&
\left[ 0.666667 \Nf - 11.000000 \right] a^2 ~+~ 
\left[ 12.666667 \Nf - 93.608510 \right] a^3 \nonumber \\
&& + \left[ - 2.658115 \Nf^3 + 54.791594 \Nf^2 + 401.565562 \Nf 
- 3543.358228 \right] a^4 \nonumber \\
&& +~ O(a^5) \nonumber \\
\beta^{\MOMmhs}(a,0) &=&
\left[ 0.666667 \Nf - 11.000000 \right] a^2 ~+~ 
\left[ 12.666667 \Nf - 108.000000 \right] a^3 \nonumber \\
&& + \left[ - 25.035332 \Nf^2 + 674.085832 \Nf - 2991.050472 \right] a^4
\nonumber \\
&& +~ O(a^5) \nonumber \\
\beta^{\MOMmqs}(a,0) &=&
\left[ 0.666667 \Nf - 11.000000 \right] a^2 ~+~ 
\left[ 12.666667 \Nf - 96.936557 \right] a^3 \nonumber \\
&& + \left[ - 22.587812 \Nf^2 + 627.275918 \Nf - 2266.490127 \right] a^4 ~+~ 
O(a^5)
\end{eqnarray} 
where we have restricted to the Landau gauge for the linear covariant gauge and
$\alpha$~$=$~$0$ for the MAG for the MOM schemes. Our notation for the
renormalization group functions is that the second argument in a MOM or iMOM
scheme is the gauge parameter. Such an argument is not needed for the $\MSbar$
$\beta$-function. In the $\MSbar$, MOM and $\iMOMx$ scheme renormalization 
group functions we recall that the coupling constant and gauge parameter 
correspond to the variable in the specific case given by the label on the left 
hand side. This is to avoid encumbrancing the variables themselves with labels.
The two sets of $\beta$-functions are 
\begin{eqnarray}
\left. \beta^{\iMOMgs}(a,0) \right|_{\omega=\half} &=&
\left[ 0.666667 \Nf - 11.000000 \right] a^2
+ \left[ 12.666667 \Nf - 102.000000 \right] a^3 \nonumber \\
&& + \left[ - 1.958625 \Nf^3 + 45.770375 \Nf^2 + 154.329226 \Nf
- 1973.775606 \right] a^4 \nonumber \\
&& +~ O(a^5) \nonumber \\
\left. \beta^{\iMOMhs}(a,0) \right|_{\omega=\half} &=&
\left[ 0.666667 \Nf - 11.000000 \right] a^2
+ \left[ 12.666667 \Nf - 102.000000 \right] a^3 \nonumber \\
&& + \left[ - 21.248801 \Nf^2 + 615.665280 \Nf - 2861.242336 \right] a^4 ~+~
O(a^5) \nonumber \\
\left. \beta^{\iMOMqs}(a,0) \right|_{\omega=\half} &=&
\left[ 0.666667 \Nf - 11.000000 \right] a^2
+ \left[ 12.666667 \Nf - 102.000000 \right] a^3 \nonumber \\
&& + \left[ - 21.559789 \Nf^2 + 599.589376 \Nf - 2133.132445 \right] a^4 ~+~
O(a^5) \nonumber \\
\left. \beta^{\iMOMmgs}(a,0) \right|_{\omega=\half} &=&
\left[ 0.666667 \Nf - 11.000000 \right] a^2 ~+~ 
\left[ 12.666667 \Nf - 96.417290 \right] a^3 \nonumber \\
&& + \left[ - 1.958625 \Nf^3 + 36.668278 \Nf^2 + 469.963542 \Nf
- 3720.350935 \right] a^4 \nonumber \\
&& +~ O(a^5) \nonumber \\
\left. \beta^{\iMOMmhs}(a,0) \right|_{\omega=\half} &=&
\left[ 0.666667 \Nf - 11.000000 \right] a^2 ~+~ 
\left[ 12.666667 \Nf - 108.504849 \right] a^3 \nonumber \\
&& + \left[ - 24.371002 \Nf^2 + 689.727288 \Nf
- 3346.349782 \right] a^4 ~+~ O(a^5) \nonumber \\
\left. \beta^{\iMOMmqs}(a,0) \right|_{\omega=\half} &=&
\left[ 0.666667 \Nf - 11.000000 \right] a^2 ~+~ 
\left[ 12.666667 \Nf - 100.990317 \right] a^3 \nonumber \\
&& + \left[ - 21.514813 \Nf^2 + 630.898042 \Nf
- 2435.486351 \right] a^4 \nonumber \\
&& +~ O(a^5) 
\end{eqnarray}
and 
\begin{eqnarray}
\left. \beta^{\iMOMgs}(a,0) \right|_{\omega=2} &=&
\left[ 0.666667 \Nf - 11.000000 \right] a^2
+ \left[ 12.666667 \Nf - 102.000000 \right] a^3 \nonumber \\
&& + \left[ - 3.752885 \Nf^3 + 99.867703 \Nf^2 - 234.213856 \Nf
- 976.833287 \right] a^4 \nonumber \\
&& +~ O(a^5) \nonumber \\
\left. \beta^{\iMOMhs}(a,0) \right|_{\omega=2} &=&
\left[ 0.666667 \Nf - 11.000000 \right] a^2
+ \left[ 12.666667 \Nf - 102.000000 \right] a^3 \nonumber \\
&& + \left[ - 21.654322 \Nf^2 + 617.879121 \Nf - 2746.474396 \right] a^4 ~+~
O(a^5) \nonumber \\
\left. \beta^{\iMOMqs}(a,0) \right|_{\omega=2} &=&
\left[ 0.666667 \Nf - 11.000000 \right] a^2
+ \left[ 12.666667 \Nf - 102.000000 \right] a^3 \nonumber \\
&& + \left[ - 23.801168 \Nf^2 + 563.445891 \Nf - 1355.780477 \right] a^4 ~+~ 
O(a^5) \nonumber \\ 
\left. \beta^{\iMOMmgs}(a,0) \right|_{\omega=2} &=&
\left[ 0.666667 \Nf - 11.000000 \right] a^2 ~+~
\left[ 12.666667 \Nf - 89.805313 \right] a^3 \nonumber \\
&& + \left[ - 3.752885 \Nf^3 + 82.563084 \Nf^2 + 297.046404 \Nf
- 3156.729291 \right] a^4 \nonumber \\
&& +~ O(a^5) \nonumber \\
\left. \beta^{\iMOMmhs}(a,0) \right|_{\omega=2} &=&
\left[ 0.666667 \Nf - 11.000000 \right] a^2 ~+~
\left[ 12.666667 \Nf - 107.331545 \right] a^3 \nonumber \\
&& + \left[ - 25.485264 \Nf^2 + 636.479467 \Nf
- 2354.843991 \right] a^4 ~+~ O(a^5)
\nonumber \\
\left. \beta^{\iMOMmqs}(a,0) \right|_{\omega=2} &=&
\left[ 0.666667 \Nf - 11.000000 \right] a^2 ~+~
\left[ 12.666667 \Nf - 91.096267 \right] a^3 \nonumber \\
&& + \left[ - 23.905680 \Nf^2 + 614.725445 \Nf
- 2055.563293 \right] a^4 \nonumber \\
&& +~ O(a^5) ~.
\end{eqnarray}
In the Landau gauge the correct scheme independent part emerges in each case
with the scheme dependence only present at three loops. However, we can now
quantify the effect of varying the parameter $\omega$ between $\half$ and $2$ 
in comparison with the symmetric point MOM scheme of \cite{20,21}. If one 
examines the $\Nf$ independent part of each three loop term, for example, it is
evident that the corresponding coefficients in the $\omega$ case lie roughly 
halfway between the coefficients for $\omega$~$=$~$\half$ and $2$ in each of 
the respective $\iMOMx$ schemes. This is consistent with our expectations. 
However, the comparison with say the $\iMOMg$ $\omega$~$=$~$2$ $\beta$-function
with the $\MSbar$ case is not appropriate as the coupling constants are not as 
similar as within an $\omega$ comparison. Moreover, $\beta$-functions are not 
physically meaningful quantities and the running of the coupling constant with 
scale is not the same at three loops in each of the schemes. This is one of the
reasons why it is more beneficial to examine critical exponents since they are 
renormalization group invariants. 

We repeat the exercise for the quark mass operator. First, the known $\MSbar$ 
and MOM scheme results for comparison are, \cite{86,87,88,89,90,31},
\begin{eqnarray}
\gamma_{\bar{\psi}\psi}^{\MSbars}(a) &=&
-~ 4.000000 a + \left[ 2.222222 \Nf - 67.333333 \right] a^2 \nonumber \\
&& +~ \left[ 1.728395 \Nf^2 + 146.183776 \Nf - 1249.000000 \right] a^3 ~+~ 
O(a^4) \nonumber \\
\gamma_{\bar{\psi}\psi}^{\MOMgs}(a,0) &=&
-~ 4.000000 a
+ \left[ -~ 11.014658 \Nf + 31.535915 \right] a^2 \nonumber \\
&& +~ \left[ -~ 48.143325 \Nf^2 + 263.855175 \Nf + 354.125435 \right] a^3 ~+~
O(a^4) \nonumber \\
\gamma_{\bar{\psi}\psi}^{\MOMhs}(a,0) &=&
-~ 4.000000 a
+ \left[ -~ 1.791876 \Nf - 0.240939 \right] a^2 \nonumber \\
&& +~ \left[ -~ 2.666667 \Nf^2 - 3.207195 \Nf + 759.902600 \right] a^3 ~+~
O(a^4) \nonumber \\
\gamma_{\bar{\psi}\psi}^{\MOMqs}(a,0) &=&
-~ 4.000000 a
+ \left[ -~ 1.791876 \Nf - 7.570942 \right] a^2 \nonumber \\
&& +~ \left[ -~ 2.666667 \Nf^2 - 16.284387 \Nf + 324.949029 \right] a^3 ~+~
O(a^4) \nonumber \\
\gamma_{\bar{\psi}\psi}^{\MOMmgs}(a,0) &=&
-~ 4.000000 a
+ \left[ -~ 11.014658 \Nf + 1.580982 \right] a^2 \nonumber \\
&& +~ \left[ -~ 48.143325 \Nf^2 + 24.289770 \Nf + 993.171684 \right] a^3 ~+~
O(a^4) \nonumber \\
\gamma_{\bar{\psi}\psi}^{\MOMmhs}(a,0) &=&
-~ 4.000000 a
+ \left[ -~ 1.791876 \Nf + 9.540363 \right] a^2 \nonumber \\
&& +~ \left[ -~ 2.666667 \Nf^2 - 16.444922 \Nf + 861.561558 \right] a^3 ~+~
O(a^4) \nonumber \\
\gamma_{\bar{\psi}\psi}^{\MOMmqs}(a,0) &=&
-~ 4.000000 a
+ \left[ -~ 1.791876 \Nf - 14.868106 \right] a^2 \nonumber \\
&& +~ \left[ -~ 2.666667 \Nf^2 - 23.628280 \Nf + 368.925469 \right] a^3 ~+~
O(a^4) 
\end{eqnarray}
where we have computed the $\omega$~$=$~$1$ expressions for the MAG as a
corollary of the $\iMOMmx$ calculation. Again the second argument corresponds
to the gauge parameter in the non-$\MSbar$ schemes. We recall that the scheme 
dependence begins at two loops for the quark mass anomalous dimension. The 
corresponding $\iMOMx$ results are 
\begin{eqnarray}
\left. \gamma_{\bar{\psi}\psi}^{\iMOMgs}(a,0) \right|_{\omega=\half}
&=& -~ 4.000000 a
+ \left[ -~ 8.524052 \Nf + 9.467706 \right] a^2
\nonumber \\
&& +~ \left[ -~ 32.491101 \Nf^2 + 140.861801 \Nf + 357.940500 \right] a^3 ~+~ 
O(a^4) \nonumber \\
\left. \gamma_{\bar{\psi}\psi}^{\iMOMhs}(a,0) \right|_{\omega=\half}
&=& -~ 4.000000 a
+ \left[ -~ 0.607233 \Nf - 19.365345 \right] a^2
\nonumber \\
&& +~ \left[ -~ 2.666667 \Nf^2 + 36.838982 \Nf + 341.949868 \right] a^3 ~+~ 
O(a^4) \nonumber \\
\left. \gamma_{\bar{\psi}\psi}^{\iMOMqs}(a,0) \right|_{\omega=\half}
&=& -~ 4.000000 a
+ \left[ -~ 0.607233 \Nf - 33.503320 \right] a^2
\nonumber \\
&& +~ \left[ -~ 2.666667 \Nf^2 + 30.680528 \Nf - 240.778923 \right] a^3 ~+~ 
O(a^4) \nonumber \\
\left. \gamma_{\bar{\psi}\psi}^{\iMOMmgs}(a,0) \right|_{\omega=\half} &=&
-~ 4.000000 a
+ \left[ -~ 8.524052 \Nf - 17.879808 \right] a^2 \nonumber \\
&& +~ \left[ -~ 32.491101 \Nf^2 - 29.514002 \Nf + 659.067463 \right] a^3 ~+~
O(a^4) \nonumber \\
\left. \gamma_{\bar{\psi}\psi}^{\iMOMmhs}(a,0) \right|_{\omega=\half} &=&
-~ 4.000000 a
+ \left[ -~ 0.607233 \Nf - 15.413838 \right] a^2 \nonumber \\
&& +~ \left[ -~ 2.666667 \Nf^2 + 20.098107 \Nf + 577.599012 \right] a^3 ~+~
O(a^4) \nonumber \\
\left. \gamma_{\bar{\psi}\psi}^{\iMOMmqs}(a,0) \right|_{\omega=\half} &=&
-~ 4.000000 a
+ \left[ -~ 0.607233 \Nf - 41.488147 \right] a^2 \nonumber \\
&& +~ \left[ -~ 2.666667 \Nf^2 + 29.318648 \Nf - 337.686951 \right] a^3 ~+~
O(a^4) 
\end{eqnarray}
for $\omega$~$=$~$\half$ and 
\begin{eqnarray}
\left. \gamma_{\bar{\psi}\psi}^{\iMOMgs}(a,0) \right|_{\omega=2}
&=& -~ 4.000000 a
+ \left[ -~ 14.510481 \Nf + 61.367526 \right] a^2
\nonumber \\
&& +~ \left[ -~ 74.6686432 \Nf^2 + 492.015439 \Nf + 158.572781 \right] a^3 ~+~ 
O(a^4) \nonumber \\
\left. \gamma_{\bar{\psi}\psi}^{\iMOMhs}(a,0) \right|_{\omega=2}
&=& -~ 4.000000 a
+ \left[ -~ 3.551816 \Nf + 27.691568 \right] a^2
\nonumber \\
&& +~ \left[ -~ 2.666667 \Nf^2 - 53.424921 \Nf + 1239.598236 \right] a^3 ~+~ 
O(a^4) \nonumber \\
\left. \gamma_{\bar{\psi}\psi}^{\iMOMqs}(a,0) \right|_{\omega=2}
&=& -~ 4.000000 a
+ \left[ -~ 3.551816 \Nf + 31.128323 \right] a^2
\nonumber \\
&& +~ \left[ -~ 2.666667 \Nf^2 - 60.202636 \Nf + 715.222060 \right] a^3 ~+~ 
O(a^4) \nonumber \\
\left. \gamma_{\bar{\psi}\psi}^{\iMOMmgs}(a,0) \right|_{\omega=2} &=&
-~ 4.000000 a
+ \left[ -~ 14.510481 \Nf + 27.838788 \right] a^2 \nonumber \\
&& +~ \left[ -~ 74.668643 \Nf^2 + 138.832502 \Nf + 1356.765556 \right] a^3 ~+~
O(a^4) \nonumber \\
\left. \gamma_{\bar{\psi}\psi}^{\iMOMmhs}(a,0) \right|_{\omega=2} &=&
-~ 4.000000 a
+ \left[ -~ 3.551816 \Nf + 45.455562 \right] a^2 \nonumber \\
&& +~ \left[ -~ 2.666667 \Nf^2 - 50.959511 \Nf + 921.958650 \right] a^3 ~+~
O(a^4) \nonumber \\
\left. \gamma_{\bar{\psi}\psi}^{\iMOMmqs}(a,0) \right|_{\omega=2} &=&
-~ 4.000000 a
+ \left[ -~ 3.551816 \Nf + 24.910097 \right] a^2 \nonumber \\
&& +~ \left[ -~ 2.666667 \Nf^2 - 77.968868 \Nf + 999.570302 \right] a^3 ~+~
O(a^4)
\end{eqnarray}
for $\omega$~$=$~$2$. Now if we compare the $\Nf$~$=$~$0$ two and three loop
terms within the same vertex scheme we find the same general trend that was
apparent in the $\beta$-function. First, at two loop the $\omega$~$=$~$1$
coefficient lies roughly halfway between the $\omega$~$=$~$\half$ and $2$
values. In this case for the ghost and quark vertex based schemes there is a
sign change in the coefficient across the range which does not affect this
overall observation. At three loops there is a slackening of the feature of the
$\omega$~$=$~$1$ coefficients lying roughly halfway between the other two
values. This is perhaps not surprising as the higher loop expressions are
teasing out the effective asymmetry in the range. It will be interesting to see
if this is evident in the critical exponent case and how pronounced it is. 

Our final results oriented remarks concern the internal checks on our 
computation. We have determined all the two loop renormalization group 
functions in two ways for each class of gauges. The first is the direct 
evaluation and renormalization of the $2$- and $3$-point functions in the 
respective schemes by using the various vertex functions as well as the 
$\iMOMx$ renormalization conditions to determine the renormalization constants.
From these we have produced the respective renormalization group functions. 
There is also an internal check at that stage in that the non-simple poles in 
$\epsilon$ in the renormalization constants are pre-determined by the simple 
poles at the previous loop orders. An error in these non-simple poles would 
have resulted in $\epsilon$ divergent renormalization group functions. With the
various parameters such as the two separate gauge parameters and the colour 
group Casimirs present this is a useful check. The second way we have 
constructed the renormalization group functions is via the conversion function 
route using the formalism of (\ref{rgemap}) once the various renormalization 
constants are available. Therefore we are able to verify that the 
renormalization group functions computed directly are consistent. Although this
method has the advantage that it automatically produces the {\em three} loop 
terms of the renormalization group functions. The three loop $\iMOMx$ coupling 
constant renormalization constants cannot be adduced from these as there is no 
point of contact with the finite part of the corresponding three loop $\iMOMx$ 
coupling constant renormalization constants. However for the wave function 
renormalizations we have directly renormalized the respective $2$-point 
functions at {\em three} loops and checked that they agree precisely with the 
three loop expressions constructed using the conversion functions. This 
represents a useful verification of the conversion function formalism. The 
final check on our results is that the $\omega$~$\to$~$1$ limit correctly 
emerges. 

\sect{Critical exponents.}

As an application of our results we now turn to the evaluation of various
critical exponents which are renormalization group invariants. Therefore the
values of the exponents in any scheme will be the same provided the
renormalization group functions are known to all orders. However, when one has
a truncated perturbative expansion the values for exponents in different
schemes will be different with the hope that the discrepancy reduces at high
loop order. This is part of our motivation for determining the renormalization
group functions in these new schemes here. Our choice of the two specific 
values of the parameter $\omega$ can be used to quantify the variation in some 
way. First, we summarize the formalism we will apply, \cite{31}, concentrating
on the Landau gauge for illustration. We define the $\beta$-function in a 
scheme ${\cal S}$ by 
\begin{equation}
\beta^{\cal S}(a,0) ~=~ \sum_{r=1}^\infty \beta_r^{\cal S} a^{r+1}
\end{equation}
where the coupling constant is understood to be in the scheme ${\cal S}$. The
associated partial sums or truncated $\beta$-functions are 
\begin{equation}
\beta^{\cal S}_n(a,0) ~=~ \sum_{r=1}^n \beta_r^{\cal S} a^{r+1} ~.
\end{equation}
where there is no $O(a)$ term since we are only considering the Banks-Zaks
fixed points and not $d$-dimensional Wilson-Fisher critical points. We denote 
the critical coupling constant at the $L$th loop order by $a_L$ and define it 
as the solution of the $L$th partial sum
\begin{equation}
\beta^{\cal S}_L(a_L,0) ~=~ 0 
\label{bzloc}
\end{equation}
in scheme ${\cal S}$. As we will be considering the critical exponent 
associated with the quark mass renormalization we formally define the anomalous
dimension in the scheme ${\cal S}$ by, \cite{31},
\begin{equation}
\gamma^{\cal S}_{\bar{\psi}\psi}(a,0) ~=~ \sum_{r=1}^\infty \gamma_r^{\cal S}
a^r
\end{equation}
in the Landau gauge and the corresponding partial sums by
\begin{equation}
\gamma^{\cal S}_{\bar{\psi}\psi \, n}(a,0) ~=~ \sum_{r=1}^n \gamma_r^{\cal S}
a^r ~.
\end{equation}
The same formalism will also apply to the case of the MAG. Then from each 
partial sum the truncated critical exponents we will evaluate in each of the 
$\iMOMx$ schemes are  
\begin{equation}
{\widetilde{\omega}}_L ~=~ 2 \beta^\prime_L(a_L,0) ~~~,~~~
\rho_L ~=~ -~ 2 \gamma_{\bar{\psi}\psi \, L}(a_L,0)
\label{expdef}
\end{equation}
in the notation of \cite{31}. Here we use the notation of
${\widetilde{\omega}}$ as the exponent corresponding to correction to scaling 
in order not to confuse it with our choice of interpolating parameter which was
introduced originally in \cite{51} in the study of the quark mass
renormalization. In defining the exponents ${\widetilde{\omega}}$ and $\rho$ 
with the factors specified in (\ref{expdef}) we have the same definition of 
\cite{43} and \cite{31}. However, since we used the $\beta$-function 
conventions of \cite{64} comparing the location of the critical couplings with 
\cite{43} there will be a difference of a factor of $4\pi$. This has been 
absorbed into our coupling constant. 

Having introduced the partial sum formalism we have solved (\ref{bzloc}) for 
the Banks-Zaks fixed point at two and three loops in each of the $\iMOMx$ 
schemes. We concentrate on the values of $\omega$~$=$~$\half$ and $2$ and
provide numerical results in a series of Tables. The critical couplings at two 
and three loops for both values of $\omega$ are given in Table 1. Since the 
$\beta$-function is scheme independent to two loops inclusive we have chosen to
present the values for ${\widetilde{\omega}}_3$ in Table 2 again for each of 
the three schemes together. The subsequent three Tables contain the results for
the quark mass exponent in the $\iMOMq$, $\iMOMh$ and $\iMOMg$ schemes 
respectively. In each we have included the two and three loop values but for 
three values of $\omega$ which are $\half$, $1$ and $2$. The symmetric point 
values corresponding to $\omega$~$=$~$1$ were computed in \cite{31}. We have 
included them here for comparison with the new values and in order to gauge, 
for instance, what the range of the exponent is when $\omega$ is varied. For 
all these Tables we have concentrated on the $SU(2)$ and $SU(3)$ colour groups 
for their two loop conformal windows which are $6$~$\leq$~$\Nf$~$\leq$~$10$ and 
$9$~$\leq$~$\Nf$~$\leq$~$16$ respectively. The subsequent tables contain the
same data but for the case of the MAG. For instance, the critical couplings for
each of $\iMOMmq$, $\iMOMmh$ and $\iMOMmh$ schemes are given in Tables $6$,
$7$ and $8$ respectively while the corresponding values of 
${\widetilde{\omega}}$ are given in Tables $9$, $10$ and $11$ for the same
three schemes. Finally Tables $12$, $13$ and $14$ record the parallel two and
three loop estimates for the quark mass exponent $\rho$ also at two and three
loops with $\omega$ values of $\half$, $1$ and $2$ for each iMOMm scheme. In
several of the $\iMOMmx$ schemes the lower end of the conformal window is at
$\Nf$~$=$~$8$ rather than $9$ for $SU(3)$ and we have included some data for
these schemes. In these cases we omitted to record two loop exponents purely
for the reason that the values were several orders of magnitude greater than
either the subsequent $\Nf$ estimates or the three loop value. This no doubt
reflects the fact that perturbation theory is probably not truly reliable at
that point. However as the three loop data for $\Nf$~$=$~$8$ is not
unreasonable compared to the $\Nf$~$=$~$9$ value we have included those for
guidance.

In order to ascertain how the exponent ${\widetilde{\omega}}$ depends on the 
scheme we have plotted the data of Table $1$ for the Landau gauge in Figure $1$
for both $SU(2)$ and $SU(3)$. While we determined the exponents for discrete 
values of $\Nf$ we have chosen to present piecewise linear connections between 
the spot values for this and the other Tables, similar to \cite{50}, in order 
to see any trends more clearly. In Figure $1$ the three schemes are shown in 
the order of $\iMOMq$, $\iMOMh$ and $\iMOMg$ and on each plot we have included 
the corresponding values for the $\MSbar$ and mini-MOM (mMOM) schemes which 
were computed in \cite{31}. As these are the same in each plot they provide a 
guide point for comparing the scheme results for each colour group. The 
definition of the mini-MOM scheme, \cite{91}, centres on the fact that the 
ghost-gluon vertex in the Landau gauge does not get renormalized, \cite{92}. 
Therefore the coupling constant renormalization constant in the mini-MOM scheme
is determined by evaluating the ghost-gluon vertex function with one external 
ghost leg nullified and then requiring that the non-renormalization condition 
is preserved away for all covariant gauges, \cite{91}. While the scheme is 
motivated by a specific property of the gauge theory the fact that the 
subtraction point is at an exceptional momentum configuration means that there 
may be infrared problems unlike the $\iMOMx$ schemes. Several general themes 
appear to emerge for the Landau gauge. First for each of the three schemes 
comparing the $SU(2)$ to the respective $SU(3)$ plot the $\MSbar$ values 
diverge from the $\iMOMx$ values at about the midpoint of the conformal window 
as $\Nf$ decreases. This is not unexpected as this is around the place where 
one would expect perturbation theory to become less reliable. The effect is 
most pronounced for the $\iMOMg$ scheme whereas for the $\iMOMh$ scheme there 
is a smaller spread across the schemes for relatively low $\Nf$. For the 
$\iMOMx$ schemes the spread over the range of $\omega$ is relatively small 
which is perhaps surprising for low values of $\Nf$ but in keeping with our 
expectations for the higher values where one is in the perturbative region. For
$SU(3)$ regarding $\Nf$~$=$~$12$ as a rough boundary of this point then 
estimates for ${\widetilde{\omega}}$ would appear to be in agreement from all 
the schemes except the $\MSbar$ one.  

One of the features of the three loop plots for ${\widetilde{\omega}}$ is the 
relatively small spread for the range of $\omega$ we took. However the two loop
value for this exponent is scheme independent and therefore we cannot say 
whether the momentum subtraction based schemes have any marked difference with 
the non-kinematic schemes. By contrast the quark mass anomalous dimension is 
scheme dependent at two loops and so we can examine scheme features over 
several loop orders. We have given plots of $\rho_2$ and $\rho_3$ in Figures
$2$ and $3$ for both $SU(2)$ and $SU(3)$. In both figures the left set are the
two loop values and the right set are the three loop ones with Figure $2$
giving the $SU(2)$ values. The order of the schemes is the same as Figure $1$.  
Some general comments are in order at the outset. First in both cases the two
loop results at the lower boundary of the conformal window for both groups are 
clearly unreliable. While this is more marked than for the exponent 
${\widetilde{\omega}}$ it is clear that there is a huge difference at this end
of the window when one compares with the three loop plots. Even for values of 
$\Nf$ above the lower boundary there is still a large discrepancy between the 
two and three loop cases as the large vertex scale at two loop camouflages the 
disparity. The other general feature is one shared with the exponent 
${\widetilde{\omega}}$ in that the $\MSbar$ scheme, and to a lesser extent the 
mini-MOM scheme, has different behaviour compared to the $\iMOMx$ schemes as 
$\Nf$ decreases. This should also be tempered by the fact that the discrepancy 
becomes apparent at around $\Nf$~$=$~$12$ for $SU(3)$ which is where 
perturbation theory is perhaps on the limit of credibility. In terms of the 
different schemes there is clearly a parallel structure when comparing each 
scheme for both groups which is reassuring. Equally for the three loop plots 
aside from the $\iMOMg$ scheme there is a slight discrepancy between the 
$\MSbar$ and mini-MOM scheme estimates and the $\iMOMx$ ones which has a more 
dramatic low $\Nf$ limit for the $\iMOMh$ case. Clearly for the $\iMOMx$ 
schemes there is a significant difference in the limit to the lower window 
boundary and therefore no significance should be placed on any estimate of 
$\rho_3$ in this case. However, compared with the dramatic change from two 
loops it would suggest that a four loop evaluation could improve the picture 
for lower $\Nf$ in the kinematic schemes. One interesting feature emerges if 
one examines the three loop plots for both exponents. For the most part the 
mini-MOM plots appear to faithfully track the $\MSbar$ ones. Both schemes are 
defined in closely similar ways. For instance, the $\MSbar$ scheme is a mass 
independent scheme and can be defined at an exceptional momentum configuration 
as a consequence. Equally the mini-MOM scheme has its origin in preserving a 
property of a vertex function at specific exceptional momentum configuration. 
However the exception to parallel behaviour for mini-MOM exponents compared to 
the $\MSbar$ scheme is the exponent ${\widetilde{\omega}}$ in the $\iMOMh$ 
scheme. For both colour groups the mini-MOM exponents are virtually on top of 
each of the $\iMOMx$ schemes for the whole range of the conformal window. While
it is premature to say that this is a general feature ahead of a four loop 
analysis. In other words it may be the fact that as the mini-MOM scheme 
preserves by definition a property of the ghost-gluon vertex then this is 
reflected in the agreement with the kinematical scheme behaviour. Indeed of the
three schemes the $\iMOMh$ ${\widetilde{\omega}}$ exponents have minimal 
spread for all $\Nf$. Again this observation needs to be balanced by noting 
that the $\iMOMg$ behaviour of $\rho_3$ is parallel to the $\MSbar$ and 
mini-MOM schemes for low $\Nf$.  

For the MAG we have provided similar plots for both exponents at three loops
in Figure $4$ for the $\iMOMmq$, $\iMOMmh$ and $\iMOMmg$ schemes respectively
for $SU(3)$. Fewer plots for this gauge have been included as there is a strong
general similarity with the Landau gauge plots at two loops. Also we have
omitted points for $\Nf$~$=$~$8$ for the two schemes where there is a window
as the relatively large values of the respective exponents would skew the 
analysis. For instance if the data from that value of $\Nf$ were included then
the plots would make it appear that for $\Nf$~$\geq$~$9$ all the scheme
estimates were equivalent. Taking the range as $9$~$\leq$~$\Nf$~$\leq$~$16$
allows the finer detail to be seen. The $\MSbar$ and mini-MOM three loop
estimates are again provided for comparison. First for ${\widetilde{\omega}}$
the three loop plots are virtually the same as for the Landau gauge. Down to
around $\Nf$~$=$~$13$ there is little difference between the two reference
schemes and the kinematic ones. For the border point of $\Nf$~$=$~$12$ the
$\iMOMmq$ and $\iMOMmh$ schemes are practically the same but backs up the
earlier observation that this is probably the place where higher order
corrections could remove scheme ambiguity. For the $\iMOMmg$ there is clearly
a discrepancy at $\Nf$~$=$~$13$ which is slightly larger than the Landau 
gauge. This could be due in part to the nature of the MAG where a subset of
gluon fields are isolated in the definition of the gauge itself. However the
closer agreement for the other two schemes would suggest that with higher
orders this discrepancy could wash out. The situation for $\rho$ is somewhat
different with a different functional behaviour for each scheme below
$\Nf$~$=$~$16$. However the general behaviour of the three $\iMOMmx$ schemes
is not dissimilar to that of the Landau gauge plots. It may be that the 
different behaviour lies in the nature of the quantity plotted which is the
quark mass operator. As an operator it does not have any gluon content where
the split colour group property would be significant. However, the plots may be
misleading in that the difference between $\rho$ exponent estimates between 
$\Nf$~$=$~$13$ and $15$ range from $5\%$ to $8\%$  Finally, what is noticeable 
in both gauges is that the behaviour of the schemes based on the triple gluon 
vertex is different from the other two schemes at the lower end of the 
conformal window. That this is the case in the MAG as well as the Landau gauge 
suggests that it is a feature of the particular vertex which has significantly 
more graphs at two loops and these are predominantly gluonic. It will not be 
until three loops that there would be a commensurate number of gluonic 
contributions to the quark- and ghost-gluon vertex functions with which to 
compare. It may be then that the behaviour at the lower end of the window 
becomes similar across all three $\iMOMx$ schemes. 

\sect{Discussion.}

We conclude with various remarks. First we have completed the full 
renormalization of QCD in a new set of kinematic schemes called iMOM which
extend the MOM schemes of Celmaster and Gonsalves, \cite{20,21}. In particular
we have derived all the renormalization group functions at three loops in the
three iMOM schemes for both the Landau and maximal abelian gauges. The schemes
depend on a parameter $\omega$, which is restricted to the range 
$0$~$<$~$\omega$~$<$~$4$. One motivation for introducing the iMOM schemes was
to provide a testing ground for evaluating quantities of physical interest in
truncated perturbation theory and seeing how far the scheme independence was
apparent. The major application of this idea here was to the Banks-Zaks fixed 
point in the conformal window of QCD. As critical exponents are the evaluation
of the renormalization group functions at a fixed point and therefore physical
quantities it was important to study the exponents in the iMOM scheme. The
conformal window is such that for values of $\Nf$ near the upper limit, 
perturbation theory should be a good tool for reliable information. By contrast
as $\Nf$ reduces inside the window perturbation will cease to be a reliable
guide. However, where the breakdown occurs is not immediately obvious without a
numerical analysis. Overall for both gauges it seems that from the three loop
results one cannot fully rely on the estimates at $\Nf$~$=$~$12$. This should
be qualified by noting that this is from the raw results without resummation
to improve convergence. 

One aspect of our results which is worth remarking on is in regard to agreement
between $\MSbar$ and $\iMOMx$ scheme results for the upper end of the window. 
Numerically the data in the plots for both sets of schemes lie on top of each 
other. This strongly suggests bona fide scheme independence. However, this 
needs to be balanced by the fact that the numerology of the $\MSbar$ and 
$\iMOMx$ schemes is different with the differences appearing first in the 
scheme dependent terms. Therefore, this ought to motivate a future analytic 
study in order to see if this can be established beyond numerical evidence. Of 
course one could extend the $\iMOMx$ schemes beyond the appearance of one 
parameter. For instance, a more general set of schemes could involve two
parameters related to the dimensionless variables $x$ and $y$ appearing in the 
underlying polylogarithms of the master one and two loop integrals. While we 
have not studied this we would expect the outcome to be the same. In other 
words there would be scheme independence. Such a more general set of schemes 
might be useful in the extension of these ideas to other quantities of physical
interest such as the $R$ ratio. There one has experimental data for which the 
truncated perturbative expansion is also available but has been computed 
primarily in the $\MSbar$ scheme. By recomputing in the $\iMOMx$ schemes one 
could systematically provide bounds on the measured value at a particular 
energy scale by using the tolerance from the values at $\omega$~$=$~$\half$ and
$2$. This would appear to be a more quantum field theory motivated approach as 
$\omega$ tracks the effect the scheme has through the Feynman diagrams 
underlying the quantity of interest. This is in contrast to what is currently 
used in terms of varying the scale itself of where the measurement is made. The
toy example of the critical exponents suggests that the scheme variation would 
be a more robust procedure. Finally, in completing the derivation of the three 
loop renormalization group functions in these classes of kinematic schemes and
gauges the natural extension is to proceed to the next loop order in future 
work.

\vspace{1cm}
\noindent
{\bf Acknowledgements.} This work was carried out in part with the support of 
the STFC through the Consolidated Grant ST/L000431/1 and a studentship (RMS). 
The authors thank J. Hackett, Prof T. Ryttov and Prof R. Shrock for useful 
conversations. 

\appendix

\sect{Tensor bases and projection matrices.}

In this appendix we record the various tensor bases for each of the $3$-point
vertex functions. While these are in effect the same as the symmetric point the
corresponding projection matrices are $\omega$ dependent. For the ghost and 
quark vertices we have the tensor bases 
\begin{equation}
{\cal P}^{\mbox{\footnotesize{ccg}}}_{(1) \sigma }(p,q) ~=~ p_\sigma ~~~,~~~
{\cal P}^{\mbox{\footnotesize{ccg}}}_{(2) \sigma }(p,q) ~=~ q_\sigma ~.
\end{equation} 
and
\begin{eqnarray}
{\cal P}^{\mbox{\footnotesize{qqg}}}_{(1) \sigma }(p,q) &=& 
\gamma_\sigma ~~~,~~~
{\cal P}^{\mbox{\footnotesize{qqg}}}_{(2) \sigma }(p,q) ~=~ 
\frac{{p}_\sigma \pslash}{\mu^2} ~~~,~~~
{\cal P}^{\mbox{\footnotesize{qqg}}}_{(3) \sigma }(p,q) ~=~ 
\frac{{p}_\sigma \qslash}{\mu^2} ~, \nonumber \\
{\cal P}^{\mbox{\footnotesize{qqg}}}_{(4) \sigma }(p,q) &=& 
\frac{{q}_\sigma \pslash}{\mu^2} ~~~,~~~
{\cal P}^{\mbox{\footnotesize{qqg}}}_{(5) \sigma }(p,q) ~=~ 
\frac{{q}_\sigma \qslash}{\mu^2} ~~~,~~~
{\cal P}^{\mbox{\footnotesize{qqg}}}_{(6) \sigma }(p,q) ~=~ 
\frac{1}{\mu^2} \Gamma_{(3) \, \sigma p q} 
\end{eqnarray}
where in the latter we use the generalized basis of $\gamma$-matrices which are
denoted by $\Gamma_{(n)}^{\mu_1 \ldots \mu_n}$ and defined earlier. We use the
convention that when an external momentum is contracted with a Lorentz index
then that index is replaced by the particular momentum itself. For the 
$3$-point gluon vertex there are fourteen independent tensors given by
\begin{eqnarray}
{\cal P}^{\mbox{\footnotesize{ggg}}}_{(1) \mu \nu \sigma }(p,q) &=& 
\eta_{\mu \nu} p_\sigma ~~,~~ 
{\cal P}^{\mbox{\footnotesize{ggg}}}_{(2) \mu \nu \sigma }(p,q) ~=~ 
\eta_{\nu \sigma} p_\mu ~~,~~ 
{\cal P}^{\mbox{\footnotesize{ggg}}}_{(3) \mu \nu \sigma }(p,q) ~=~ 
\eta_{\sigma \mu} p_\nu \nonumber \\
{\cal P}^{\mbox{\footnotesize{ggg}}}_{(4) \mu \nu \sigma }(p,q) &=& 
\eta_{\mu \nu} q_\sigma ~~,~~ 
{\cal P}^{\mbox{\footnotesize{ggg}}}_{(5) \mu \nu \sigma }(p,q) ~=~ 
\eta_{\nu \sigma} q_\mu ~~,~~ 
{\cal P}^{\mbox{\footnotesize{ggg}}}_{(6) \mu \nu \sigma }(p,q) ~=~ 
\eta_{\sigma \mu} q_\nu \nonumber \\
{\cal P}^{\mbox{\footnotesize{ggg}}}_{(7) \mu \nu \sigma }(p,q) &=& 
\frac{1}{\mu^2} p_\mu p_\nu p_\sigma ~~,~~ 
{\cal P}^{\mbox{\footnotesize{ggg}}}_{(8) \mu \nu \sigma }(p,q) ~=~ 
\frac{1}{\mu^2} p_\mu p_\nu q_\sigma ~~,~~ 
{\cal P}^{\mbox{\footnotesize{ggg}}}_{(9) \mu \nu \sigma }(p,q) ~=~ 
\frac{1}{\mu^2} p_\mu q_\nu p_\sigma \nonumber \\ 
{\cal P}^{\mbox{\footnotesize{ggg}}}_{(10) \mu \nu \sigma }(p,q) &=& 
\frac{1}{\mu^2} q_\mu p_\nu p_\sigma ~~,~~ 
{\cal P}^{\mbox{\footnotesize{ggg}}}_{(11) \mu \nu \sigma }(p,q) ~=~ 
\frac{1}{\mu^2} p_\mu q_\nu q_\sigma ~~,~~ 
{\cal P}^{\mbox{\footnotesize{ggg}}}_{(12) \mu \nu \sigma }(p,q) ~=~ 
\frac{1}{\mu^2} q_\mu p_\nu q_\sigma \nonumber \\ 
{\cal P}^{\mbox{\footnotesize{ggg}}}_{(13) \mu \nu \sigma }(p,q) &=& 
\frac{1}{\mu^2} q_\mu q_\nu p_\sigma ~~,~~ 
{\cal P}^{\mbox{\footnotesize{ggg}}}_{(14) \mu \nu \sigma }(p,q) ~=~ 
\frac{1}{\mu^2} q_\mu q_\nu q_\sigma ~. 
\end{eqnarray}
Finally, for the mass operator Green's function there are two tensors in the
basis which are
\begin{equation}
{\cal P}^{\bar{\psi}\psi}_{(1)}(p,q) ~=~ \Gamma_{(0)} ~~~,~~~
{\cal P}^{\bar{\psi}\psi}_{(2)}(p,q) ~=~ \frac{1}{\mu^2} \Gamma_{(2)}^{pq}
\end{equation}
where $\Gamma_{(0)}$ is the unit matrix in spinor space.  

From these bases the construction of the projection matrices is straightforward
and uses the relations (\ref{imomdef1}) and (\ref{imomdef2}). Consequently the
matrices are $\omega$ dependent. For instance in the ghost-gluon vertex case
the matrix is
\begin{equation}
{\cal M}^{ccg} ~=~ \frac{2}{\omega[\omega-4]} \left(
\begin{array}{cc}
2 & (2-\omega) \\
(2-\omega) & 2 \\
\end{array}
\right) ~.
\end{equation}
For the other two cases we have chosen to record the diagonal and upper
triangle entries as the matrices like ${\cal M}^{ccg}$ are diagonal. First,
factoring off a common factor by defining 
\begin{equation}
{\cal M}^{qqg} ~=~ \frac{1}{4(d-2)\omega^2[\omega-4]^2} 
\widetilde{\cal M}^{qqg} ~~~,~~~ 
{\cal M}^{ggg} ~=~ \frac{1}{(d-2)\omega^3[\omega-4]^3} \widetilde{\cal M}^{ggg}
\end{equation}
then the entries for the quark-gluon vertex case are 
\begin{eqnarray}
\widetilde{\cal M}^{qqg}_{1,1} &=& [\omega - 4]^2 \omega^2 ~~,~~
\widetilde{\cal M}^{qqg}_{1,2} ~=~ -~ 4 [\omega - 4] \omega ~~,~~
\widetilde{\cal M}^{qqg}_{1,3} ~=~ 2 [\omega - 2] [\omega - 4] \omega \nonumber \\
\widetilde{\cal M}^{qqg}_{1,4} &=& 2 [\omega - 2] [\omega - 4] \omega ~~,~~
\widetilde{\cal M}^{qqg}_{1,5} ~=~ -~ 4 [\omega - 4] \omega ~~,~~
\widetilde{\cal M}^{qqg}_{1,6} ~=~ 0 ~~,~~
\widetilde{\cal M}^{qqg}_{2,2} ~=~ 16 [d - 1] \nonumber \\
\widetilde{\cal M}^{qqg}_{2,3} &=& -~ 8 [d - 1] [\omega - 2] ~~,~~
\widetilde{\cal M}^{qqg}_{2,4} ~=~ -~ 8 [d - 1] [\omega - 2] \nonumber \\
\widetilde{\cal M}^{qqg}_{2,5} &=& -~ 4 [2 [\omega^2 - 4 \omega + 2] - [\omega - 2]^2 d] ~~,~~
\widetilde{\cal M}^{qqg}_{2,6} ~=~ 0 ~~,~~
\widetilde{\cal M}^{qqg}_{3,3} ~=~ 4 [\omega^2 - 4 \omega - 4 + 4 d] \nonumber \\
\widetilde{\cal M}^{qqg}_{3,4} &=& 4 [d - 1] [\omega - 2]^2 ~~,~~
\widetilde{\cal M}^{qqg}_{3,5} ~=~ -~ 8 [d - 1] [\omega - 2] ~~,~~
\widetilde{\cal M}^{qqg}_{3,6} ~=~ 0 \nonumber \\
\widetilde{\cal M}^{qqg}_{4,4} &=& 4 [\omega^2 - 4 \omega - 4 + 4 d] ~~,~~
\widetilde{\cal M}^{qqg}_{4,5} ~=~ -~ 8 [d - 1] [\omega - 2] ~~,~~
\widetilde{\cal M}^{qqg}_{4,6} ~=~ 0 \nonumber \\
\widetilde{\cal M}^{qqg}_{5,5} &=& 16 [d - 1] ~~,~~ 
\widetilde{\cal M}^{qqg}_{5,6} ~=~ 0 ~~,~~
\widetilde{\cal M}^{qqg}_{6,6} ~=~ 4 [\omega - 4] \omega
\end{eqnarray}
and  
\begin{eqnarray}
\widetilde{\cal M}^{ggg}_{1,1} &=& 4 [\omega - 4]^2 \omega^2 ~~,~~
\widetilde{\cal M}^{ggg}_{1,2} ~=~ 0
~~,~~
\widetilde{\cal M}^{ggg}_{1,3} ~=~ 0
~~,~~
\widetilde{\cal M}^{ggg}_{1,4} ~=~ -~ 2 [\omega - 2] [\omega - 4]^2 \omega^2
\nonumber \\
\widetilde{\cal M}^{ggg}_{1,5} &=& 0
~~,~~
\widetilde{\cal M}^{ggg}_{1,6} ~=~ 0
~~,~~
\widetilde{\cal M}^{ggg}_{1,7} ~=~ -~ 16 [\omega - 4] \omega
~~,~~
\widetilde{\cal M}^{ggg}_{1,8} ~=~ 8 [\omega - 2] [\omega - 4] \omega
\nonumber \\
\widetilde{\cal M}^{ggg}_{1,9} &=& 8 [\omega - 2] [\omega - 4] \omega
~~,~~
\widetilde{\cal M}^{ggg}_{1,10} ~=~ 8 [\omega - 2] [\omega - 4] \omega
\nonumber \\
\widetilde{\cal M}^{ggg}_{1,11} &=& -~ 4 [\omega - 2]^2 [\omega - 4] \omega
~~,~~
\widetilde{\cal M}^{ggg}_{1,12} ~=~ -~ 4 [\omega - 2]^2 [\omega - 4] \omega
\nonumber \\
\widetilde{\cal M}^{ggg}_{1,13} &=& -~ 16 [\omega - 4] \omega
~~,~~
\widetilde{\cal M}^{ggg}_{1,14} ~=~ 8 [\omega - 2] [\omega - 4] \omega
~~,~~
\widetilde{\cal M}^{ggg}_{2,2} ~=~ 4 [\omega - 4]^2 \omega^2
\nonumber \\
\widetilde{\cal M}^{ggg}_{2,3} &=& 0
~~,~~
\widetilde{\cal M}^{ggg}_{2,4} ~=~ 0
~~,~~
\widetilde{\cal M}^{ggg}_{2,5} ~=~ -~ 2 [\omega - 2] [\omega - 4]^2 \omega^2
~~,~~
\widetilde{\cal M}^{ggg}_{2,6} ~=~ 0
\nonumber \\
\widetilde{\cal M}^{ggg}_{2,7} &=& -~ 16 [\omega - 4] \omega
~~,~~
\widetilde{\cal M}^{ggg}_{2,8} ~=~ 8 [\omega - 2] [\omega - 4] \omega
~~,~~
\widetilde{\cal M}^{ggg}_{2,9} ~=~ 8 [\omega - 2] [\omega - 4] \omega
\nonumber \\
\widetilde{\cal M}^{ggg}_{2,10} &=& 8 [\omega - 2] [\omega - 4] \omega
~~,~~
\widetilde{\cal M}^{ggg}_{2,11} ~=~ -~ 16 [\omega - 4] \omega
\nonumber \\
\widetilde{\cal M}^{ggg}_{2,12} &=& -~ 4 [\omega - 2]^2 [\omega - 4] \omega
~~,~~
\widetilde{\cal M}^{ggg}_{2,13} ~=~ -~ 4 [\omega - 2]^2 [\omega - 4] \omega
\nonumber \\
\widetilde{\cal M}^{ggg}_{2,14} &=& 8 [\omega - 2] [\omega - 4] \omega
~~,~~
\widetilde{\cal M}^{ggg}_{3,3} ~=~ 4 [\omega - 4]^2 \omega^2
~~,~~
\widetilde{\cal M}^{ggg}_{3,4} ~=~ 0
~~,~~
\widetilde{\cal M}^{ggg}_{3,5} ~=~ 0
\nonumber \\
\widetilde{\cal M}^{ggg}_{3,6} &=& -~ 2 [\omega - 2] [\omega - 4]^2 \omega^2
~~,~~
\widetilde{\cal M}^{ggg}_{3,7} ~=~ -~ 16 [\omega - 4] \omega
\nonumber \\
\widetilde{\cal M}^{ggg}_{3,8} &=& 8 [\omega - 2] [\omega - 4] \omega
~~,~~
\widetilde{\cal M}^{ggg}_{3,9} ~=~ 8 [\omega - 2] [\omega - 4] \omega
\nonumber \\
\widetilde{\cal M}^{ggg}_{3,10} &=& 8 [\omega - 2] [\omega - 4] \omega
~~,~~
\widetilde{\cal M}^{ggg}_{3,11} ~=~ -~ 4 [\omega - 2]^2 [\omega - 4] \omega
\nonumber \\
\widetilde{\cal M}^{ggg}_{3,12} &=& -~ 16 [\omega - 4] \omega
~~,~~
\widetilde{\cal M}^{ggg}_{3,13} ~=~ -~ 4 [\omega - 2]^2 [\omega - 4] \omega
\nonumber \\
\widetilde{\cal M}^{ggg}_{3,14} &=& 8 [\omega - 2] [\omega - 4] \omega
~~,~~
\widetilde{\cal M}^{ggg}_{4,4} ~=~ 4 [\omega - 4]^2 \omega^2
~~,~~
\widetilde{\cal M}^{ggg}_{4,5} ~=~ 0
~~,~~
\widetilde{\cal M}^{ggg}_{4,6} ~=~ 0
\nonumber \\
\widetilde{\cal M}^{ggg}_{4,7} &=& 8 [\omega - 2] [\omega - 4] \omega
~~,~~
\widetilde{\cal M}^{ggg}_{4,8} ~=~ -~ 16 [\omega - 4] \omega
\nonumber \\
\widetilde{\cal M}^{ggg}_{4,9} &=& -~ 4 [\omega - 2]^2 [\omega - 4] \omega
~~,~~
\widetilde{\cal M}^{ggg}_{4,10} ~=~ -~ 4 [\omega - 2]^2 [\omega - 4] \omega
\nonumber \\
\widetilde{\cal M}^{ggg}_{4,11} &=& 8 [\omega - 2] [\omega - 4] \omega
~~,~~
\widetilde{\cal M}^{ggg}_{4,12} ~=~ 8 [\omega - 2] [\omega - 4] \omega
\nonumber \\
\widetilde{\cal M}^{ggg}_{4,13} &=& 8 [\omega - 2] [\omega - 4] \omega
~~,~~
\widetilde{\cal M}^{ggg}_{4,14} ~=~ -~ 16 [\omega - 4] \omega
~~,~~
\widetilde{\cal M}^{ggg}_{5,5} ~=~ 4 [\omega - 4]^2 \omega^2
\nonumber \\
\widetilde{\cal M}^{ggg}_{5,6} &=& 0
~~,~~
\widetilde{\cal M}^{ggg}_{5,7} ~=~ 8 [\omega - 2] [\omega - 4] \omega
~~,~~
\widetilde{\cal M}^{ggg}_{5,8} ~=~ -~ 4 [\omega - 2]^2 [\omega - 4] \omega
\nonumber \\
\widetilde{\cal M}^{ggg}_{5,9} &=& -~ 4 [\omega - 2]^2 [\omega - 4] \omega
~~,~~
\widetilde{\cal M}^{ggg}_{5,10} ~=~ -~ 16 [\omega - 4] \omega
\nonumber \\
\widetilde{\cal M}^{ggg}_{5,11} &=& 8 [\omega - 2] [\omega - 4] \omega
~~,~~
\widetilde{\cal M}^{ggg}_{5,12} ~=~ 8 [\omega - 2] [\omega - 4] \omega
\nonumber \\
\widetilde{\cal M}^{ggg}_{5,13} &=& 8 [\omega - 2] [\omega - 4] \omega
~~,~~
\widetilde{\cal M}^{ggg}_{5,14} ~=~ -~ 16 [\omega - 4] \omega
~~,~~
\widetilde{\cal M}^{ggg}_{6,6} ~=~ 4 [\omega - 4]^2 \omega^2
\nonumber \\
\widetilde{\cal M}^{ggg}_{6,7} &=& 8 [\omega - 2] [\omega - 4] \omega
~~,~~
\widetilde{\cal M}^{ggg}_{6,8} ~=~ -~ 4 [\omega - 2]^2 [\omega - 4] \omega
\nonumber \\
\widetilde{\cal M}^{ggg}_{6,9} &=& -~ 16 [\omega - 4] \omega
~~,~~
\widetilde{\cal M}^{ggg}_{6,10} ~=~ -~ 4 [\omega - 2]^2 [\omega - 4] \omega
\nonumber \\
\widetilde{\cal M}^{ggg}_{6,11} &=& 8 [\omega - 2] [\omega - 4] \omega
~~,~~
\widetilde{\cal M}^{ggg}_{6,12} ~=~ 8 [\omega - 2] [\omega - 4] \omega
~~,~~
\widetilde{\cal M}^{ggg}_{6,13} ~=~ 8 [\omega - 2] [\omega - 4] \omega
\nonumber \\
\widetilde{\cal M}^{ggg}_{6,14} &=& -~ 16 [\omega - 4] \omega
~~,~~
\widetilde{\cal M}^{ggg}_{7,7} ~=~ 64 [d + 1]
~~,~~
\widetilde{\cal M}^{ggg}_{7,8} ~=~ -~ 32 [d + 1] [\omega - 2]
\nonumber \\
\widetilde{\cal M}^{ggg}_{7,9} &=& -~ 32 [d + 1] [\omega - 2]
~~,~~
\widetilde{\cal M}^{ggg}_{7,10} ~=~ -~ 32 [d + 1] [\omega - 2]
\nonumber \\
\widetilde{\cal M}^{ggg}_{7,11} &=& 16 [d \omega^2 - 4 d \omega + 4 d + 4]
~~,~~
\widetilde{\cal M}^{ggg}_{7,12} ~=~ 16 [d \omega^2 - 4 d \omega + 4 d + 4]
\nonumber \\
\widetilde{\cal M}^{ggg}_{7,13} &=& 16 [d \omega^2 - 4 d \omega + 4 d + 4]
\nonumber \\
\widetilde{\cal M}^{ggg}_{7,14} &=& -~ 8 [d \omega^2 - 4 d \omega + 4 d - 2 \omega^2 + 8 \omega + 4] [\omega - 2]
~~,~~
\widetilde{\cal M}^{ggg}_{8,8} ~=~ 32 [2 d + \omega^2 - 4 \omega + 2]
\nonumber \\
\widetilde{\cal M}^{ggg}_{8,9} &=& 16 [d + 1] [\omega - 2]^2
~~,~~
\widetilde{\cal M}^{ggg}_{8,10} ~=~ 16 [d + 1] [\omega - 2]^2
\nonumber \\
\widetilde{\cal M}^{ggg}_{8,11} &=& -~ 8 [4 d + \omega^2 - 4 \omega + 4] [\omega - 2]
~~,~~
\widetilde{\cal M}^{ggg}_{8,12} ~=~ -~ 8 [4 d + \omega^2 - 4 \omega + 4] [\omega - 2]
\nonumber \\
\widetilde{\cal M}^{ggg}_{8,13} &=& -~ 8 [d \omega^2 - 4 d \omega + 4 d + 4] [\omega - 2]
~~,~~
\widetilde{\cal M}^{ggg}_{8,14} ~=~ 16 [d \omega^2 - 4 d \omega + 4 d + 4]
\nonumber \\
\widetilde{\cal M}^{ggg}_{9,9} &=& 32 [2 d + \omega^2 - 4 \omega + 2]
~~,~~
\widetilde{\cal M}^{ggg}_{9,10} ~=~ 16 [d + 1] [\omega - 2]^2
\nonumber \\
\widetilde{\cal M}^{ggg}_{9,11} &=& -~ 8 [4 d + \omega^2 - 4 \omega + 4] [\omega - 2]
~~,~~
\widetilde{\cal M}^{ggg}_{9,12} ~=~ -~ 8 [d \omega^2 - 4 d \omega + 4 d + 4] [\omega - 2]
\nonumber \\
\widetilde{\cal M}^{ggg}_{9,13} &=& -~ 8 [4 d + \omega^2 - 4 \omega + 4] [\omega - 2]
~~,~~
\widetilde{\cal M}^{ggg}_{9,14} ~=~ 16 [d \omega^2 - 4 d \omega + 4 d + 4]
\nonumber \\
\widetilde{\cal M}^{ggg}_{10,10} &=& 32 [2 d + \omega^2 - 4 \omega + 2]
~~,~~
\widetilde{\cal M}^{ggg}_{10,11} ~=~ -~ 8 [d \omega^2 - 4 d \omega + 4 d + 4] [\omega - 2]
\nonumber \\
\widetilde{\cal M}^{ggg}_{10,12} &=& -~ 8 [4 d + \omega^2 - 4 \omega + 4] [\omega - 2]
~~,~~
\widetilde{\cal M}^{ggg}_{10,13} ~=~ -~ 8 [4 d + \omega^2 - 4 \omega + 4] [\omega - 2]
\nonumber \\
\widetilde{\cal M}^{ggg}_{10,14} &=& 16 [d \omega^2 - 4 d \omega + 4 d + 4]
~~,~~
\widetilde{\cal M}^{ggg}_{11,11} ~=~ 32 [2 d + \omega^2 - 4 \omega + 2]
\nonumber \\
\widetilde{\cal M}^{ggg}_{11,12} &=& 16 [d + 1] [\omega - 2]^2
~~,~~
\widetilde{\cal M}^{ggg}_{11,13} ~=~ 16 [d + 1] [\omega - 2]^2
\nonumber \\
\widetilde{\cal M}^{ggg}_{11,14} &=& -~ 32 [d + 1] [\omega - 2]
~~,~~
\widetilde{\cal M}^{ggg}_{12,12} ~=~ 32 [2 d + \omega^2 - 4 \omega + 2]
\nonumber \\
\widetilde{\cal M}^{ggg}_{12,13} &=& 16 [d + 1] [\omega - 2]^2
~~,~~
\widetilde{\cal M}^{ggg}_{12,14} ~=~ -~ 32 [d + 1] [\omega - 2]
\nonumber \\
\widetilde{\cal M}^{ggg}_{13,13} &=& 32 [2 d + \omega^2 - 4 \omega + 2]
~~,~~
\widetilde{\cal M}^{ggg}_{13,14} ~=~ -~ 32 [d + 1] [\omega - 2]
\nonumber \\
\widetilde{\cal M}^{ggg}_{14,14} &=& 64 [d + 1]
\end{eqnarray}
for the triple gluon vertex. For the mass operator the projection matrix is
diagonal since 
\begin{equation}
{\cal M}^{\bar{\psi}\psi} ~=~ \frac{1}{4\omega[\omega-4]} \left(
\begin{array}{cc}
\omega [\omega-4] & 0 \\
0 & 4 \\
\end{array}
\right) ~.
\end{equation}
For each of the three vertices and the operator we have checked that the 
symmetric point matrices of \cite{22} emerge in the $\omega$~$\to$~$1$ limit.

\clearpage

{\begin{table}[ht]
\begin{center}
\begin{tabular}{|c|c||r|r|r||r|r|r|}
\hline
$\Nc$ & $\Nf$ & $\iMOMq$ & $\iMOMh$ & $\iMOMg$ & $\iMOMq$ & $\iMOMh$ & $\iMOMg$
\\
\hline
2 & 6 & 0.082180 & 0.102178 & 0.079931 & 0.075759 & 0.097485 & 0.069611 \\
2 & 7 & 0.060932 & 0.070282 & 0.054313 & 0.058801 & 0.068282 & 0.047987 \\
2 & 8 & 0.044235 & 0.048765 & 0.037734 & 0.044086 & 0.047873 & 0.033758 \\
2 & 9 & 0.029369 & 0.031291 & 0.024820 & 0.029912 & 0.030949 & 0.022590 \\
2 & 10 & 0.014858 & 0.015344 & 0.013027 & 0.015226 & 0.015270 & 0.012204 \\
\hline
3 & 9 & 0.053594 & 0.066233 & 0.050891 & 0.049640 & 0.063241 & 0.044376 \\
3 & 10 & 0.043677 & 0.051213 & 0.039222 & 0.041718 & 0.049554 & 0.034557 \\
3 & 11 & 0.035531 & 0.040203 & 0.030790 & 0.034749 & 0.039239 & 0.027353 \\
3 & 12 & 0.028408 & 0.031285 & 0.024117 & 0.028304 & 0.030730 & 0.021614 \\
3 & 13 & 0.021854 & 0.023529 & 0.018438 & 0.022984 & 0.023223 & 0.016704 \\
3 & 14 & 0.015572 & 0.016421 & 0.013274 & 0.015876 & 0.016283 & 0.012202 \\
3 & 15 & 0.009362 & 0.009671 & 0.008248 & 0.009557 & 0.009627 & 0.007743 \\
3 & 16 & 0.003123 & 0.003158 & 0.002949 & 0.003156 & 0.003154 & 0.002865 \\
\hline
\end{tabular}
\end{center}
\begin{center}
{Table $1$. Three loop critical couplings for the three schemes $\iMOMq$, 
$\iMOMg$ and $\iMOMh$ for $\omega$~$=$~$\half$ (left bank) and $\omega$~$=$~$2$ 
(right bank).}
\end{center}
\end{table}}

{\begin{table}[ht]
\begin{center}
\begin{tabular}{|c|c||r|r|r||r|r|r|}
\hline
$\Nc$ & $\Nf$ & $\iMOMq$ & $\iMOMh$ & $\iMOMg$ & $\iMOMq$ & $\iMOMh$ & $\iMOMg$
\\
\hline
2 & 6 & 1.046201 & 1.285814 & 1.018897 & 0.968029 & 1.230113 & 0.892618 \\
2 & 7 & 0.562075 & 0.632772 & 0.509522 & 0.545385 & 0.618001 & 0.457363 \\
2 & 8 & 0.275610 & 0.295000 & 0.244918 & 0.274943 & 0.291309 & 0.224482 \\
2 & 9 & 0.108045 & 0.111728 & 0.097670 & 0.109128 & 0.111102 & 0.091733 \\
2 & 10 & 0.023580 & 0.023809 & 0.022407 & 0.023756 & 0.023776 & 0.021721 \\
\hline 
3 & 9 & 1.002950 & 1.219382 & 0.955671 & 0.933669 & 1.168830 & 0.840254 \\
3 & 10 & 0.662954 & 0.758306 & 0.603957 & 0.637256 & 0.737797 & 0.540073 \\
3 & 11 & 0.426862 & 0.468961 & 0.380803 & 0.419495 & 0.460538 & 0.345318 \\
3 & 12 & 0.260195 & 0.277545 & 0.231243 & 0.259538 & 0.274324 & 0.212649 \\
3 & 13 & 0.144113 & 0.150218 & 0.129481 & 0.144993 & 0.149178 & 0.120933 \\
3 & 14 & 0.067279 & 0.068846 & 0.061947 & 0.067865 & 0.068606 & 0.058914 \\
3 & 15 & 0.022022 & 0.022223 & 0.021019 & 0.022153 & 0.022196 & 0.020420 \\
3 & 16 & 0.002200 & 0.002203 & 0.002181 & 0.002203 & 0.002202 & 0.002167 \\
\hline
\end{tabular}
\end{center}
\begin{center}
{Table $2$. Three loop exponent ${\widetilde{\omega}}$ for the three schemes 
$\iMOMq$, $\iMOMg$ and $\iMOMh$ for $\omega$~$=$~$\half$ (left bank) and 
$\omega$~$=$~$2$ (right bank).}
\end{center}
\end{table}}

{\begin{table}[ht]
\begin{center}
\begin{tabular}{|c|c||r|r|r||r|r|r|}
\hline
$\Nc$ & $\Nf$ & $\omega$~$=$~$\half$ & $\omega$~$=$~$1$ & $\omega$~$=$~$2$ & 
$\omega$~$=$~$\half$ & $\omega$~$=$~$1$ & $\omega~$=$~$2$$ \\
\hline
2 & 6 & 27.239861 & 17.262397 & 2.285966 & 0.600572 & 0.461381 & 0.337700 \\
2 & 7 & 2.471236 & 1.925820 & 1.106095 & 0.401204 & 0.346755 & 0.290809 \\
2 & 8 & 0.743765 & 0.649692 & 0.508076 & 0.266242 & 0.247039 & 0.225155 \\
2 & 9 & 0.280376 & 0.262282 & 0.234985 & 0.161626 & 0.156674 & 0.150586 \\
2 & 10 & 0.092919 & 0.090649 & 0.087215 & 0.074218 & 0.073686 & 0.073000 \\
\hline
3 & 9 & 16.864034 & 11.561746 & 3.624314 & 0.708237 & 0.553462 & 0.411676 \\
3 & 10 & 3.848168 & 2.978729 & 1.676364 & 0.533844 & 0.452897 & 0.371313 \\
3 & 11 & 1.560926 & 1.312033 & 0.938925 & 0.405984 & 0.364656 & 0.319805 \\
3 & 12 & 0.773689 & 0.689329 & 0.562753 & 0.304867 & 0.285218 & 0.262590 \\
3 & 13 & 0.412657 & 0.383454 & 0.339589 & 0.220487 & 0.212345 & 0.202509 \\
3 & 14 & 0.218111 & 0.208960 & 0.195196 & 0.147470 & 0.144860 & 0.141586 \\
3 & 15 & 0.101914 & 0.099806 & 0.096629 & 0.082990 & 0.082504 & 0.081877 \\
3 & 16 & 0.027438 & 0.027285 & 0.027054 & 0.025855 & 0.025840 & 0.025821 \\
\hline
\end{tabular}
\end{center}
\begin{center}
{Table $3$. Exponent $\rho$ for the $\iMOMq$ scheme for $\omega$~$=$~$\half$, 
$1$ and $2$ at two loop (left bank) and three loop (right bank).}
\end{center}
\end{table}}

{\begin{table}[ht]
\begin{center}
\begin{tabular}{|c|c||r|r|r||r|r|r|}
\hline
$\Nc$ & $\Nf$ & $\omega$~$=$~$\half$ & $\omega$~$=$~$1$ & $\omega$~$=$~$2$ & 
$\omega$~$=$~$\half$ & $\omega$~$=$~$1$ & $\omega~$=$~$2$$ \\
\hline
2 & 6 & 19.223415 & 13.977991 & 6.482302 & 0.397351 & 0.305679 & 0.207404 \\
2 & 7 & 1.978641 & 1.724000 & 1.363952 & 0.331401 & 0.304515 & 0.274502 \\
2 & 8 & 0.646766 & 0.609951 & 0.558852 & 0.241931 & 0.234852 & 0.226952 \\
2 & 9 & 0.258640 & 0.253377 & 0.246363 & 0.155029 & 0.153796 & 0.152490 \\
2 & 10 & 0.089653 & 0.089311 & 0.088924 & 0.073435 & 0.073373 & 0.073325 \\
\hline
3 & 9 & 11.955015 & 9.016606 & 4.817632 & 0.481863 & 0.375534 & 0.265557 \\
3 & 10 & 2.975518 & 2.526293 & 1.888493 & 0.423952 & 0.377682 & 0.328109 \\
3 & 11 & 1.288175 & 1.170622 & 1.005228 & 0.350646 & 0.330763 & 0.309163 \\
3 & 12 & 0.671895 & 0.636553 & 0.587498 & 0.277955 & 0.270097 & 0.261584 \\
3 & 13 & 0.373456 & 0.363130 & 0.349118 & 0.208770 & 0.206167 & 0.203416 \\
3 & 14 & 0.204271 & 0.201785 & 0.198561 & 0.143428 & 0.142818 & 0.142216 \\
3 & 15 & 0.098263 & 0.097913 & 0.097517 & 0.082160 & 0.082094 & 0.082043 \\
3 & 16 & 0.027128 & 0.027124 & 0.027129 & 0.025826 & 0.025826 & 0.025827 \\
\hline
\end{tabular}
\end{center}
\begin{center}
{Table $4$. Exponent $\rho$ for the $\iMOMh$ scheme for $\omega$~$=$~$\half$, 
$1$ and $2$ at two loop (left bank) and three loop (right bank).}
\end{center}
\end{table}}

{\begin{table}[ht]
\begin{center}
\begin{tabular}{|c|c||r|r|r||r|r|r|}
\hline
$\Nc$ & $\Nf$ & $\omega$~$=$~$\half$ & $\omega$~$=$~$1$ & $\omega$~$=$~$2$ & 
$\omega$~$=$~$\half$ & $\omega$~$=$~$1$ & $\omega~$=$~$2$$ \\
\hline
2 & 6 & 45.515760 & 45.730994 & 46.741820 & 0.920418 & 0.861479 & 0.796716 \\
2 & 7 & 4.046556 & 4.202074 & 4.463909 & 0.554462 & 0.542352 & 0.527694 \\
2 & 8 & 1.143032 & 1.201675 & 1.292562 & 0.336918 & 0.336642 & 0.335932 \\
2 & 9 & 0.389802 & 0.409221 & 0.438400 & 0.187578 & 0.189326 & 0.191453 \\
2 & 10 & 0.112358 & 0.116219 & 0.121927 & 0.078876 & 0.079602 & 0.080609 \\
\hline
3 & 9 & 26.683597 & 26.804168 & 27.370418 & 1.020688 & 0.954324 & 0.883068 \\
3 & 10 & 6.082398 & 6.257561 & 6.573996 & 0.725820 & 0.701362 & 0.673553 \\
3 & 11 & 2.411977 & 2.514774 & 2.681119 & 0.524318 & 0.516777 & 0.507645 \\
3 & 12 & 1.148311 & 1.204608 & 1.291859 & 0.374713 & 0.374024 & 0.372958 \\
3 & 13 & 0.578876 & 0.607462 & 0.650756 & 0.257811 & 0.259356 & 0.261181 \\
3 & 14 & 0.284546 & 0.297076 & 0.315784 & 0.163806 & 0.165375 & 0.167404 \\
3 & 15 & 0.121486 & 0.125435 & 0.131273 & 0.088730 & 0.088262 & 0.089284 \\
3 & 16 & 0.029273 & 0.029663 & 0.030236 & 0.026094 & 0.026154 & 0.026247 \\
\hline
\end{tabular}
\end{center}
\begin{center}
{Table $5$. Exponent $\rho$ for the $\iMOMg$ scheme for $\omega$~$=$~$\half$, 
$1$ and $2$ at two loop (left bank) and three loop (right bank).}
\end{center}
\end{table}}

{\begin{table}[ht]
\begin{center}
\begin{tabular}{|c|c||r|r|r||r|r|r|}
\hline
$\Nc$ & $\Nf$ & $\omega=\half$ & $\omega=1$ & $\omega=2$ & 
$\omega=\half$ & $\omega=1$ & $\omega=2$
\\
\hline
2 & 6 & 0.799071 & 0.537774 & 0.365556 & 0.085019 & 0.080231 & 0.081241 \\
2 & 7 & 0.216131 & 0.185636 & 0.154275 & 0.061121 & 0.059714 & 0.059190 \\
2 & 8 & 0.097538 & 0.088764 & 0.078600 & 0.043575 & 0.042749 & 0.042380 \\
2 & 9 & 0.046504 & 0.043433 & 0.039661 & 0.028639 & 0.028092 & 0.027695 \\
2 & 10 & 0.018097 & 0.017153 & 0.015955 & 0.014474 & 0.014104 & 0.013695 \\
\hline
3 & 8 & 16.520099 & 1.288823 & 0.553544 & 0.067606 & 0.064207 & 0.061489 \\
3 & 9 & 0.384329 & 0.293024 & 0.218305 & 0.053563 & 0.051567 & 0.050019 \\
3 & 10 & 0.168767 & 0.145756 & 0.121824 & 0.043260 & 0.042012 & 0.041088 \\
3 & 11 & 0.095628 & 0.086485 & 0.076013 & 0.034976 & 0.034165 & 0.033589 \\
3 & 12 & 0.058812 & 0.054483 & 0.049258 & 0.027846 & 0.027301 & 0.026914 \\
3 & 13 & 0.036644 & 0.034450 & 0.031716 & 0.021368 & 0.020986 & 0.020687 \\
3 & 14 & 0.021831 & 0.020731 & 0.019327 & 0.015219 & 0.014936 & 0.014671 \\
3 & 15 & 0.011235 & 0.010745 & 0.010111 & 0.009169 & 0.008966 & 0.008733 \\
3 & 16 & 0.003278 & 0.003153 & 0.002988 & 0.003078 & 0.002985 & 0.002865 \\
\hline
\end{tabular}
\end{center}
\begin{center}
{Table $6$. Critical couplings for the MAG in the $\iMOMmq$ scheme for 
$\omega$~$=$~$\half$, $1$ and $2$ at two loop (left bank) and three loop (right
bank).}
\end{center}
\end{table}}

\clearpage

{\begin{table}[ht]
\begin{center}
\begin{tabular}{|c|c||r|r|r||r|r|r|}
\hline
$\Nc$ & $\Nf$ & $\omega=\half$ & $\omega=1$ & $\omega=2$ & 
$\omega=\half$ & $\omega=1$ & $\omega=2$
\\
\hline
2 & 7 & 0.352911 & 0.348066 & 0.342161 & 0.078831 & 0.073482 & 0.066826 \\
2 & 8 & 0.127203 & 0.126358 & 0.125311 & 0.053505 & 0.051563 & 0.048882 \\
2 & 9 & 0.055812 & 0.055568 & 0.055263 & 0.034175 & 0.033613 & 0.032775 \\
2 & 10 & 0.020797 & 0.020729 & 0.020644 & 0.016844 & 0.016771 & 0.016655 \\
\hline
3 & 9 & 0.909893 & 0.833333 & 0.749799 & 0.072040 & 0.066273 & 0.059306 \\
3 & 10 & 0.238596 & 0.232143 & 0.224117 & 0.054752 & 0.051951 & 0.048238 \\
3 & 11 & 0.118938 & 0.117021 & 0.114577 & 0.042665 & 0.041258 & 0.039270 \\
3 & 12 & 0.068973 & 0.068182 & 0.067161 & 0.033125 & 0.032451 & 0.031448 \\
3 & 13 & 0.041547 & 0.041176 & 0.040696 & 0.024931 & 0.024651 & 0.024211 \\
3 & 14 & 0.024215 & 0.024038 & 0.023809 & 0.017440 & 0.017353 & 0.017207 \\
3 & 15 & 0.012271 & 0.012195 & 0.012097 & 0.010298 & 0.010279 & 0.010245 \\
3 & 16 & 0.003540 & 0.003521 & 0.003496 & 0.003366 & 0.003356 & 0.003343 \\
\hline
\end{tabular}
\end{center}
\begin{center}
{Table $7$. Critical couplings for the MAG in the $\iMOMmh$ scheme for 
$\omega$~$=$~$\half$, $1$ and $2$ at two loop (left bank) and three loop (right
bank).}
\end{center}
\end{table}}

{\begin{table}[ht]
\begin{center}
\begin{tabular}{|c|c||r|r|r||r|r|r|}
\hline
$\Nc$ & $\Nf$ & $\omega=\half$ & $\omega=1$ & $\omega=2$ & 
$\omega=\half$ & $\omega=1$ & $\omega=2$
\\
\hline
2 & 6 & 0.456502 & 0.368962 & 0.293214 & 0.068321 & 0.058604 & 0.048399 \\
2 & 7 & 0.172393 & 0.155029 & 0.136508 & 0.046244 & 0.041077 & 0.035270 \\
2 & 8 & 0.084619 & 0.078840 & 0.072198 & 0.032787 & 0.029333 & 0.025796 \\
2 & 9 & 0.041926 & 0.039760 & 0.037173 & 0.021571 & 0.019852 & 0.017805 \\
2 & 10 & 0.016680 & 0.015987 & 0.015140 & 0.011542 & 0.010807 & 0.009909 \\
\hline
3 & 8 & 1.152688 & 0.733566 & 0.491556 & 0.062251 & 0.053790 & 0.044666 \\
3 & 9 & 0.284370 & 0.245200 & 0.206657 & 0.045268 & 0.040311 & 0.034640 \\
3 & 10 & 0.143254 & 0.131082 & 0.117558 & 0.034919 & 0.031600 & 0.027699 \\
3 & 11 & 0.085438 & 0.080190 & 0.074032 & 0.027530 & 0.025184 & 0.022383 \\
3 & 12 & 0.053974 & 0.051377 & 0.048236 & 0.021688 & 0.020014 & 0.017991 \\
3 & 13 & 0.034188 & 0.032837 & 0.031169 & 0.016696 & 0.015532 & 0.014111 \\
3 & 14 & 0.020597 & 0.019906 & 0.019041 & 0.012119 & 0.011373 & 0.010500 \\
3 & 15 & 0.010686 & 0.010374 & 0.009981 & 0.007610 & 0.007219 & 0.006726 \\
3 & 16 & 0.003137 & 0.003056 & 0.002953 & 0.002766 & 0.002665 & 0.002534 \\
\hline
\end{tabular}
\end{center}
\begin{center}
{Table $8$. Critical couplings for the MAG in the $\iMOMmg$ scheme for 
$\omega$~$=$~$\half$, $1$ and $2$ at two loop (left bank) and three loop (right
bank).}
\end{center}
\end{table}}

\clearpage 

{\begin{table}[ht]
\begin{center}
\begin{tabular}{|c|c||r|r|r||r|r|r|}
\hline
$\Nc$ & $\Nf$ & $\omega=\half$ & $\omega=1$ & $\omega=2$ &
$\omega=\half$ & $\omega=1$ & $\omega=2$
\\
\hline
2 & 6 & 5.327143 & 3.585161 & 2.437039 & 1.073281 & 1.013502 & 0.962845 \\
2 & 7 & 1.152701 & 0.990057 & 0.822798 & 0.559772 & 0.534507 & 0.510243 \\
2 & 8 & 0.390152 & 0.355055 & 0.314319 & 0.270730 & 0.259641 & 0.247612 \\
2 & 9 & 0.124010 & 0.115822 & 0.105762 & 0.105709 & 0.101371 & 0.096136 \\
2 & 10 & 0.024130 & 0.022871 & 0.021273 & 0.023162 & 0.022148 & 0.020846 \\
\hline
3 & 8 &          &          &          & 1.529262 & 1.419098 & 1.316340 \\
3 & 9 & 3.843291 & 2.930241 & 2.183050 & 0.996607 & 0.940596 & 0.885769 \\
3 & 10 & 1.462652 & 1.263216 & 1.055809 & 0.653734 & 0.623259 & 0.592086 \\
3 & 11 & 0.701272 & 0.634220 & 0.557432 & 0.419166 & 0.402108 & 0.383796 \\
3 & 12 & 0.352874 & 0.326896 & 0.295548 & 0.255045 & 0.245527 & 0.234734 \\
3 & 13 & 0.171004 & 0.160769 & 0.148006 & 0.141288 & 0.136210 & 0.130111 \\
3 & 14 & 0.072771 & 0.068102 & 0.064422 & 0.066093 & 0.063703 & 0.060684 \\
3 & 15 & 0.022469 & 0.021491 & 0.020222 & 0.021710 & 0.020901 & 0.019846 \\
3 & 16 & 0.002186 & 0.002102 & 0.001992 & 0.002177 & 0.002096 & 0.001988 \\
\hline
\end{tabular}
\end{center}
\begin{center}
{Table $9$. Critical exponent ${\widetilde{\omega}}$ for the MAG in the 
$\iMOMmq$ scheme for $\omega$~$=$~$\half$, $1$ and $2$ at two loop (left bank) 
and three loop (right bank).}
\end{center}
\end{table}}

{\begin{table}[hb]
\begin{center}
\begin{tabular}{|c|c||r|r|r||r|r|r|}
\hline
$\Nc$ & $\Nf$ & $\omega=\half$ & $\omega=1$ & $\omega=2$ &
$\omega=\half$ & $\omega=1$ & $\omega=2$
\\
\hline
2 & 7 & 1.882193 & 1.856353 & 1.824856 & 0.746948 & 0.701073 & 0.643199 \\
2 & 8 & 0.508813 & 0.505432 & 0.501244 & 0.338018 & 0.328334 & 0.314785 \\
2 & 9 & 0.148833 & 0.148181 & 0.147369 & 0.126465 & 0.125049 & 0.122966 \\
2 & 10 & 0.027729 & 0.027638 & 0.027525 & 0.026727 & 0.026631 & 0.026498 \\
\hline
3 & 9 & 9.098932 & 8.333333 & 7.497988 & 1.383771 & 1.272760 & 1.139203 \\
3 & 10 & 2.067830 & 2.011905 & 1.942349 & 0.840139 & 0.799725 & 0.746148 \\
3 & 11 & 0.872209 & 0.858156 & 0.840231 & 0.513518 & 0.498442 & 0.477261 \\
3 & 12 & 0.413839 & 0.409091 & 0.402969 & 0.302051 & 0.296744 & 0.289024 \\
3 & 13 & 0.193884 & 0.192157 & 0.189917 & 0.162873 & 0.161206 & 0.158755 \\
3 & 14 & 0.080716 & 0.080128 & 0.079363 & 0.074398 & 0.073931 & 0.073261 \\
3 & 15 & 0.024541 & 0.024390 & 0.024193 & 0.023907 & 0.023788 & 0.023626 \\
3 & 16 & 0.002360 & 0.022347 & 0.002331 & 0.002354 & 0.002342 & 0.002326 \\
\hline
\end{tabular}
\end{center}
\begin{center}
{Table $10$. Critical exponent ${\widetilde{\omega}}$ for the MAG in the 
$\iMOMmh$ scheme for $\omega$~$=$~$\half$, $1$ and $2$ at two loop (left bank) 
and three loop (right bank).}
\end{center}
\end{table}}

\clearpage 

{\begin{table}[ht]
\begin{center}
\begin{tabular}{|c|c||r|r|r||r|r|r|}
\hline
$\Nc$ & $\Nf$ & $\omega=\half$ & $\omega=1$ & $\omega=2$ &
$\omega=\half$ & $\omega=1$ & $\omega=2$
\\
\hline
2 & 6 & 3.043344 & 2.459744 & 1.954763 & 0.842774 & 0.719335 & 0.592056 \\
2 & 7 & 0.919427 & 0.826824 & 0.728041 & 0.427109 & 0.380106 & 0.327615 \\
2 & 8 & 0.338478 & 0.315362 & 0.288792 & 0.209472 & 0.191012 & 0.169503 \\
2 & 9 & 0.111803 & 0.106027 & 0.099128 & 0.085449 & 0.079444 & 0.072219 \\
2 & 10 & 0.022240 & 0.021316 & 0.020186 & 0.020130 & 0.019078 & 0.017776 \\
\hline
3 & 8 &          &          &          & 3.123205 & 2.479437 & 1.868995 \\
3 & 9 & 2.843703 & 2.452003 & 2.066569 & 0.833292 & 0.739952 & 0.634737 \\
3 & 10 & 1.241532 & 1.136045 & 1.018833 & 0.531501 & 0.481717 & 0.423555 \\
3 & 11 & 0.626546 & 0.588059 & 0.542903 & 0.338716 & 0.311363 & 0.278660 \\
3 & 12 & 0.323842 & 0.308264 & 0.289414 & 0.207970 & 0.193387 & 0.175627 \\
3 & 13 & 0.159546 & 0.153239 & 0.145454 & 0.117778 & 0.110682 & 0.101890 \\
3 & 14 & 0.068658 & 0.066355 & 0.063472 & 0.057024 & 0.054161 & 0.050550 \\
3 & 15 & 0.021371 & 0.020749 & 0.019961 & 0.019601 & 0.018829 & 0.017838 \\
3 & 16 & 0.002092 & 0.002038 & 0.001969 & 0.002062 & 0.002004 & 0.001929 \\
\hline
\end{tabular}
\end{center}
\begin{center}
{Table $11$. Critical exponent ${\widetilde{\omega}}$ for the MAG in the 
$\iMOMmg$ scheme for $\omega$~$=$~$\half$, $1$ and $2$ at two loop (left bank) 
and three loop (right bank).}
\end{center}
\end{table}}

{\begin{table}[hb]
\vspace{-0.3cm}
\begin{center}
\begin{tabular}{|c|c||r|r|r||r|r|r|}
\hline
$\Nc$ & $\Nf$ & $\omega=\half$ & $\omega=1$ & $\omega=2$ &
$\omega=\half$ & $\omega=1$ & $\omega=2$
\\
\hline
2 & 6 & 27.151323 & 9.421916 & 2.329600 & 0.727579 & 0.529330 & 0.340358 \\
2 & 7 & 2.727786 & 1.739164 & 0.911281 & 0.448441 & 0.372273 & 0.295462 \\
2 & 8 & 0.802887 & 0.621964 & 0.434591 & 0.282612 & 0.252537 & 0.221015 \\
2 & 9 & 0.293480 & 0.252532 & 0.205388 & 0.165442 & 0.154232 & 0.141942 \\
2 & 10 & 0.094414 & 0.086687 & 0.077168 & 0.074136 & 0.070596 & 0.066387 \\
\hline
3 & 8 &          &          &          & 1.133708 & 0.749544 & 0.423023 \\
3 & 9 & 16.945454 & 7.666846 & 2.419001 & 0.786991 & 0.593762 & 0.414971 \\
3 & 10 & 4.059420 & 2.559142 & 1.289464 & 0.574469 & 0.471709 & 0.371009 \\
3 & 11 & 1.645984 & 1.209145 & 0.771741 & 0.426561 & 0.371078 & 0.314365 \\
3 & 12 & 0.807914 & 0.651783 & 0.480014 & 0.314378 & 0.284772 & 0.253448 \\
3 & 13 & 0.425766 & 0.366189 & 0.296502 & 0.223991 & 0.208688 & 0.191928 \\
3 & 14 & 0.222301 & 0.200185 & 0.173150 & 0.148067 & 0.140545 & 0.131936 \\
3 & 15 & 0.102650 & 0.095603 & 0.086687 & 0.082627 & 0.079281 & 0.075218 \\
3 & 16 & 0.027328 & 0.026087 & 0.024471 & 0.025626 & 0.024694 & 0.023492 \\
\hline
\end{tabular}
\end{center}
\begin{center}
{Table $12$. Critical exponent $\rho$ for the MAG in the $\iMOMmq$ scheme for
$\omega$~$=$~$\half$, $1$ and $2$ at two loop (left bank) and three loop (right
bank).}
\end{center}
\end{table}}

\clearpage 

{\begin{table}[ht]
\begin{center}
\begin{tabular}{|c|c||r|r|r||r|r|r|}
\hline
$\Nc$ & $\Nf$ & $\omega=\half$ & $\omega=1$ & $\omega=2$ &
$\omega=\half$ & $\omega=1$ & $\omega=2$
\\
\hline
2 & 7 & 3.747300 & 2.473367 & 0.815222 & 0.319711 & 0.282781 & 0.262300 \\
2 & 8 & 0.863985 & 0.720339 & 0.529470 & 0.252724 & 0.234933 & 0.220454 \\
2 & 9 & 0.309416 & 0.285623 & 0.254192 & 0.166848 & 0.160562 & 0.154358 \\
2 & 10 & 0.101970 & 0.099095 & 0.095369 & 0.080456 & 0.079152 & 0.077654 \\
\hline
3 & 9 & 41.850719 & 15.814616 & - 9.168835 & 0.387396 & 0.298384 & 0.276381 \\
3 & 10 & 4.355094 & 2.760172 & 0.794656 & 0.398791 & 0.340127 & 0.306951 \\
3 & 11 & 1.576575 & 1.214713 & 0.748957 & 0.347815 & 0.314402 & 0.289775 \\
3 & 12 & 0.767773 & 0.656673 & 0.511728 & 0.283215 & 0.265652 & 0.250555 \\
3 & 13 & 0.412837 & 0.376052 & 0.327950 & 0.216388 & 0.208016 & 0.199960 \\
3 & 14 & 0.221764 & 0.210274 & 0.195312 & 0.150500 & 0.147053 & 0.143388 \\
3 & 15 & 0.105550 & 0.102718 & 0.099061 & 0.086974 & 0.085862 & 0.084554 \\
3 & 15 & 0.028950 & 0.028643 & 0.028250 & 0.027477 & 0.027288 & 0.027045 \\
\hline
\end{tabular}
\end{center}
\begin{center}
{Table $13$. Critical exponent $\rho$ for the MAG in the $\iMOMmh$ scheme for
$\omega$~$=$~$\half$, $1$ and $2$ at two loop (left bank) and three loop (right
bank).}
\end{center}
\end{table}}

{\begin{table}[ht]
\begin{center}
\begin{tabular}{|c|c||r|r|r||r|r|r|}
\hline
$\Nc$ & $\Nf$ & $\omega=\half$ & $\omega=1$ & $\omega=2$ &
$\omega=\half$ & $\omega=1$ & $\omega=2$
\\
\hline
2 & 6 & 18.889955 & 13.115654 & 9.026188 & 1.137715 & 0.904094 & 0.686813 \\
2 & 7 & 3.461711 & 3.017882 & 2.588851 & 0.596908 & 0.522030 & 0.439606 \\
2 & 8 & 1.096596 & 1.031878 & 0.962335 & 0.338991 & 0.311343 & 0.277765 \\
2 & 9 & 0.381247 & 0.370717 & 0.358819 & 0.181255 & 0.171390 & 0.158654 \\
2 & 10 & 0.108207 & 0.106115 & 0.103640 & 0.074094 & 0.071156 & 0.067335 \\
\hline
3 & 8 &          &          &          & 2.311115 & 1.831479 & 1.380792 \\
3 & 9 & 17.574292 & 13.681752 & 10.430036 & 1.165083 & 0.991664 & 0.809849 \\
3 & 10 & 5.378418 & 4.779519 & 4.181646 & 0.776540 & 0.695456 & 0.602230 \\
3 & 11 & 2.313440 & 2.179419 & 2.036729 & 0.539557 & 0.498978 & 0.449066 \\
3 & 12 & 1.131927 & 1.100464 & 1.066610 & 0.375802 & 0.355280 & 0.328634 \\
3 & 13 & 0.574349 & 0.568081 & 0.561776 & 0.253855 & 0.243752 & 0.230073 \\
3 & 14 & 0.281210 & 0.280212 & 0.279458 & 0.159044 & 0.154304 & 0.147767 \\
3 & 15 & 0.118769 & 0.095603 & 0.086687 & 0.084019 & 0.079281 & 0.079108 \\
3 & 16 & 0.028135 & 0.026087 & 0.027193 & 0.024828 & 0.024694 & 0.023407 \\
\hline
\end{tabular}
\end{center}
\begin{center}
{Table $14$. Critical exponent $\rho$ for the MAG in the $\iMOMmg$ scheme for
$\omega$~$=$~$\half$, $1$ and $2$ at two loop (left bank) and three loop (right
bank).}
\end{center}
\end{table}}

\newpage
{\begin{figure}[ht]
\includegraphics[width=7cm,height=7cm]{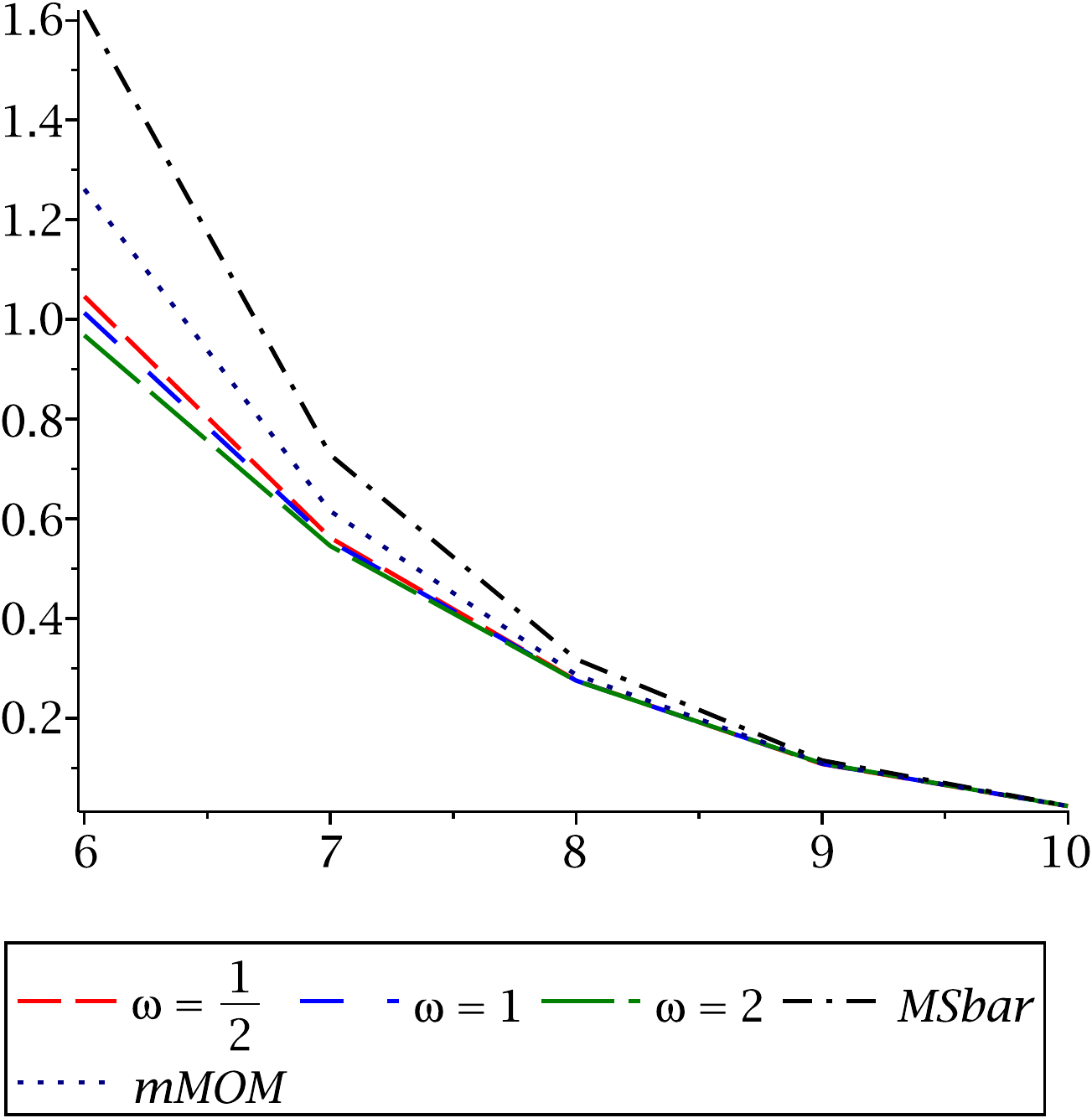}
\quad \quad \quad
\includegraphics[width=7cm,height=7cm]{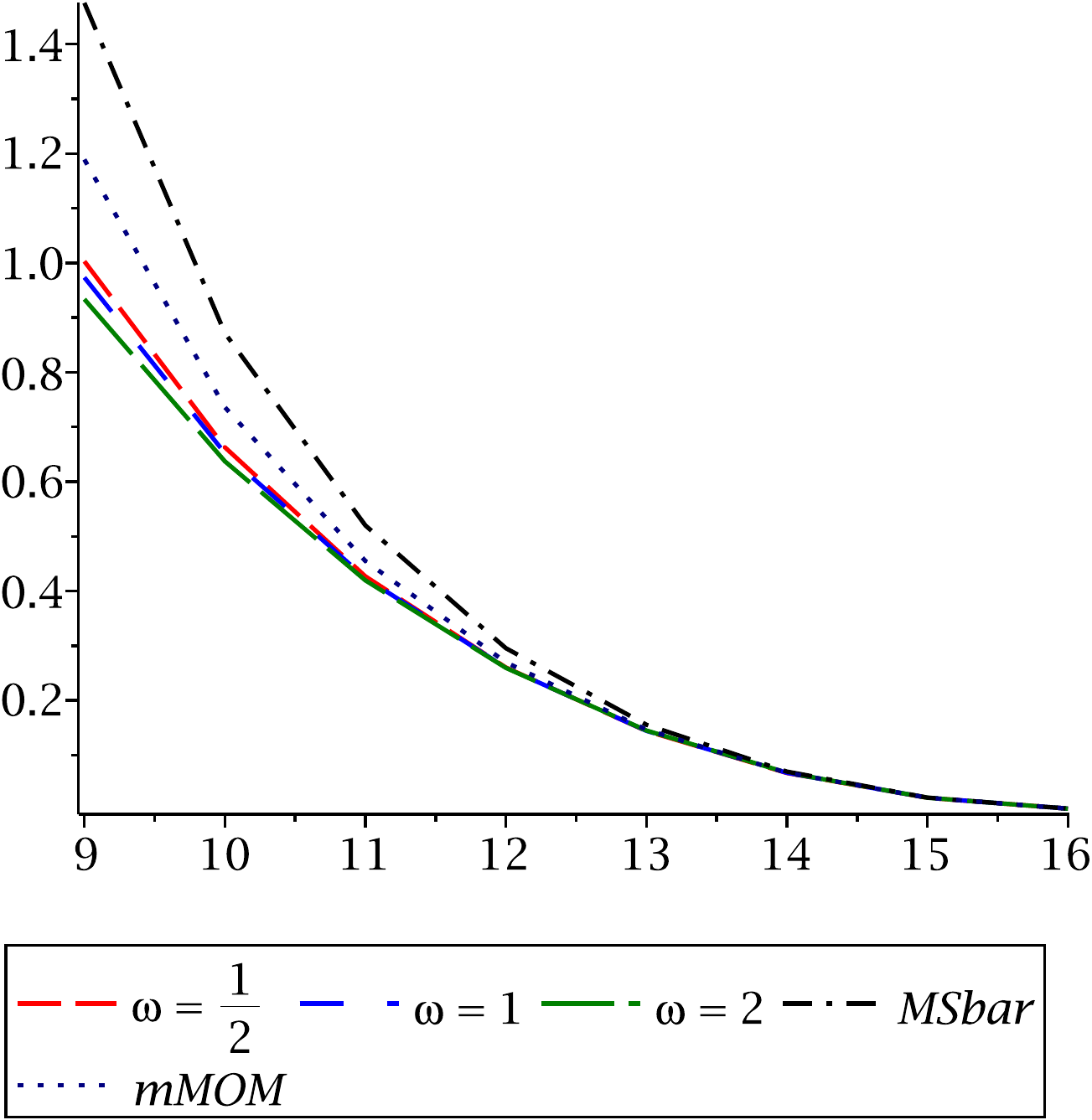}

\vspace{0.5cm}
\includegraphics[width=7cm,height=7cm]{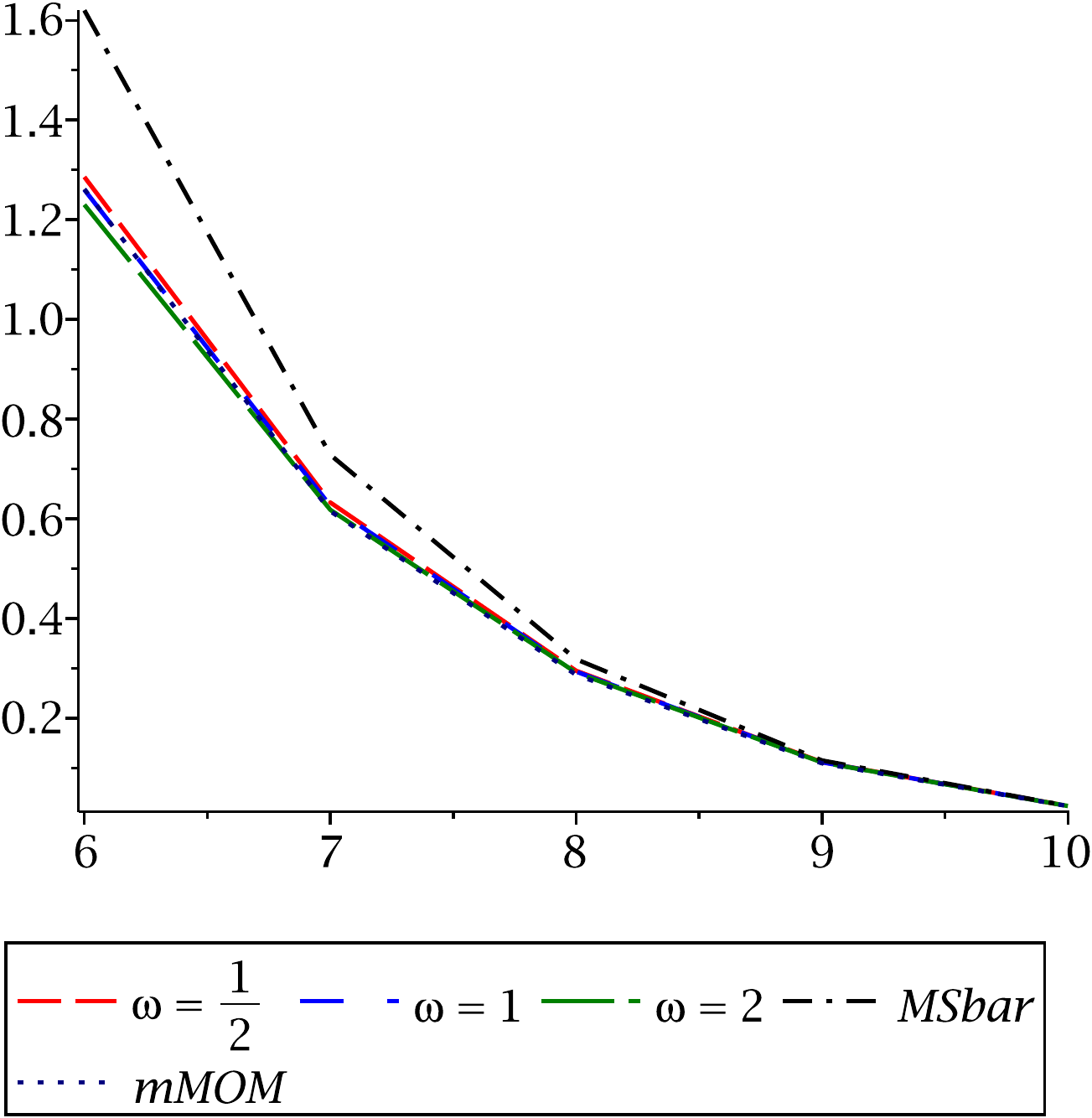}
\quad \quad \quad
\includegraphics[width=7cm,height=7cm]{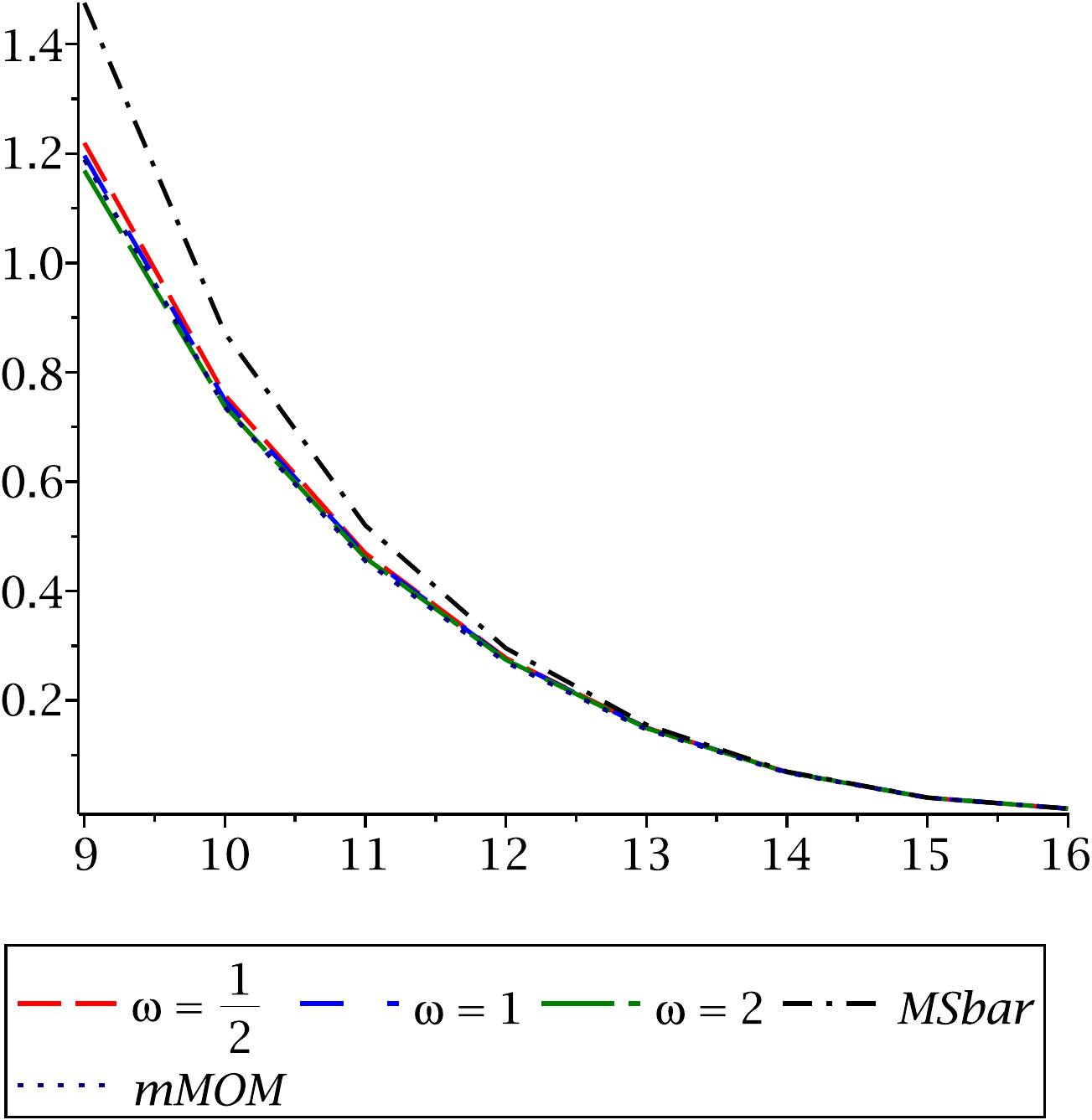}

\vspace{0.5cm}
\includegraphics[width=7cm,height=7cm]{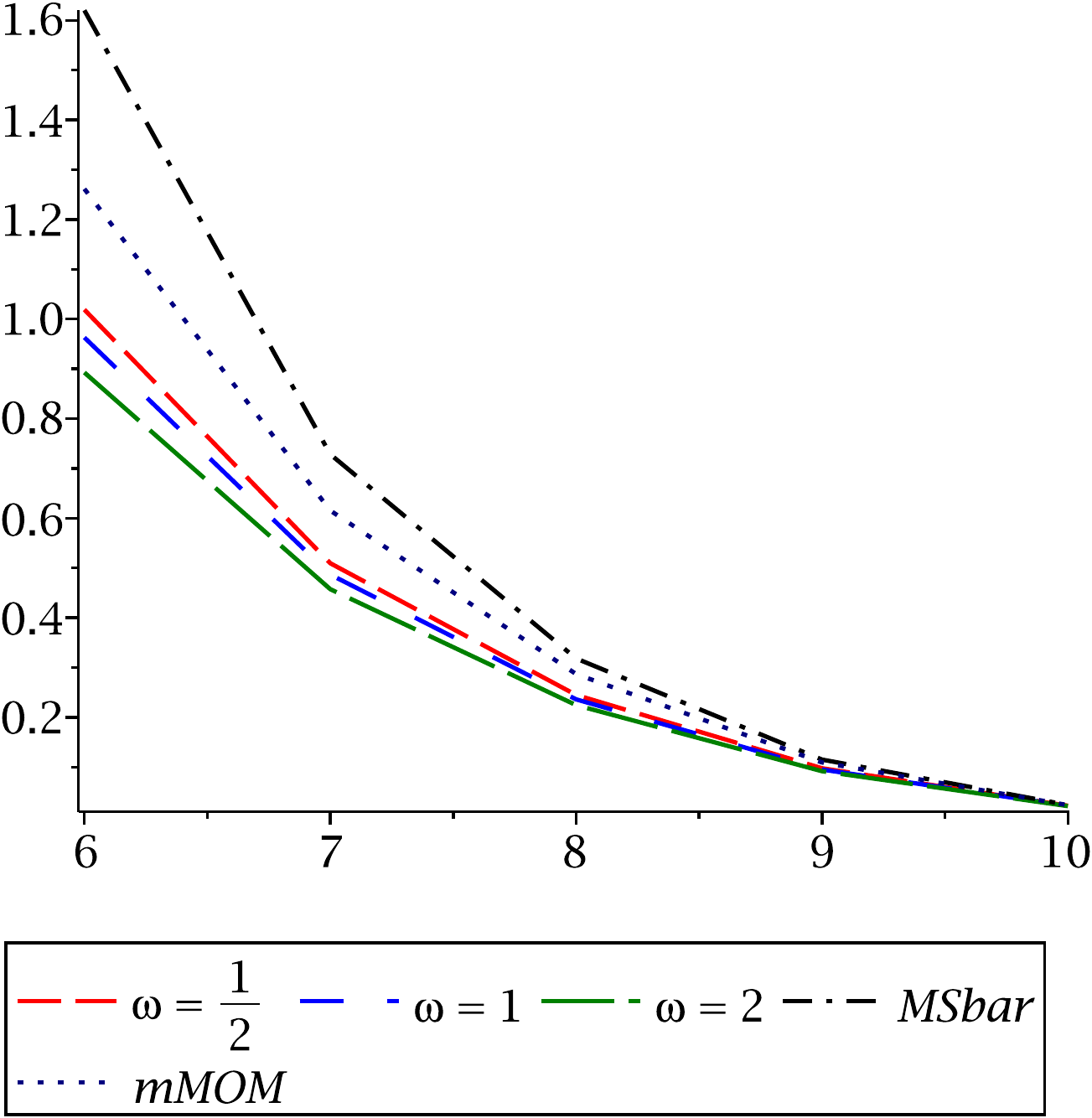}
\quad \quad \quad
\includegraphics[width=7cm,height=7cm]{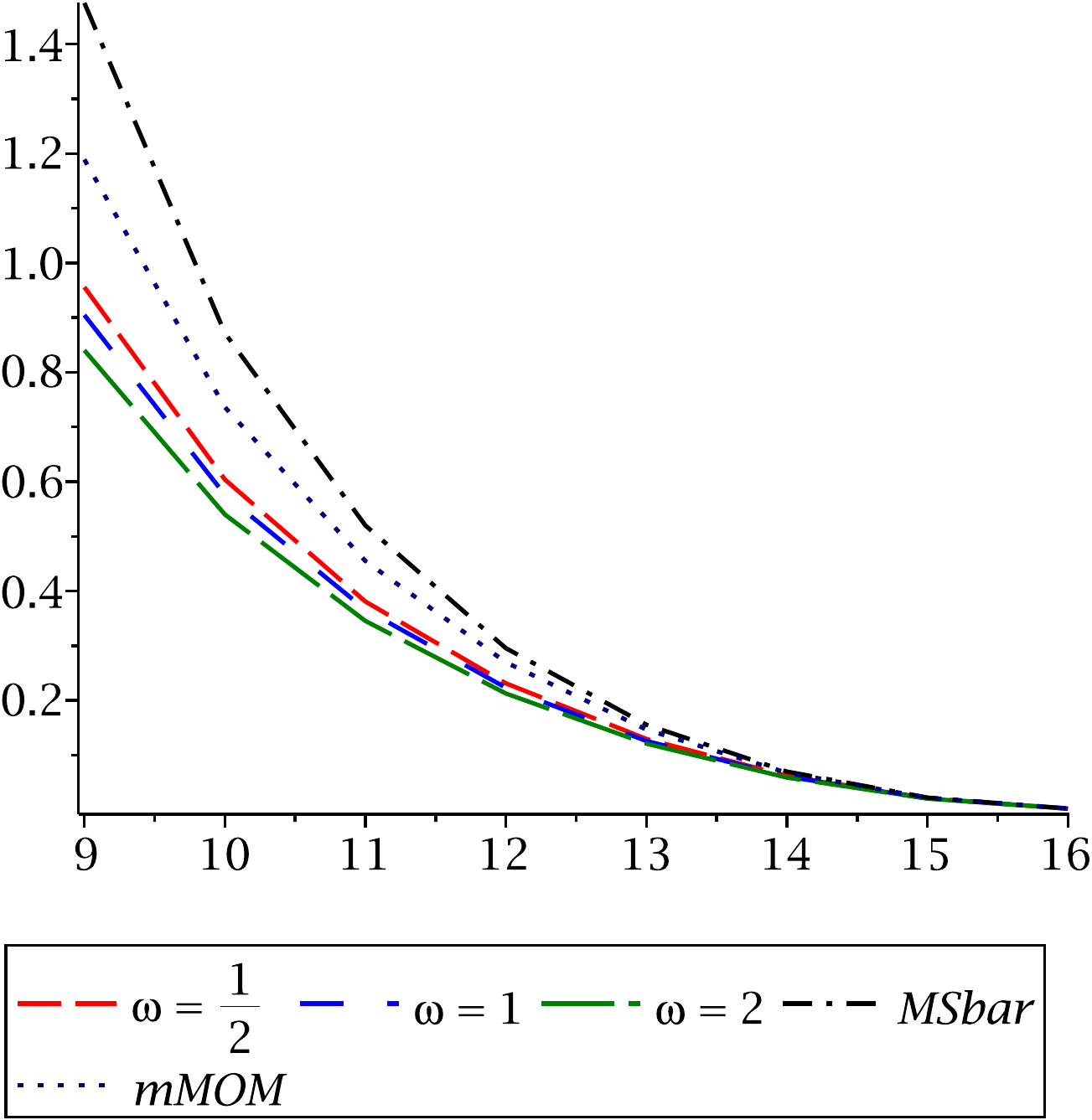}

\caption{Critical exponent ${\widetilde{\omega}}$ at three loops for $SU(2)$ 
(left panel) and $SU(3)$ (right panel) for the respective $\iMOMq$, $\iMOMh$ 
and $\iMOMg$ schemes.}
\end{figure}}

\newpage
{\begin{figure}[ht]
\includegraphics[width=7cm,height=7cm]{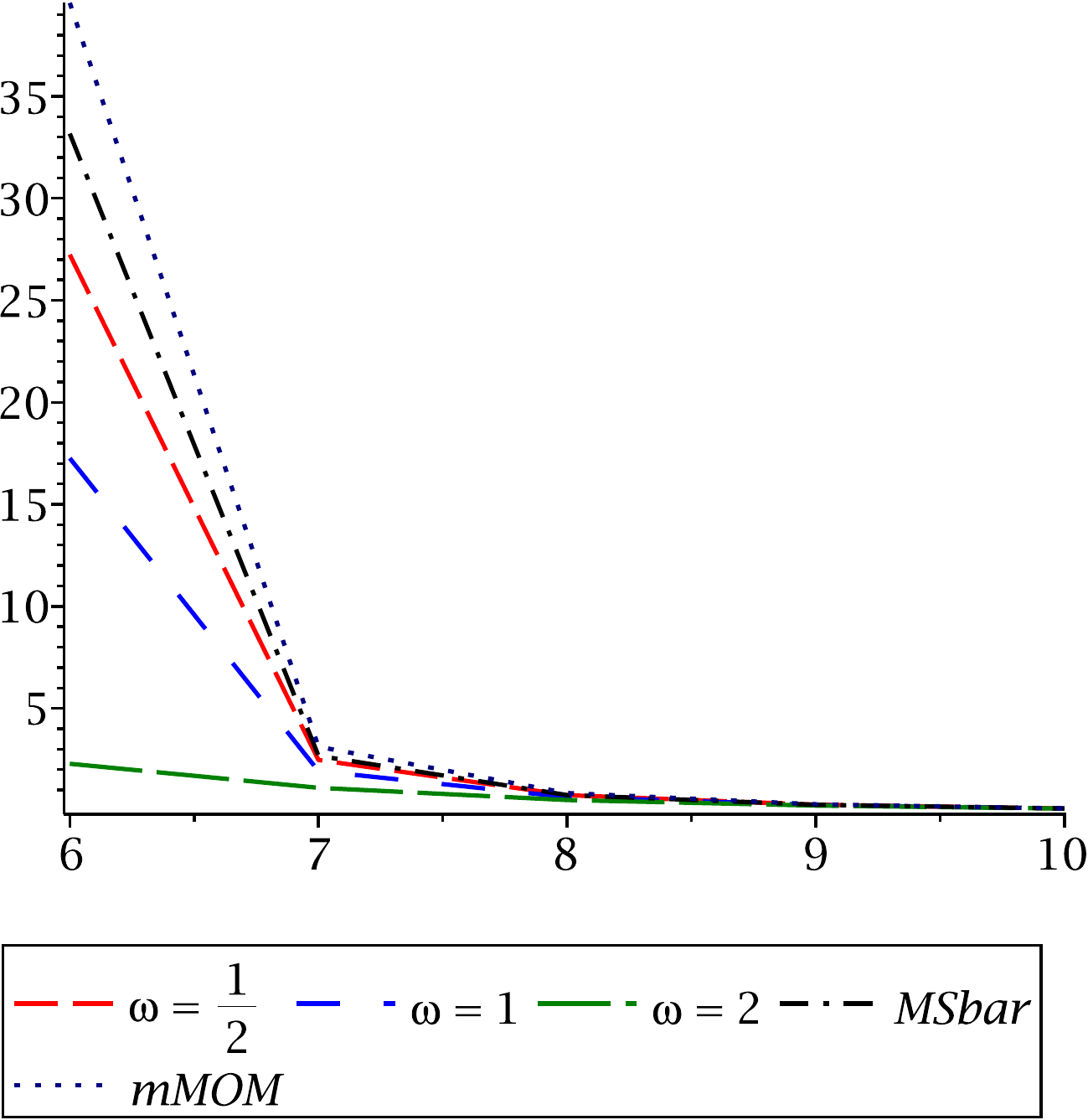}
\quad \quad \quad
\includegraphics[width=7cm,height=7cm]{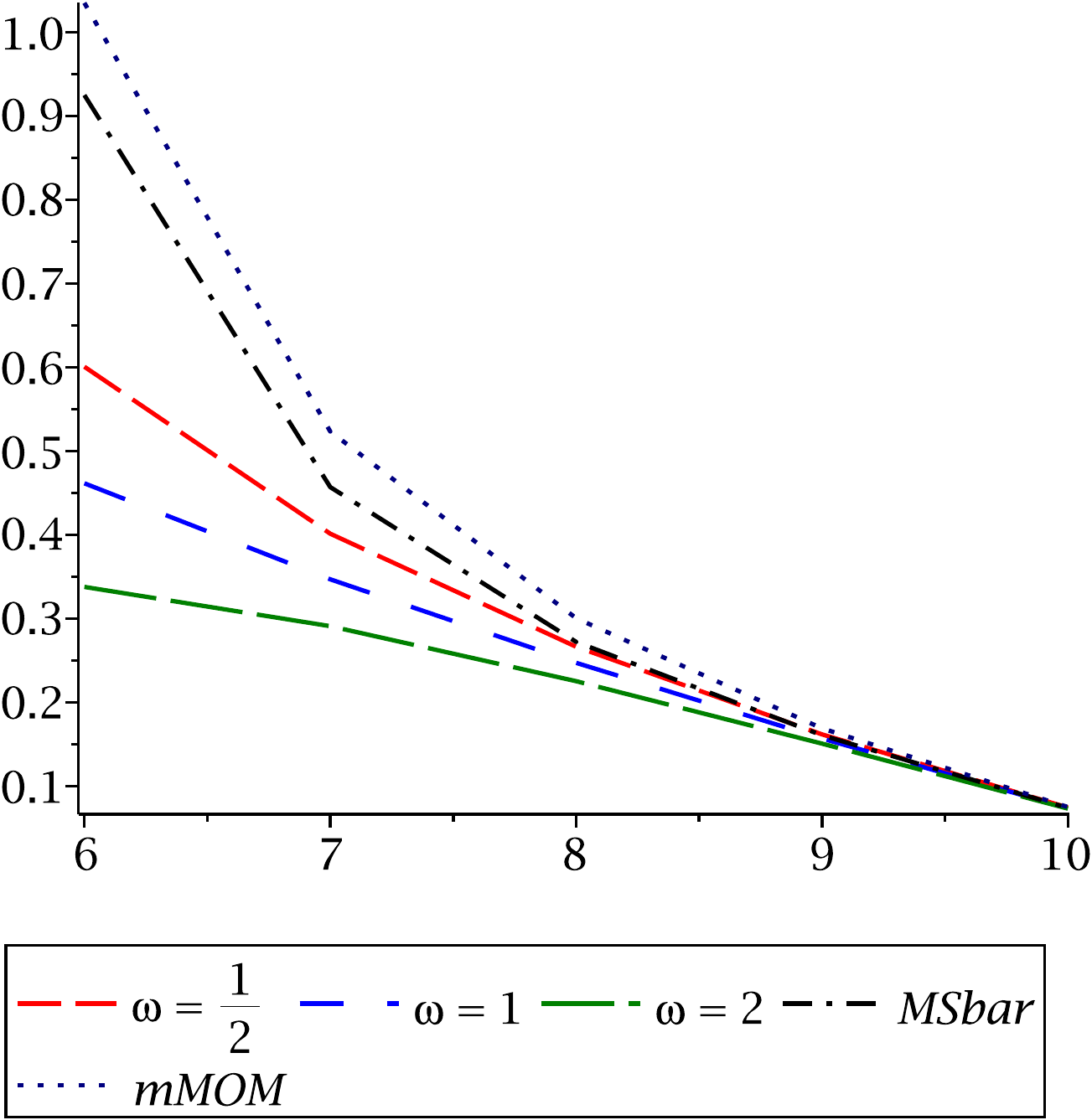}

\vspace{0.5cm}
\includegraphics[width=7cm,height=7cm]{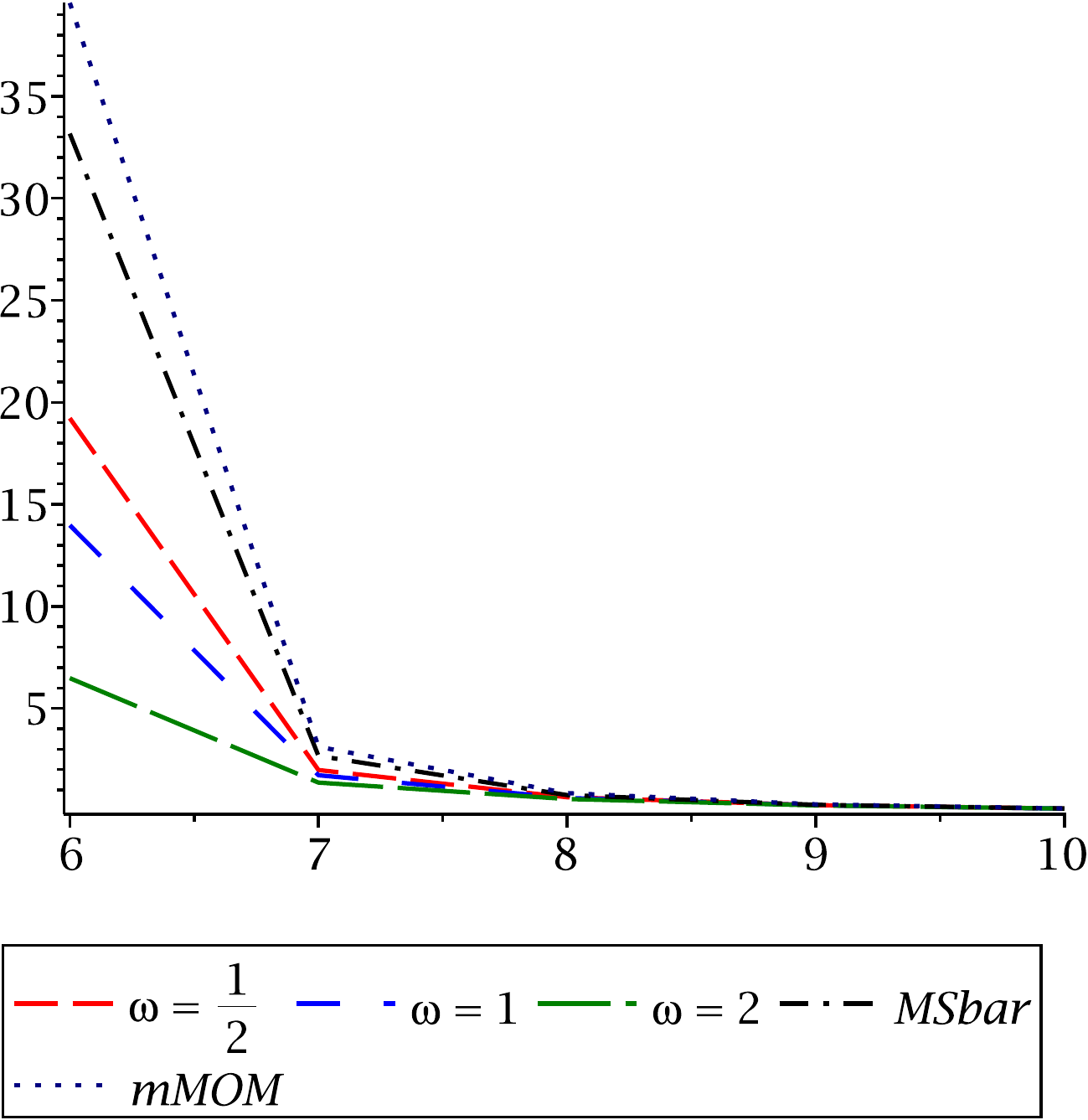}
\quad \quad \quad
\includegraphics[width=7cm,height=7cm]{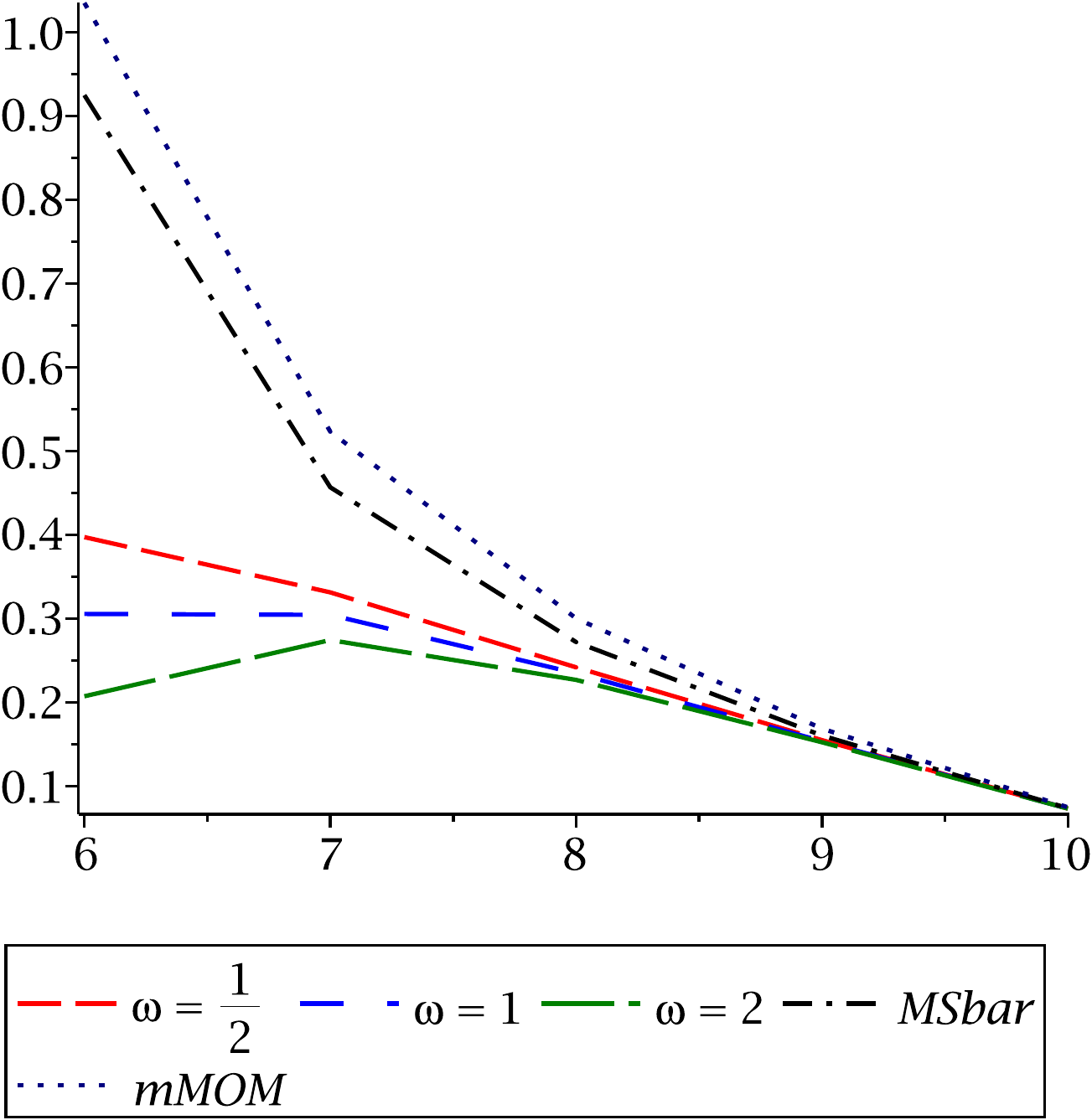}

\vspace{0.5cm}
\includegraphics[width=7cm,height=7cm]{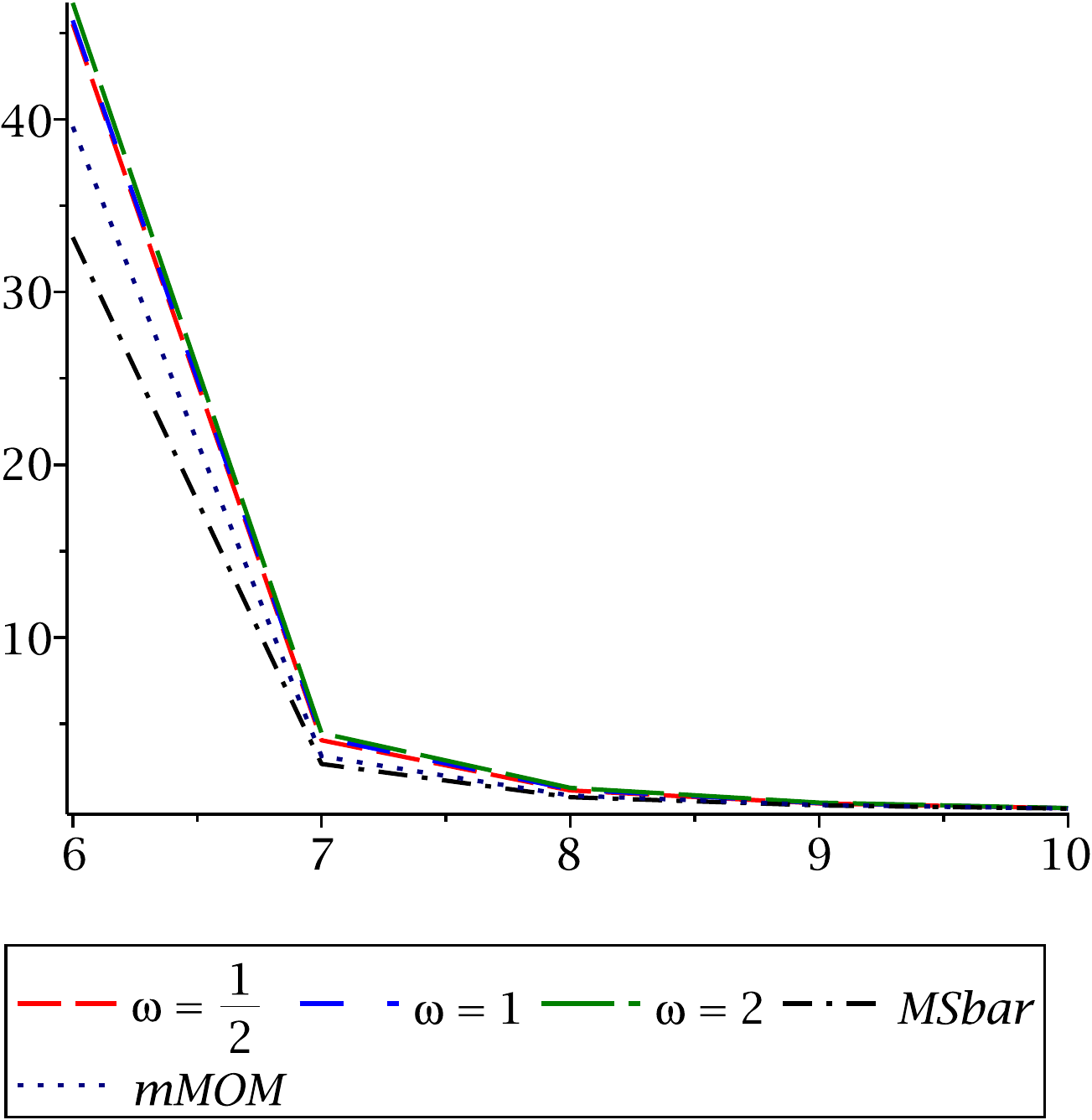}
\quad \quad \quad
\includegraphics[width=7cm,height=7cm]{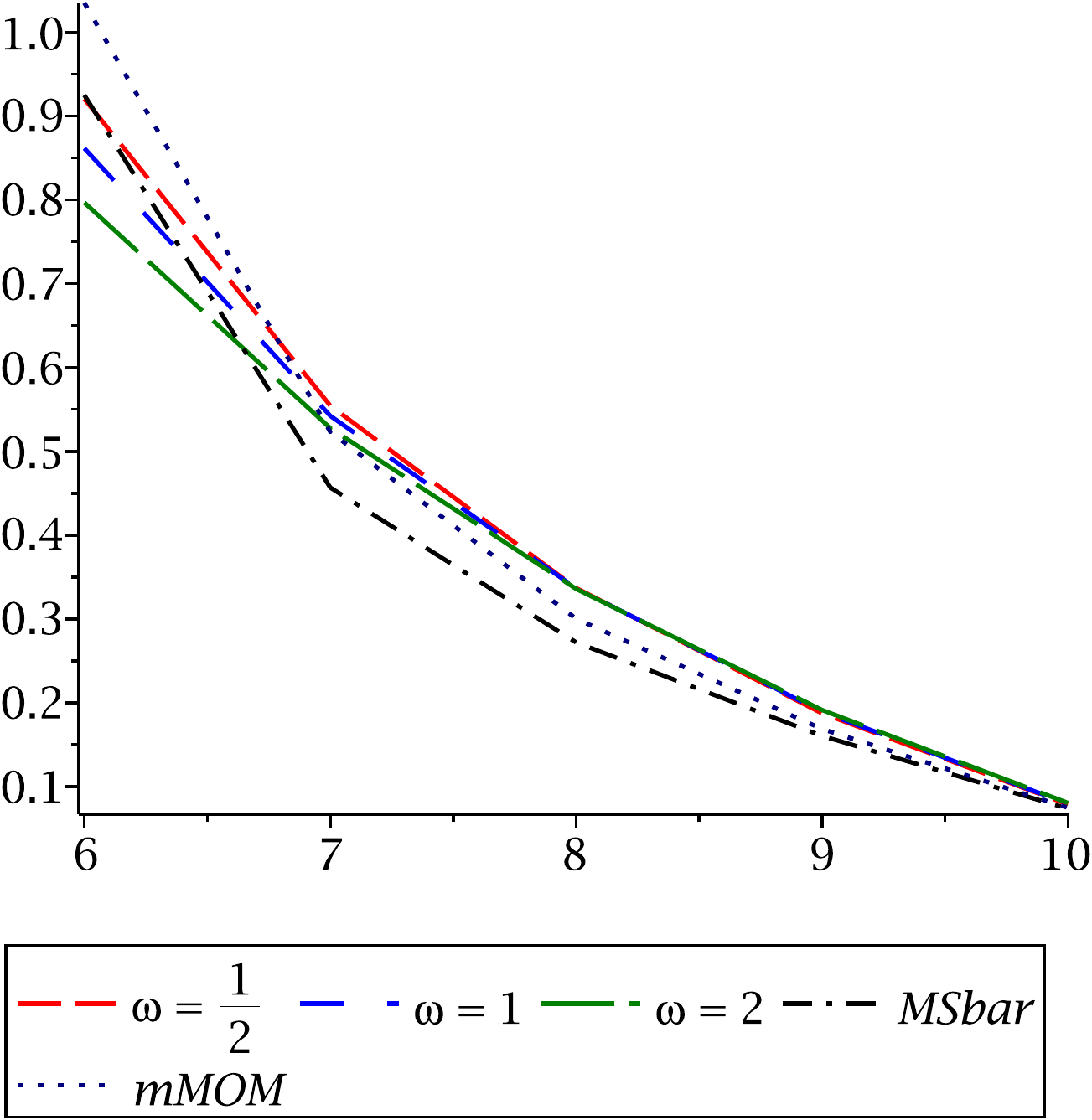}

\caption{Critical exponent $\rho$ for $SU(2)$ at two (left panel) and three 
loops (right panel) for the respective $\iMOMq$, $\iMOMh$ and $\iMOMg$ 
schemes.}
\end{figure}}

\newpage
{\begin{figure}[ht]
\includegraphics[width=7cm,height=7cm]{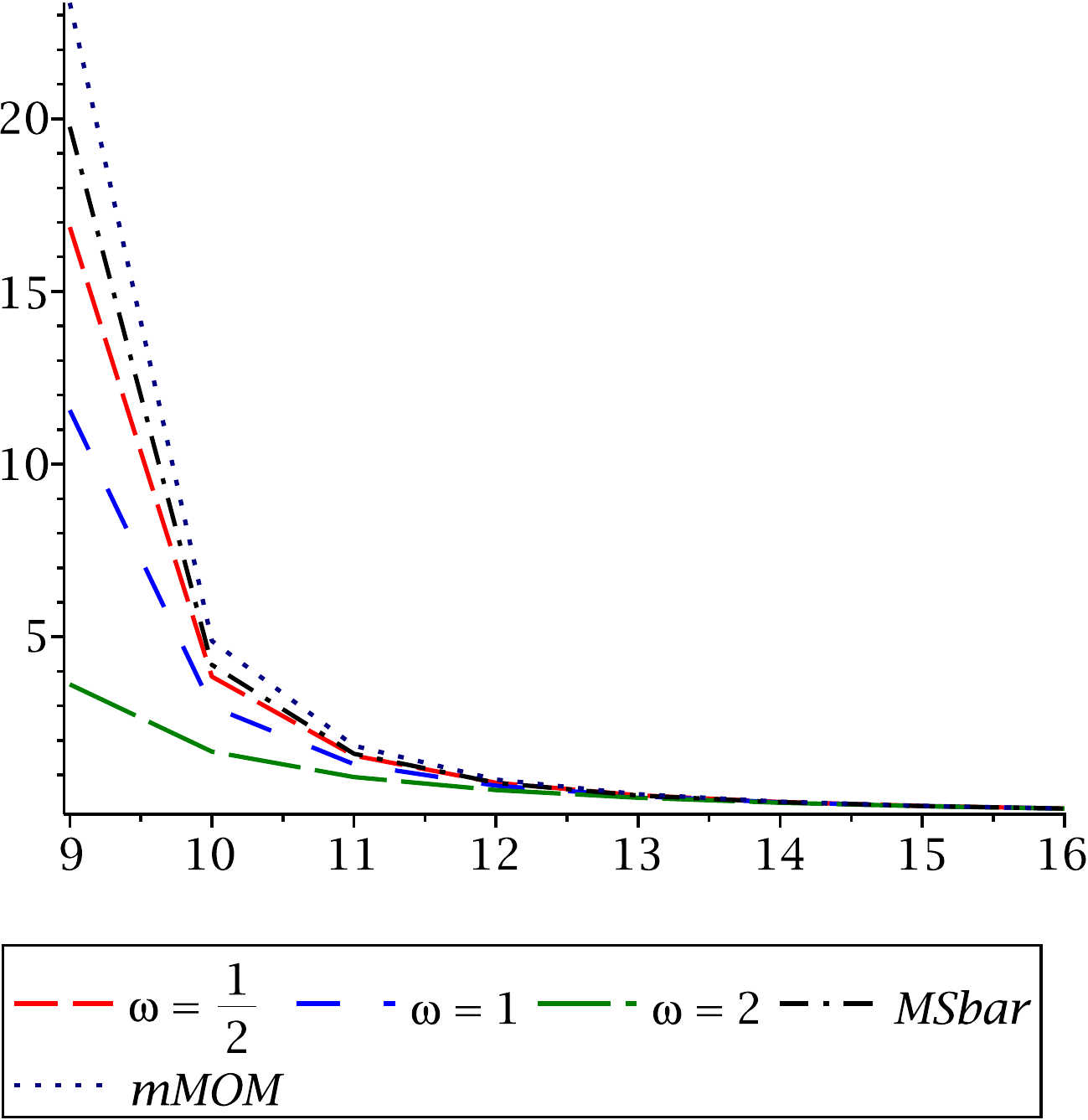}
\quad \quad \quad
\includegraphics[width=7cm,height=7cm]{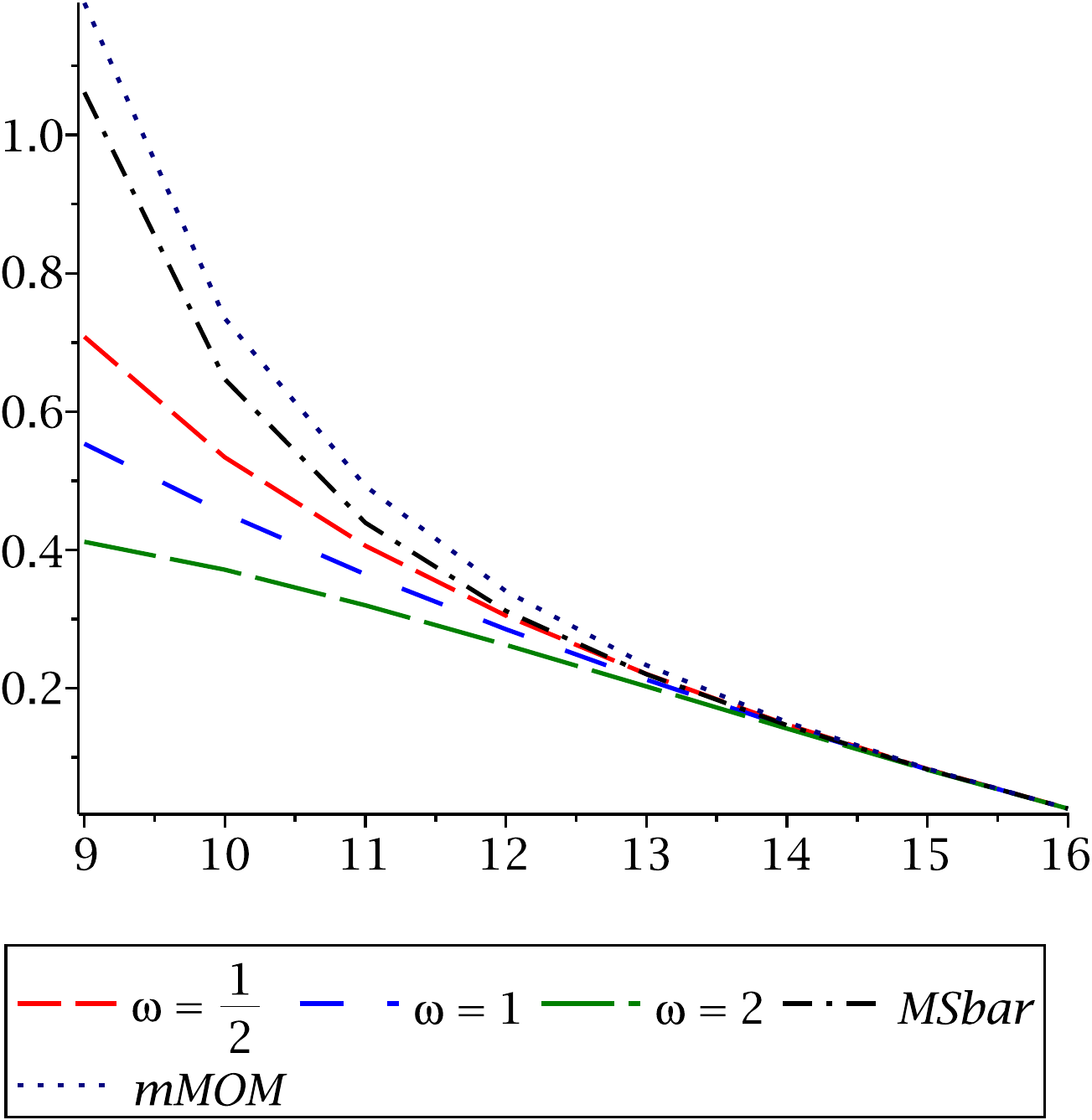}

\vspace{0.5cm}
\includegraphics[width=7cm,height=7cm]{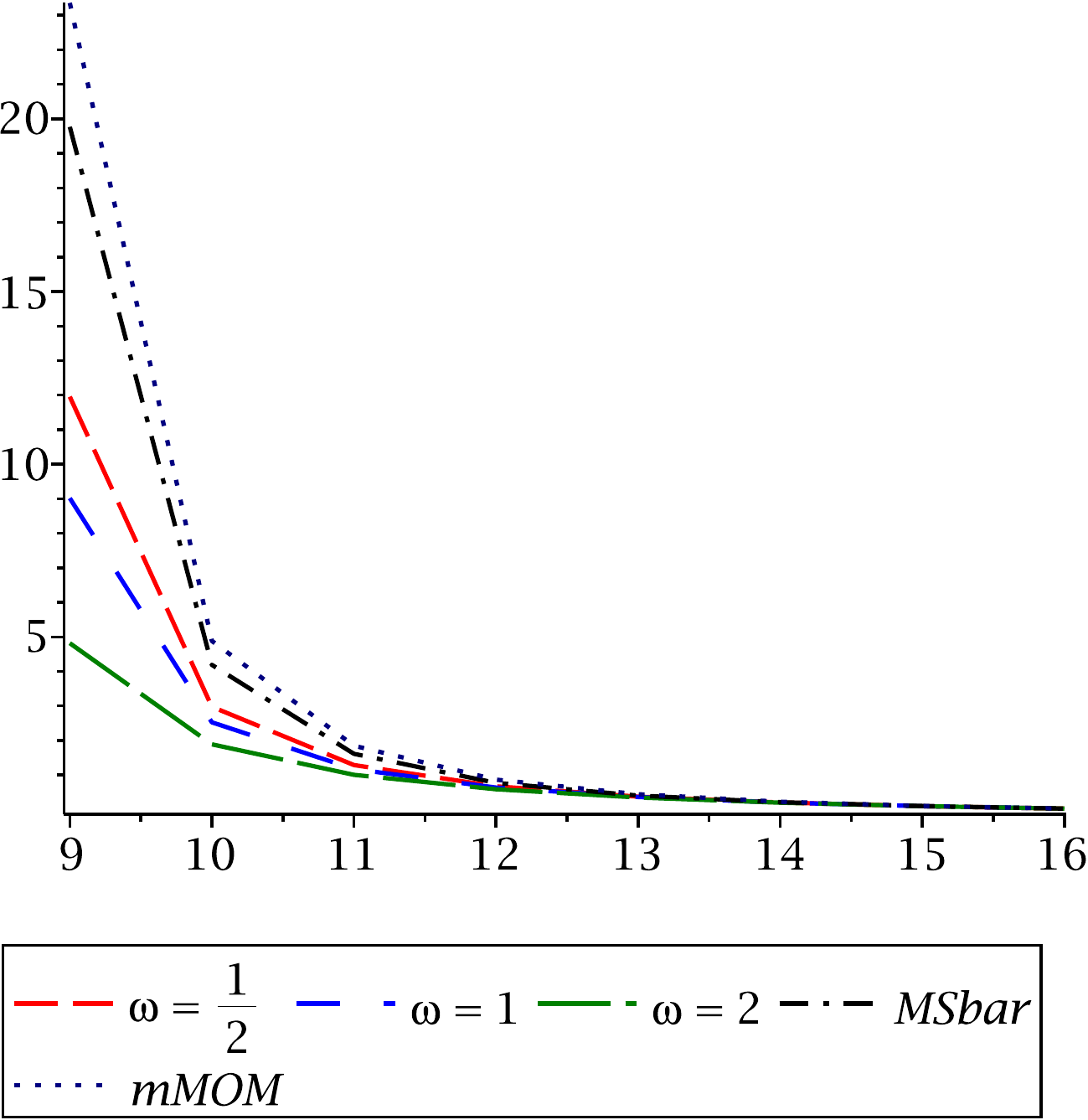}
\quad \quad \quad
\includegraphics[width=7cm,height=7cm]{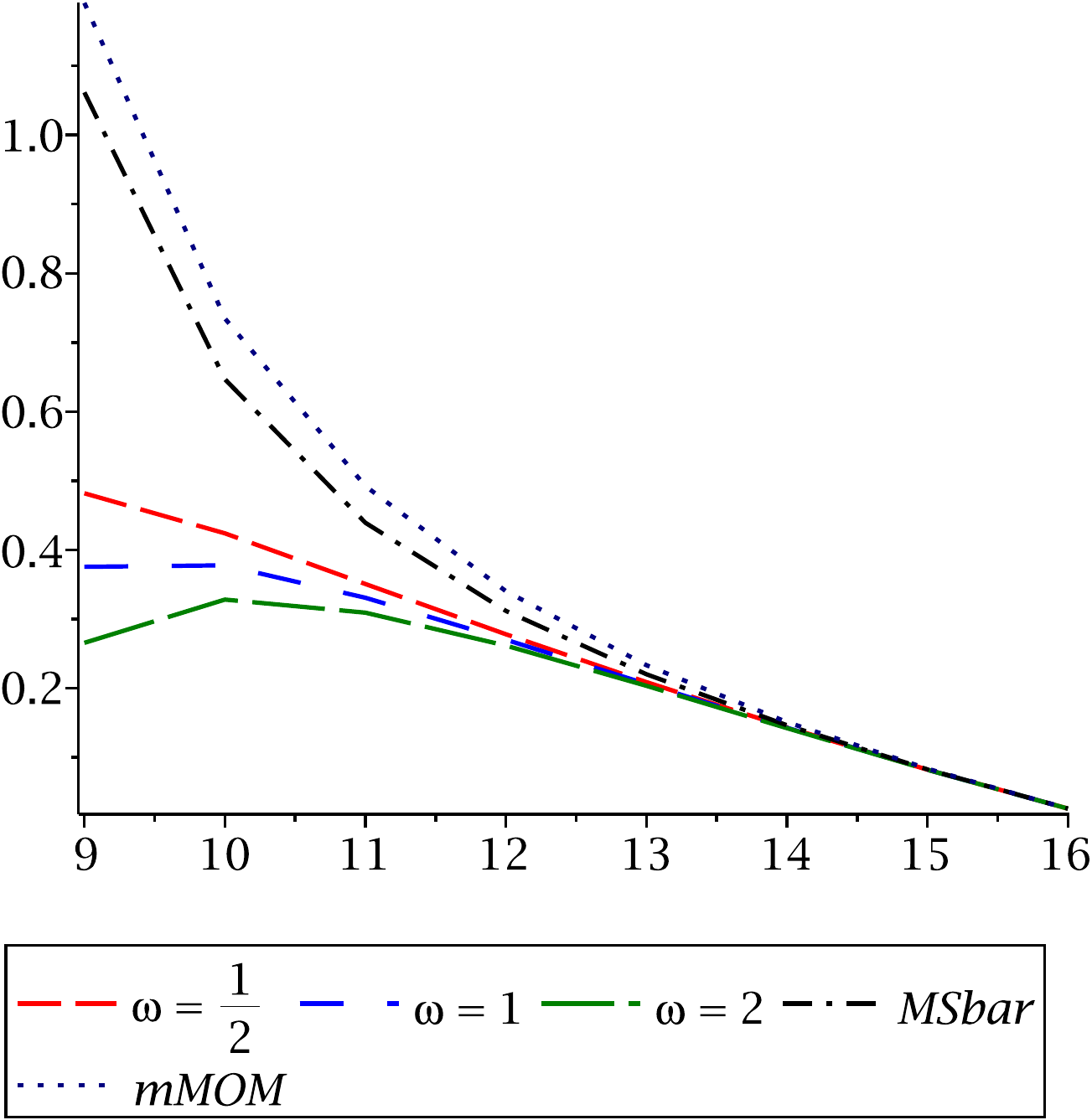}

\vspace{0.5cm}
\includegraphics[width=7cm,height=7cm]{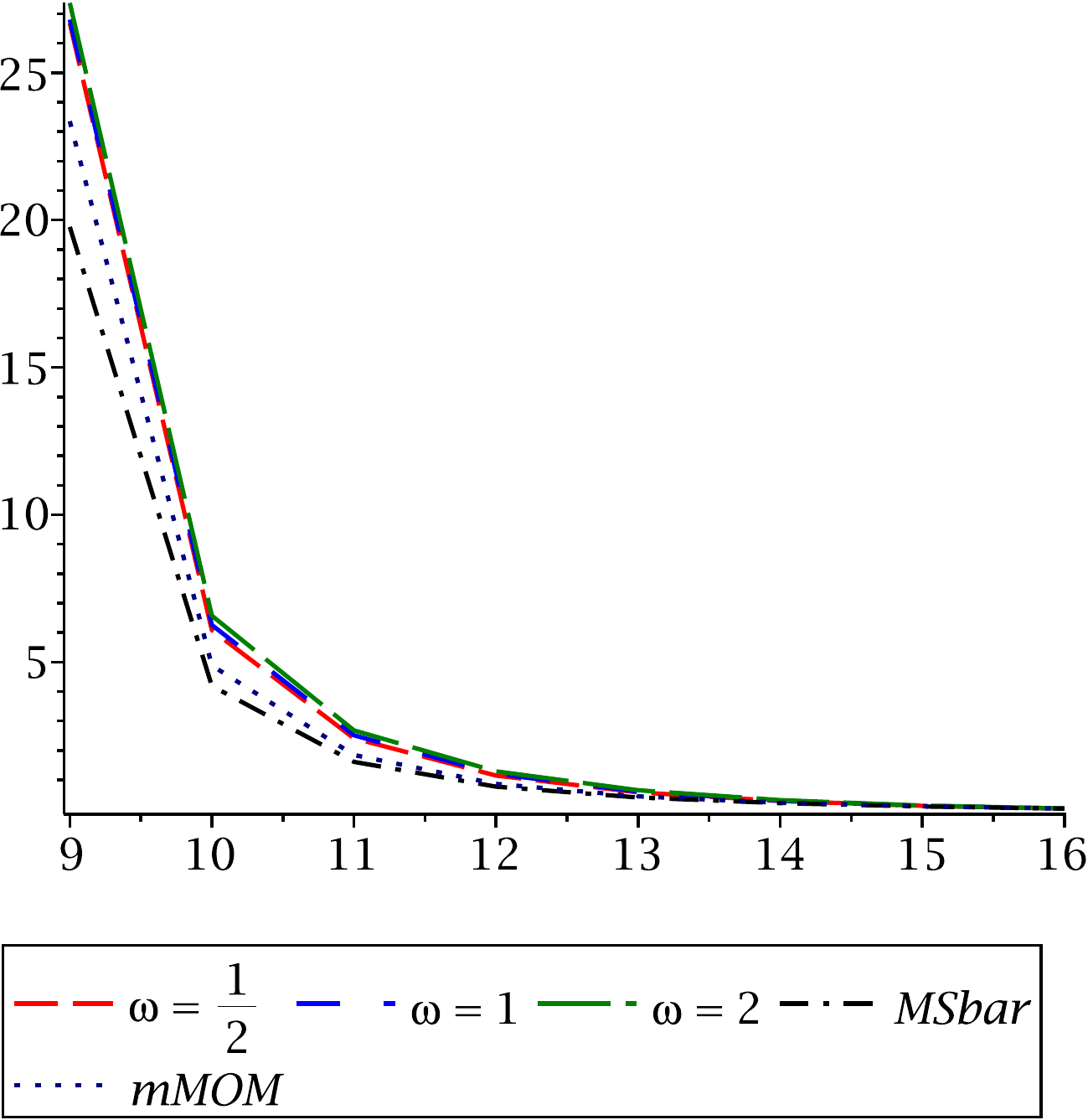}
\quad \quad \quad
\includegraphics[width=7cm,height=7cm]{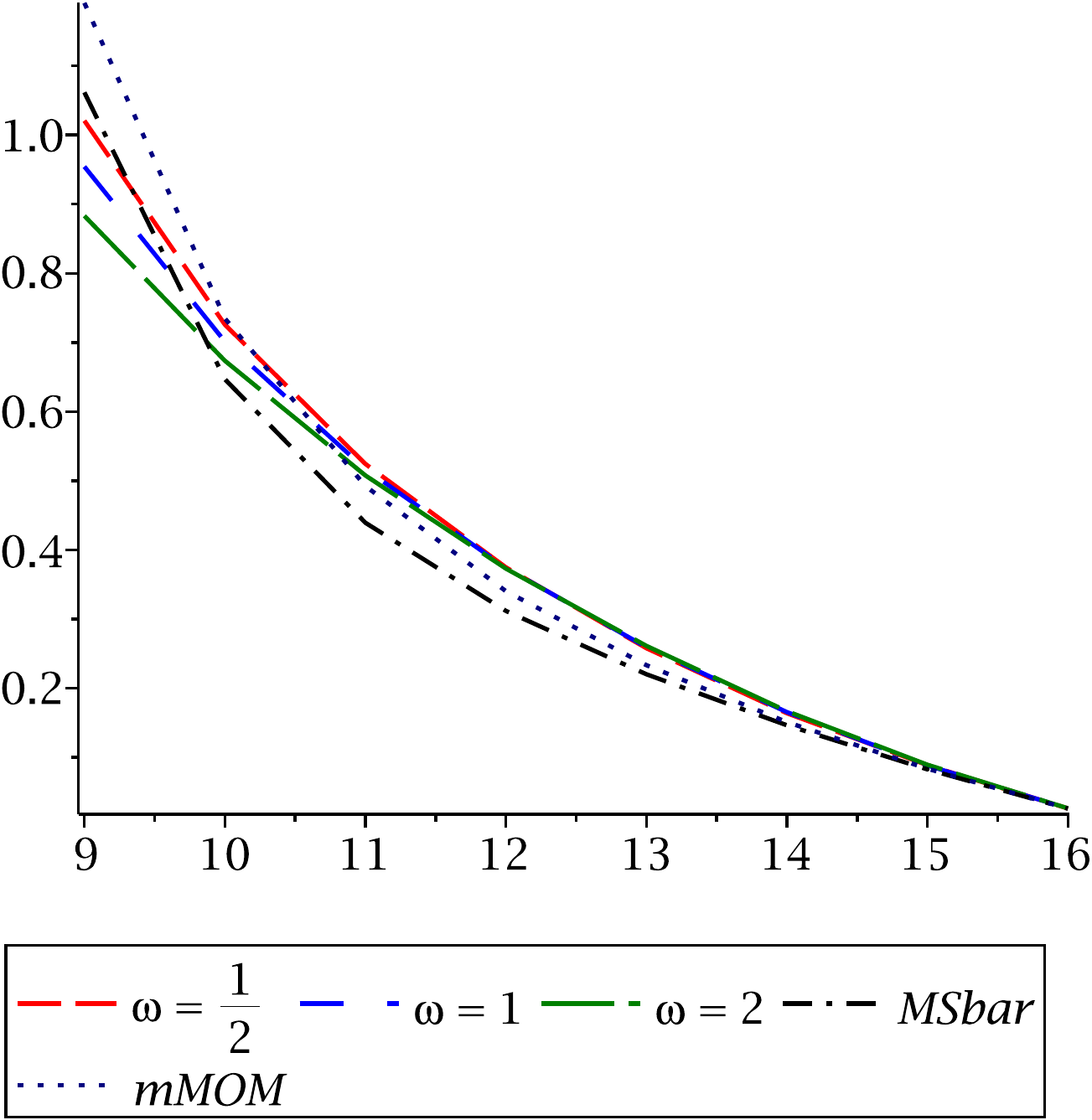}

\caption{Critical exponent $\rho$ for $SU(3)$ at two (left panel) and three 
loops (right panel) for the respective $\iMOMq$, $\iMOMh$ and $\iMOMg$ 
schemes.}
\end{figure}}

\newpage
{\begin{figure}[ht]
\includegraphics[width=7cm,height=7cm]{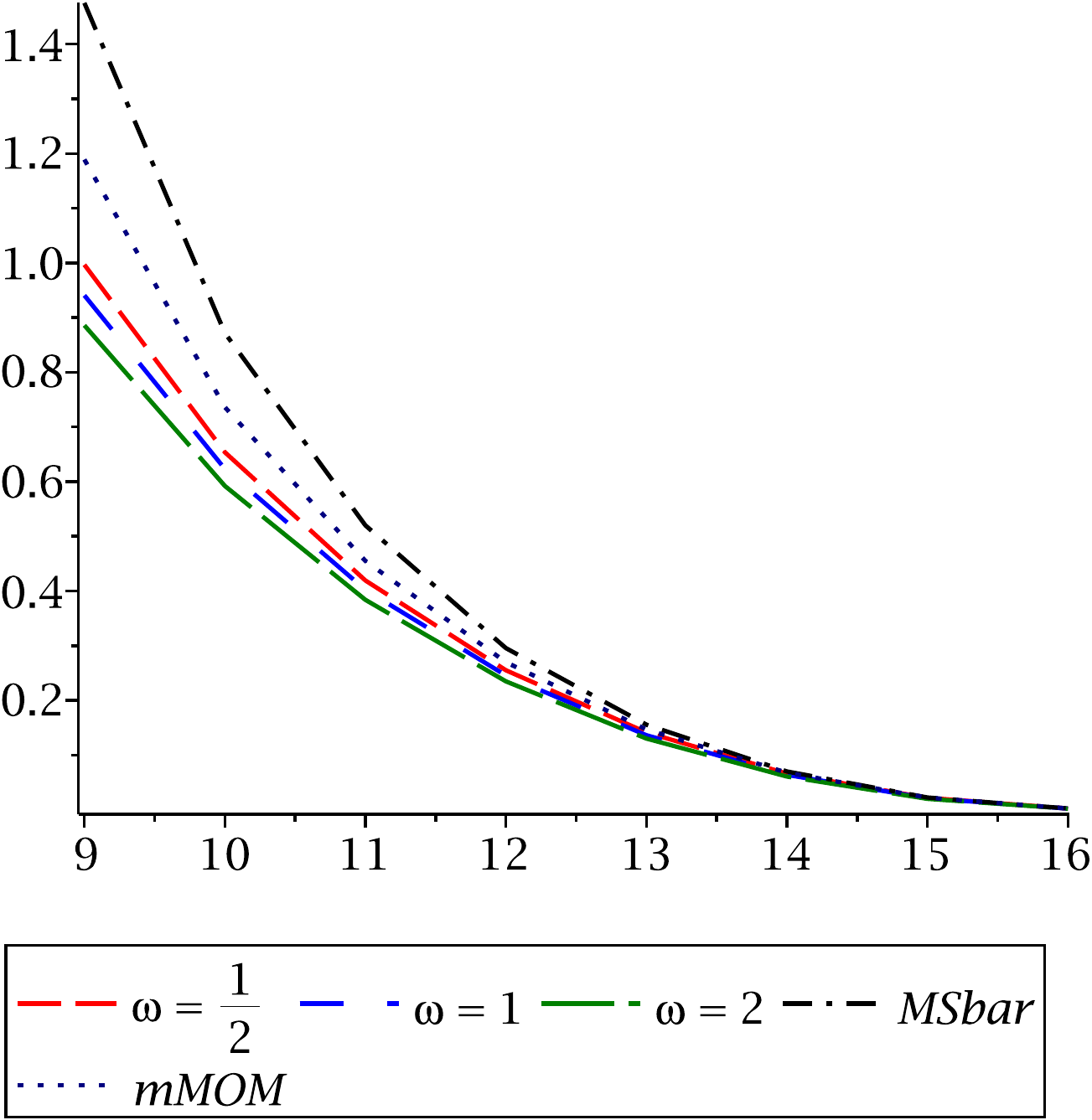}
\quad \quad \quad
\includegraphics[width=7cm,height=7cm]{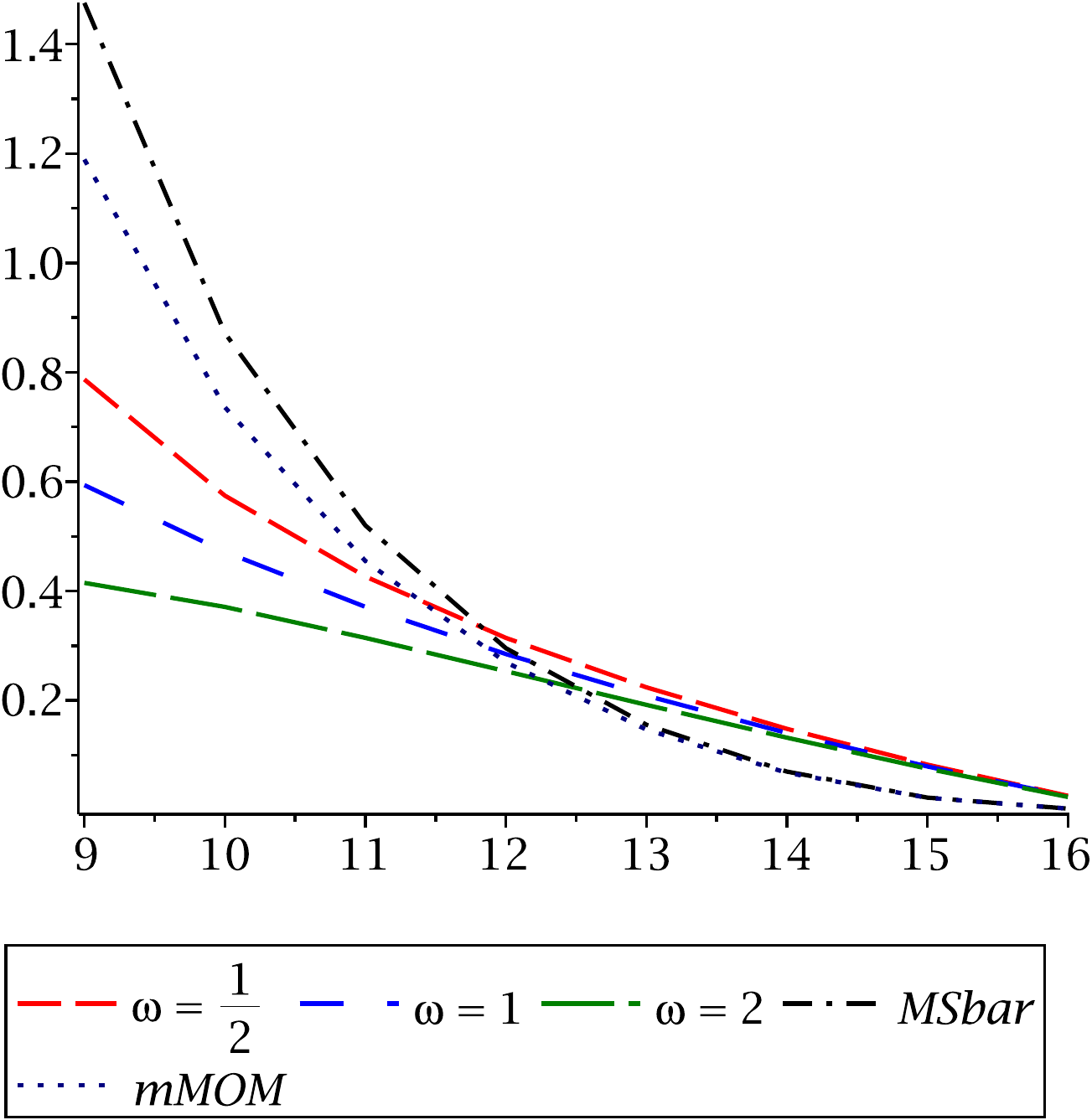}

\vspace{0.5cm}
\includegraphics[width=7cm,height=7cm]{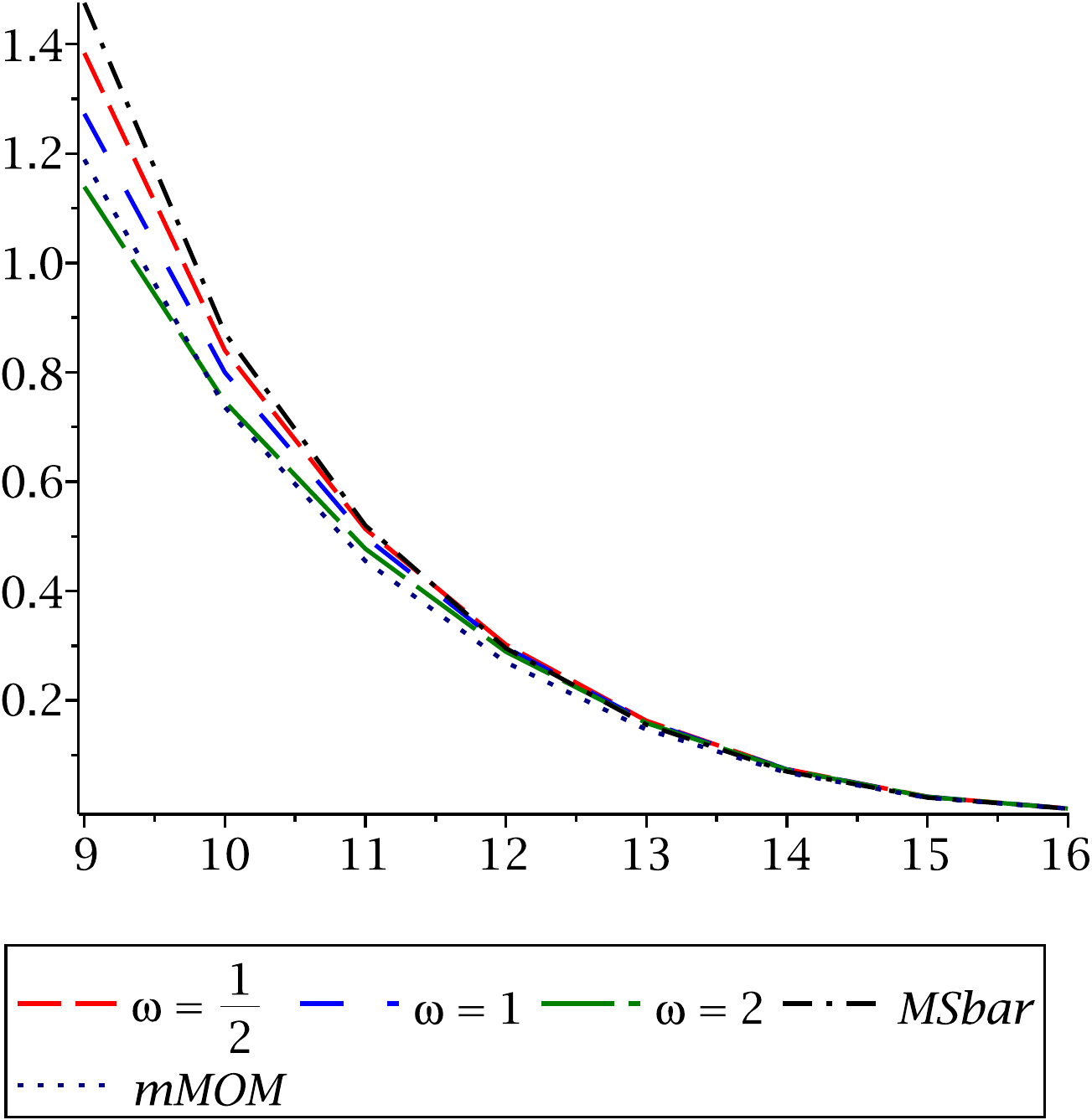}
\quad \quad \quad
\includegraphics[width=7cm,height=7cm]{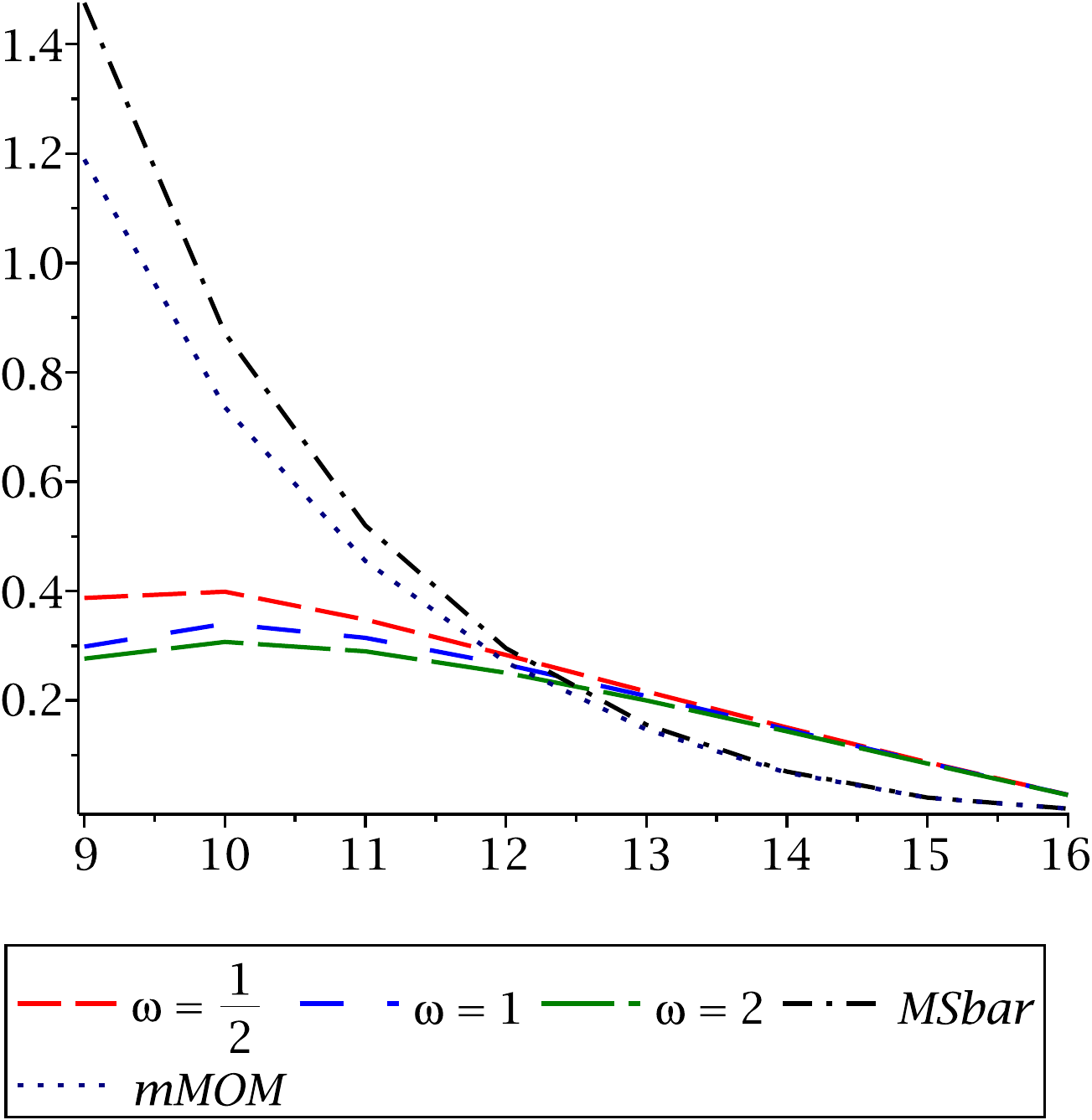}

\vspace{0.5cm}
\includegraphics[width=7cm,height=7cm]{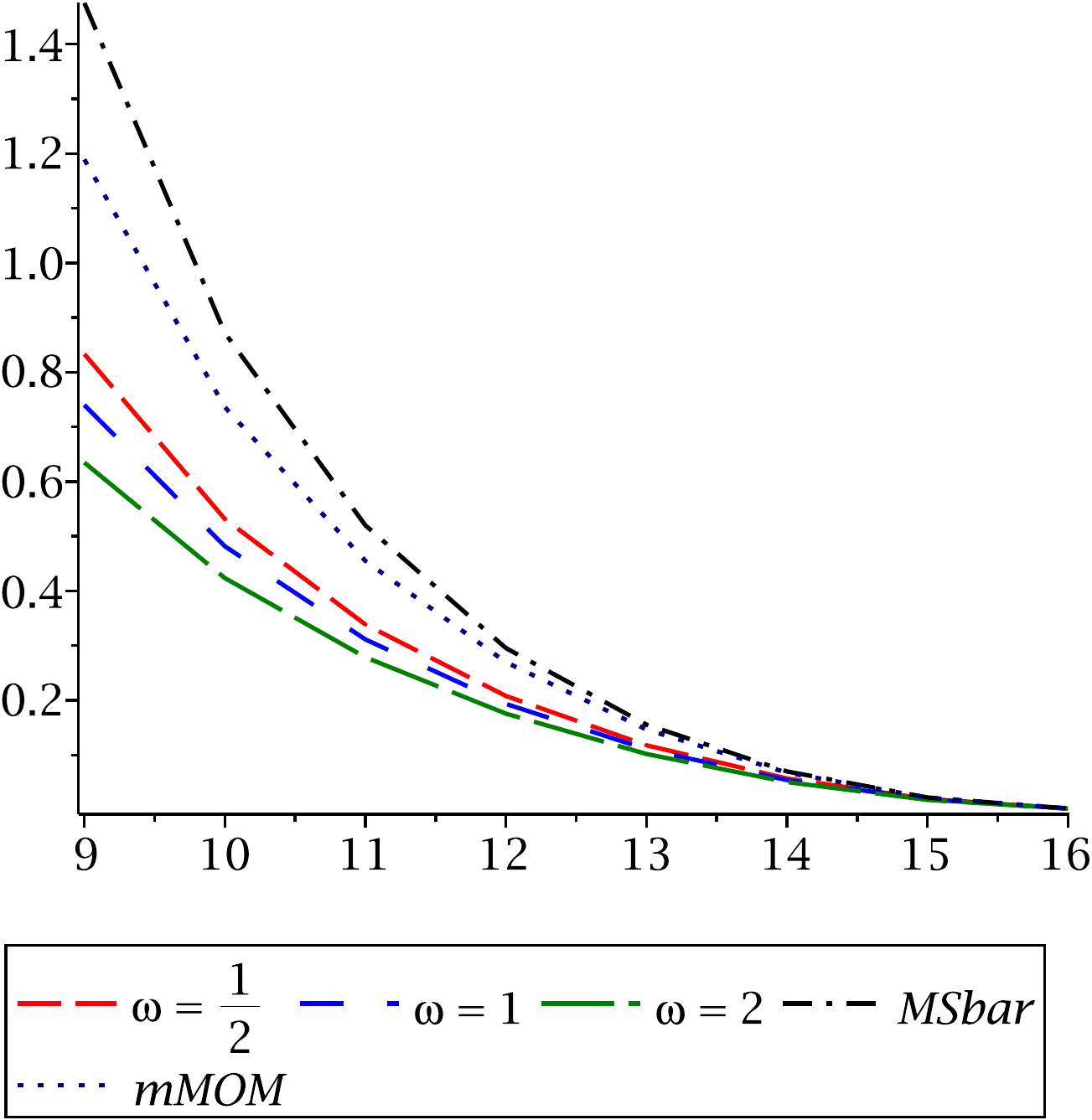}
\quad \quad \quad
\includegraphics[width=7cm,height=7cm]{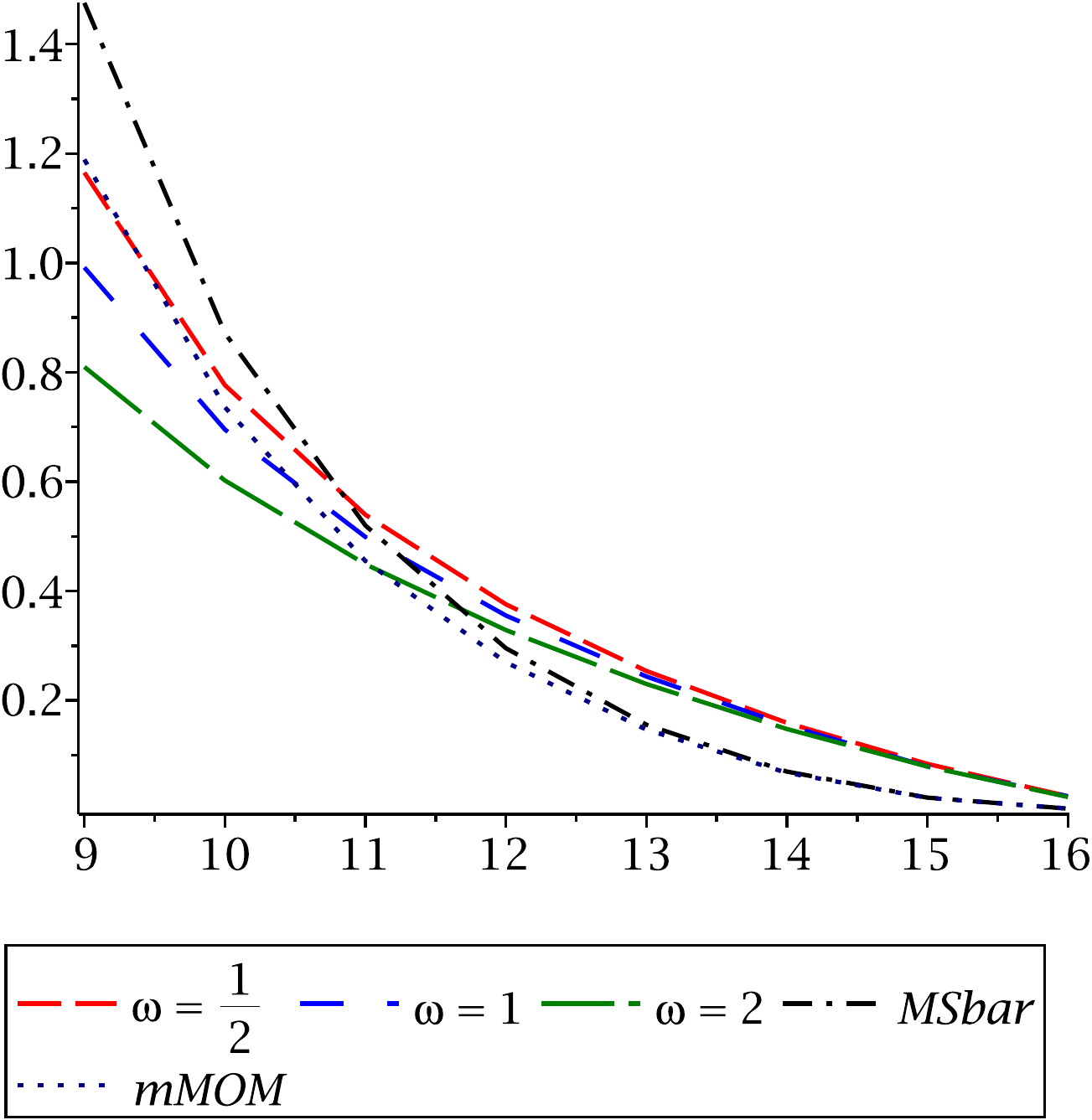}

\caption{Critical exponents ${\widetilde{\omega}}$ (left panel) and $\rho$
(right panel) for $SU(3)$ in the MAG at three loops for the respective 
$\iMOMmq$, $\iMOMmh$ and $\iMOMmg$ schemes.}
\end{figure}}


\begin{thebibliography}{99} 
\bibitem{1} W.A. Bardeen, A.J. Buras, D.W. Duke \& T. Muta, Phys. Rev.
{\bf D18} (1978), 3998.
\bibitem{2} G. 't Hooft \& M. Veltman, Nucl. Phys. {\bf B44} (1972), 189.
\bibitem{3} G. 't Hooft, Nucl. Phys. {\bf B61} (1973), 455.
\bibitem{4} P.A. Baikov, K.G. Chetyrkin \& J.H. K\"{u}hn, Phys. Rev. Lett.
{\bf 118} (2017), 082002.
\bibitem{5} F. Herzog, B. Ruijl, T. Ueda, J.A.M. Vermaseren \& A. Vogt, JHEP
{\bf 1702} (2017), 090.
\bibitem{6} T. Luthe, A. Maier, P. Marquard \& Y. Schr\"{o}der, JHEP
{\bf 1710} (2017), 166.
\bibitem{7} K.G. Chetyrkin, G. Falcioni, F. Herzog \& J.A.M. Vermaseren, JHEP
{\bf 1710} (2017), 179.
\bibitem{8} P.A. Baikov, K.G. Chetyrkin \& J.H. K\"{u}hn, JHEP {\bf 1410}
(2014), 76.
\bibitem{9} T. Luthe, A. Maier, P. Marquard \& Y. Schr\"{o}der, JHEP
{\bf 1701} (2017), 081.
\bibitem{10} T. Luthe, A. Maier, P. Marquard \& Y. Schr\"{o}der, JHEP
{\bf 1703} (2017), 020.
\bibitem{11} P.A. Baikov, K.G. Chetyrkin \& J.H. K\"{u}hn, JHEP {\bf 1704}
(2017), 119.
\bibitem{12} D.J. Gross \& F.J. Wilczek, Phys. Rev. Lett. {\bf 30} (1973), 1343.
\bibitem{13} H.D. Politzer, Phys. Rev. Lett. {\bf 30} (1973), 1346.
\bibitem{14} W.E. Caswell, Phys. Rev. Lett. {\bf 33} (1974), 244.
\bibitem{15} D.R.T. Jones, Nucl. Phys. {\bf B75} (1974), 531.
\bibitem{16} O.V. Tarasov, A.A. Vladimirov \& A.Yu. Zharkov, Phys. Lett.
{\bf B93} (1980), 429.
\bibitem{17} T. van Ritbergen, J.A.M. Vermaseren \& S.A. Larin, Phys. Lett.
{\bf B400} (1997), 379.
\bibitem{18} M. Czakon, Nucl. Phys. {\bf B710} (2005), 485.
\bibitem{19} M. B\"{o}hm, H. Spiesberger \& W. Hollik, Fortsch. Phys. {\bf 34}
(1986), 687.
\bibitem{20} W. Celmaster \& R.J. Gonsalves, Phys. Rev. Lett. {\bf 42} (1979),
1435.
\bibitem{21} W. Celmaster \& R.J. Gonsalves, Phys. Rev. {\bf D20} (1979), 1420.
\bibitem{22} J.A. Gracey, Phys. Rev. {\bf D84} (2011), 085011.
\bibitem{23} A.I. Davydychev, J. Phys. {\bf A25} (1992), 5587.
\bibitem{24} N.I. Usyukina \& A.I. Davydychev, Phys. Atom. Nucl. {\bf 56}
(1993), 1553.
\bibitem{25} N.I. Usyukina \& A.I. Davydychev, Phys. Lett. {\bf B332} (1994),
159.
\bibitem{26} T.G. Birthwright, E.W.N. Glover \& P. Marquard, JHEP {\bf 0409}
(2004), 042.
\bibitem{27} K.G. Wilson \& M.E. Fisher, Phys. Rev. Lett. {\bf 28} (1972), 240.
\bibitem{28} K.G. Wilson, Phys. Rev. {\bf B4} (1971), 3174.
\bibitem{29} K.G. Wilson, Phys. Rev. {\bf B4} (1971), 3184.
\bibitem{30} K.G. Wilson, Phys. Rev. Lett. {\bf 28} (1972), 548.
\bibitem{31} J.A. Gracey \& R.M. Simms, Phys. Rev. {\bf D91} (2015), 085037.
\bibitem{32} T. Banks \& A. Zaks, Nucl. Phys. {\bf B196} (1982), 189.
\bibitem{33} T. Ryttov \& F. Sannino, Phys. Rev. {\bf D76} (2007), 105004.
\bibitem{34} T. Ryttov \& F. Sannino, Int. J. Mod. Phys. {\bf A25} (2010), 
4603.
\bibitem{35} C. Pica \& F. Sannino, Phys. Rev. {\bf D83} (2011), 116001.
\bibitem{36} T. Ryttov \& R. Shrock, Phys. Rev. {\bf D83} (2011), 056011.
\bibitem{37} C. Pica \& F. Sannino, Phys. Rev. {\bf D83} (2011), 035013.
\bibitem{38} T. Ryttov \& R. Shrock, Phys. Rev. {\bf D86} (2012), 065032.
\bibitem{39} T. Ryttov \& R. Shrock, Phys. Rev. {\bf D86} (2012), 085005.
\bibitem{40} R. Shrock, Phys. Rev. {\bf D88} (2013), 036003.
\bibitem{41} T.A. Ryttov, Phys. Rev. {\bf D89} (2014), 016013.
\bibitem{42} T.A. Ryttov, Phys. Rev. {\bf D89} (2014), 056001.
\bibitem{43} T.A. Ryttov, Phys. Rev. {\bf D90} (2014), 056007; Phys. Rev.
{\bf D91} (2015), 039906.
\bibitem{44} G. Choi \& R. Shrock, Phys. Rev. {\bf D90} (2014), 125029.
\bibitem{45} T. Ryttov, Phys. Rev. Lett. {\bf 117} (2016), 071601.
\bibitem{46} T. Ryttov \& R. Shrock, Phys. Rev. {\bf D94} (2016), 105015.
\bibitem{47} T. Ryttov \& R. Shrock, Phys. Rev. {\bf D94} (2016), 105014.
\bibitem{48} T. Ryttov \& R. Shrock, Phys. Rev. {\bf D94} (2016), 125005.
\bibitem{49} T. Ryttov \& R. Shrock, Phys. Rev. {\bf D95} (2017), 105004.
\bibitem{50} T. Ryttov \& R. Shrock, Phys. Rev. {\bf D96} (2017), 105018.
\bibitem{51} M. Gorbahn \& S. J\"{a}ger, Phys. Rev. {\bf D82} (2010), 114001.
\bibitem{52} G. 't Hooft, Nucl. Phys. {\bf B190} (1981), 455.
\bibitem{53} A.S. Kronfeld, G. Schierholz \& U.J. Wiese, Nucl. Phys. {\bf B293}
(1987), 461.
\bibitem{54} A.S. Kronfeld, M.L. Laursen, G. Schierholz \& U.J. Wiese, Phys.
Lett. {\bf B198} (1987), 516.
\bibitem{55} Y. Nambu, Phys. Rev. {\bf D10} (1974), 4262.
\bibitem{56} G. 't Hooft, in High Energy Physics, (Editorice Compositori,
Bologna, 1975).
\bibitem{57} Z.F. Ezawa \& A. Iwazaki, Phys. Rev. {\bf D25} (1982), 2681.
\bibitem{58} H. Min, T. Lee \& P.Y. Pac, Phys. Rev. {\bf D32} (1985), 440.
\bibitem{59} A.R. Fazio, V.E.R. Lemes, M.S. Sarandy \& S.P. Sorella, Phys. Rev.
{\bf D64} (2001), 085003.
\bibitem{60} K.-I. Kondo \& T. Shinohara, Prog. Theor. Phys. {\bf 105} (2001),
649.
\bibitem{61} T. Shinohara, Mod. Phys. Lett. {\bf A18} (2003), 1398.
\bibitem{62} T. Shinohara, T. Imai \& K.-I. Kondo, Int. J. Mod. Phys. {\bf A18}
(2003), 5733.
\bibitem{63} D. Dudal, J.A. Gracey, V.E.R. Lemes, M.S. Sarandy, R.F. Sobreiro,
S.P. Sorella \& H.Verschelde, Phys. Rev. {\bf D70} (2004), 114038.
\bibitem{64} J.A. Gracey, JHEP {\bf 0504} (2005), 012.
\bibitem{65} J.M. Bell \& J.A. Gracey, Phys. Rev. {\bf D88} (2013), 085027.
\bibitem{66} J.M. Bell \& J.A. Gracey, Phys. Rev. {\bf D92} (2015), 125001.
\bibitem{67} S.G. Gorishny, S.A. Larin, L.R. Surguladze \& F.K. Tkachov,
Comput. Phys. Commun. {\bf 55} (1989), 381.
\bibitem{68} S.A. Larin, F.V. Tkachov \& J.A.M. Vermaseren, ``The Form version
of Mincer'', NIKHEF-H-91-18.
\bibitem{69} A.D. Kennedy, J. Math. Phys. {\bf 22} (1981), 1330.
\bibitem{70} A. Bondi, G. Curci, G. Paffuti \& P. Rossi, Ann. Phys. {\bf 199}
(1990), 268.
\bibitem{71} A.N. Vasil'ev, S.\'{E}. Derkachov \& N.A. Kivel, Theor. Math.
Phys. {\bf 103} (1995), 487.
\bibitem{72} A.N. Vasil'ev, M.I. Vyazovskii, S.\'{E}. Derkachov \& N.A. Kivel,
Theor. Math. Phys. {\bf 107} (1996), 441.
\bibitem{73} A.N. Vasil'ev, M.I. Vyazovskii, S.\'{E}. Derkachov \& N.A. Kivel,
Theor. Math. Phys. {\bf 107} (1996), 710.
\bibitem{74} P. Nogueira, J. Comput. Phys. {\bf 105} (1993), 279. 
\bibitem{75} S. Laporta, Int. J. Mod. Phys. {\bf A15} (2000), 5087.
\bibitem{76} C. Studerus, Comput. Phys. Commun. {\bf 181} (2010), 1293.
\bibitem{77} A. von Manteuffel \& C. Studerus, arXiv:1201.4330.
\bibitem{78} J.A.M. Vermaseren, math-ph/0010025.
\bibitem{79} M. Tentyukov \& J.A.M. Vermaseren, Comput. Phys. Commun. {\bf 181}
(2010), 1419.
\bibitem{80} S.A. Larin \& J.A.M. Vermaseren, Phys. Lett. {\bf B303} (1993), 
334. 
\bibitem{81} G. 't Hooft, Acta Universitatis Wratislaviensis {\bf 368} (1976),
345, Proceedings of the 1975 Winter School of Theoretical Physics held in
Karpacz.
\bibitem{82} B.S. DeWitt, in Proceedings of Quantum Gravity II, eds C. Isham,
R. Penrose \& S. Sciama, (Oxford, 1980), 449.
\bibitem{83} D.G. Boulware, Phys. Rev. {\bf D23} (1981), 389.
\bibitem{84} L.F. Abbott, Nucl. Phys. {\bf B185} (1981), 189.
\bibitem{85} D.M. Capper \& A. MacLean, Nucl. Phys. {\bf B203} (1982), 413.
\bibitem{86} O. Nachtmann \& W. Wetzel, Nucl. Phys. {\bf B187} (1981), 333.
\bibitem{87} R. Tarrach, Nucl. Phys. {\bf B183} (1981), 384.
\bibitem{88} O.V. Tarasov, JINR preprint P2-82-900.
\bibitem{89} K.G. Chetyrkin, Phys. Lett. {\bf B404} (1997), 161.
\bibitem{90} J.A.M. Vermaseren, S.A. Larin \& T. van Ritbergen, Phys. Lett.
{\bf B405} (1997), 327.
\bibitem{91} L. von Smekal, K. Maltman \& A. Sternbeck, Phys. Lett. {\bf B681}
(2009), 336.
\bibitem{92} J.C. Taylor, Nucl. Phys. {\bf B33} (1971), 436.
\end{thebibliography}
\end{document}